%% file: main.tex
\documentclass[11pt]{article}
\usepackage[utf8]{inputenc}
\usepackage{hyperref,amsmath,amssymb,natbib,enumerate,amsbsy,amsfonts, xcolor, graphics,graphicx, comment, algpseudocode, algorithm, bbm, booktabs, multirow,url, makecell, float, setspace}
\usepackage[tableposition=above]{caption}
\usepackage{pifont}
\usepackage{tikz}

\newcommand{\bb}{{\bf b}}

\newcommand{\tp}{\tilde p}
\newcommand{\hp}{\hat p}

\newcommand{\bfeta}{\boldsymbol \eta}

\newtheorem{lemma}{Lemma}
\newtheorem{proposition}{Proposition}
\newtheorem{theorem}{Theorem}
\newtheorem{remark}{Remark}
\newtheorem{corollary}{Corollary}

\makeatother

\usepackage{cleveref}

\usepackage{xcolor}
\hypersetup{
    colorlinks,
    linkcolor={black},
    citecolor={black},
    urlcolor={black}
}

\begin{document}

\title{Irrational Exuberance: Correcting Bias in Probability Estimates}
\author{Gareth M. James$^1$, Peter Radchenko$^2$ and Bradley Rava$^{1,3}$} 
\date{}

\footnotetext[1]{Department of Data Sciences and Operations, University of Southern California.}
\footnotetext[2]{University of Sydney.}
\footnotetext[3]{Research is generously supported by the NSF GRFP in Mathematical Statistics.}

\maketitle

\begin{abstract}

We consider the common setting where one observes probability estimates for a large number of events, such as default risks for numerous bonds. Unfortunately, even with unbiased estimates, selecting events corresponding to the most extreme probabilities can result in systematically underestimating the true level of uncertainty. We develop an empirical Bayes approach ``Excess Certainty Adjusted Probabilities" (ECAP), using a variant of Tweedie's formula, which updates probability estimates to correct for selection bias. ECAP is a flexible non-parametric method, which directly estimates the score function associated with the probability estimates, so it does not need to make any restrictive assumptions about the prior on the true probabilities. ECAP also works well in settings where the probability estimates are biased. We demonstrate through theoretical results, simulations, and an analysis of two real world data sets, that ECAP can provide significant improvements over the original probability estimates.  
\end{abstract}


\section{Introduction}

We are increasingly facing a world where automated algorithms are used to generate probabilities, often in real time, for thousands of different events. Just a small handful of examples include finance where rating agencies provide default probabilities on thousands of different risky assets \citep{Keal03, Hull05}; sporting events where each season ESPN and other sites estimate win probabilities for all the games occurring in a given sport \citep{LEUNG2014710}; politics where pundits estimate the probabilities of candidates winning in congressional and state races during a given election season \citep{Silver18, Soumbatiants2006}; or medicine where researchers estimate the survival probabilities of patients undergoing a given medical procedure \citep{Poses97, Smeenk07}. Moreover, with the increasing availability of enormous quantities of data, there are more and more automated probability estimates being generated and consumed by the general public.

Many of these probabilities have significant real world implications. For example, the rating given to a company's bonds will impact their cost of borrowing, or the estimated risk of a medical procedure will affect the patient's likelihood of undertaking the operation. This leads us to question the accuracy of these probability estimates. Let $p_i$ and $\tp_i$ respectively represent the true and estimated probability of $A_i$ occurring for a series of events $A_1,\ldots, A_n$. Then, we often seek an unbiased estimator such that $E(\tilde p_i|p_i)=p_i,$
so $\tp_i$ is neither systematically too high nor too low. Of course, there are many recent examples where this unbiasedness assumption has not held. For example, prior to the financial crisis of 2008 rating agencies systematically under estimated the risk of default for mortgage backed securities so $E(\tp_i|p_i)<p_i$. Similarly, in the lead up to the 2016 US presidential election political pundits significantly underestimated the uncertainty in which candidate would win.

However, even when unbiasedness does hold, using $\tp_i$ as an estimate for $p_i$ can cause significant problems. Consider, for example, a conservative investor who only purchases bonds with extremely low default risk. When presented with $n$ estimated bond default probabilities $\tp_1,\ldots, \tilde p_n$ from a rating agency, she only invests when $\tp_i=0.001$. Let us suppose that the rating agency has done a careful risk assessment, so their probability estimates are unbiased for all $n$ bonds. What then is the fraction of the investor's bonds which will actually default? Given that the estimates are unbiased, one might imagine (and the investor is certainly hoping) that the rate would be close to $0.001$. Unfortunately, the true default rate may be much higher. 

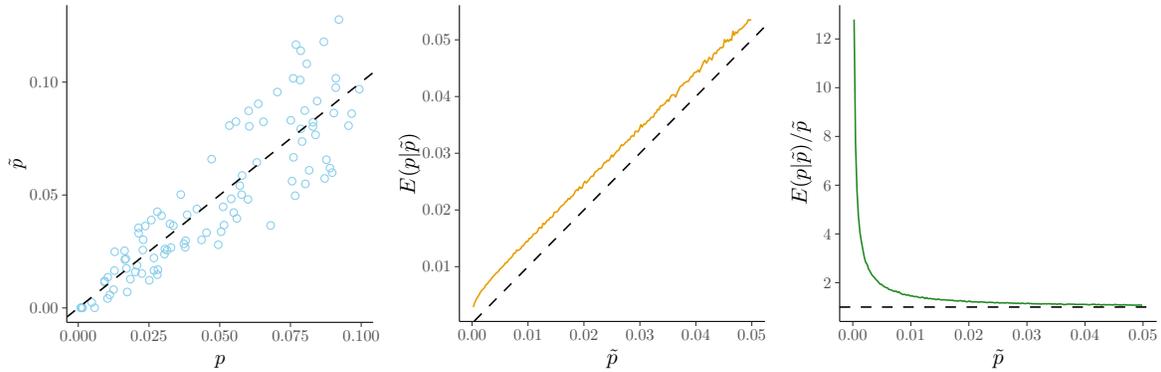
\begin{figure}[tp]
    \centering 
    \scalebox{1}{%
    \input{Figures/intro_plt1.tex}\hfill
    \input{Figures/intro_plt2.tex}\hfill
    \input{Figures/intro_plt3.tex}}
    	\caption{\label{intro.plot}\small Left: Simulated $p_i$ and associated $\tp_i$. The probability estimates are unbiased. Center: The average value of $p_i$, as a function of $\tilde p_i$ i.e. $E(p_i|\tp_i)$ (orange line) is systematically higher than $\tilde p_i$ (dashed line). Right: The ratio of $E(p_i|\tp_i)$ relative to $\tilde p_i$, as a function of $\tilde p_i$. An ideal ratio would be one (dashed line).}
\end{figure}

Figure~\ref{intro.plot} provides an illustration. We first generated a large number of probabilities $p_i$ from a uniform distribution and then produced corresponding $\tp_i$ in such a way that $E(\tp_i|p_i)=p_i$ for $i=1,\ldots, n$. In the left panel of Figure~\ref{intro.plot} we plotted a random sample of $100$ of these probabilities, concentrating on values less than 10\%. While there is some variability in the estimates, there is no evidence of bias in $\tp_i$. In the middle panel we used the simulated data to compute the average value of $p_i$ for any given value of $\tp_i$ i.e. $E(p_i|\tp_i)$. A curious effect is observed. At every point the average value of $p_i$ (orange line) is systematically higher than $\tp_i$ (dashed line) i.e. $E(p_i|\tp_i)>\tp_i$. Finally, in the right panel we have plotted the ratio of $E(p_i|\tp_i)$ to $\tp_i$. Ideally this ratio should be approximately one, which would, for example, correspond to the true risk of a set of bonds equalling the estimated risk. However, for small values of $\tp_i$ we observe ratios far higher than one. So, for example, our investor who only purchases bonds with an estimated default risk of  $\tp_i=0.001$ will in fact find that $0.004$ of her bonds end up defaulting, a 400\% higher risk level than she intended to take!


These somewhat surprising results are not a consequence of this particular simulation setting. It is in fact an instance of selection bias, a well known issue which occurs when the selection of observations is made in such a way, e.g. selecting the most extreme observations, that they can no longer be considered random samples from the underlying population. If this bias is not taken into account then any future analyses will provide a distorted estimate of the population. Consider the setting where we observe $X_1,\ldots, X_n$ with $E(X_i)=\mu_i$ and wish to estimate $\mu_i$ based on an observed $X_i$. Then it is well known that the conditional expectation $E(\mu_i|X_i)$ corrects for any selection bias associated with choosing $X_i$ in a non-random fashion \citep{efron2011}. Numerous approaches have been suggested to address selection bias, with most methods imposing some form of shrinkage to either explicitly, or implicitly, estimate $E(\mu_i|X_i)$. Among linear shrinkage methods, the James-Stein estimator \citep{james1961} is the most well known, although many others exist \citep{efron1975, ikeda2016}. There are also other popular classes of methods, including: non-linear approaches utilizing sparse priors \citep{donoho1994, abramovich2006, bickel2008, ledoit2012}, Bayesian estimators \citep{gelman2012} and empirical Bayes methods \citep{jiang2009, brown2009nonparametric, petrone2014}. 

For Gaussian data, Tweedie's formula \citep{Rob1956} provides an elegant empirical Bayes estimate for $E(\mu_i|X_i)$, using only the marginal distribution of $X_i$. While less well known than the James-Stein estimator, it has been shown to be an effective non-parametric approach for addressing selection bias \citep{efron2011}. The approach can be automatically adjusted to lean more heavily on parametric assumptions when little data is available, but in settings such as ours, where large quantities of data have been observed, it provides a highly flexible non-parametric shrinkage method \citep{benjamini2005, henderson2015}. 

However, the standard implementation of Tweedie's formula assumes that, conditional on $\mu_i$, the observed data follow a Gaussian distribution. Most shrinkage methods make similar distributional assumptions or else model the data as unbounded, which makes little sense for probabilities.
What then would be a better estimator for low probability events? In this paper we propose an empirical Bayes approach, called ``Excess Certainty Adjusted Probability" (ECAP), specifically designed for probability estimation in settings with a large number of observations. ECAP uses a variant of Tweedie's formula which models $\tp_i$ as coming from a beta distribution, automatically ensuring the estimate is bounded between $0$ and $1$. We provide theoretical and empirical evidence demonstrating that the ECAP estimate is generally significantly more accurate than $\tp_i$.



This paper makes three key contributions. First, we convincingly demonstrate that even an unbiased estimator $\tp_i$ can provide a systematically sub-optimal estimate for $p_i$, especially in situations where large numbers of probability estimates have been generated. This leads us to develop the oracle estimator for $p_i$, which results in a substantial improvement in expected loss. Second, we introduce the ECAP method which estimates the oracle. ECAP does not need to make any assumptions about the distribution of $p_i$. Instead, it relies on estimating the marginal distribution of $\tp_i$, a relatively easy problem in the increasingly common situation where we observe a large number of probability estimates. Finally, we extend ECAP to the biased data setting where $\tp_i$ represents a biased observation of $p_i$ and show that even in this setting we are able to recover systematically superior estimates of $p_i$.

The paper is structured as follows. In Section~\ref{method.sec} we first formulate a model for $\tp_i$ and a loss function for estimating $p_i$. We then provide a closed form expression for the corresponding oracle estimator and its associated reduction in expected loss. 
We conclude Section~\ref{method.sec} by proposing the ECAP estimator for the oracle and deriving its theoretical properties.
Section~\ref{extension.sec} provides two extensions. First, we propose a bias corrected version of ECAP, which can detect situations where $\tp_i$ is a biased estimator for $p_i$ and automatically adjust for the bias. Second, we generalize the ECAP model from Section~\ref{method.sec}.
Next, Section~\ref{sim.sec} contains results from an extensive simulation study that examines how well ECAP works to estimate $p_i$, in both the unbiased and biased settings. Section~\ref{emp.sec} illustrates ECAP on two interesting real world data sets. The first is a unique set of probabilities from ESPN predicting, in real time, the winner of various NCAA football games, and the second contains the win probabilities of all candidates in the 2018 US midterm elections. We conclude with a discussion and possible future extensions in Section~\ref{discussion.sec}. Proofs of all theorems are provided in the appendix.

\section{Methodology}
\label{method.sec}

Let $\tp_1,\ldots, \tp_n$ represent initial estimates of events $A_1,\ldots, A_n$ occurring. In practice, we assume that $\tp_1,\ldots, \tp_n$ have already been generated, by previous analysis or externally, say, by an outside rating agency in the case of the investment example. 
Our goal is to construct estimators $\hp_1(\tp_1), \ldots, \hp_n(\tp_n)$ which provide more accurate estimates for $p_1,\ldots, p_n$.
In order to derive the estimator we first choose a model for $\tp_i$ and select a loss function for $\hp_i$, which allows us to compute the corresponding oracle estimator $p_{i0}$. Finally, we provide an approach for generating an estimator for the oracle $\hp_i$. In this section we only consider the setting where $\tp_i$ is assumed to be an unbiased estimator for $p_i$. We extend our approach to the more general setting where $\tp_i$ may be a biased estimator in Section~\ref{biased.sec}.

\subsection{Modeling $\tp_i$ and Selecting a Loss Function}

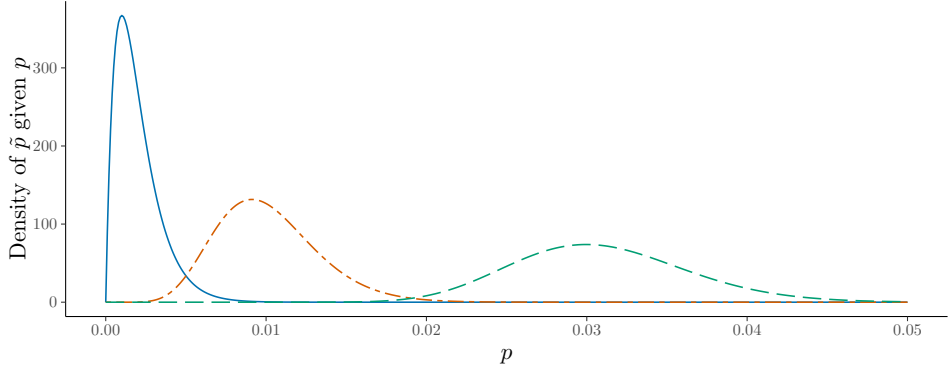
\begin{figure}[t!]
	\centering
	 \scalebox{1}{%
    \input{Figures/density_plt.tex}}
	\caption{\label{beta.plot}\small Density functions for $\tp_i$ given $p_i=0.002$ (blue / solid), $p_i=0.01$ (orange / dot-dashed), and $p_i=0.03$ (green / dashed). In all three cases $\gamma^*=0.001$.}
\end{figure}

Given that $\tilde p_i$ is a probability, we model its conditional distribution using the beta distribution\footnote{We consider a more general class of distributions for $\tp_i$ in Section~\ref{sec.mixture}}. In particular, we model
\begin{equation}
\label{beta.model}
    \tp_i|p_i \sim Beta(\alpha_i, \beta_i),\quad\text{where} \quad \alpha_i=\frac{p_i}{\gamma^*}, \quad\beta_i=\frac{1-p_i}{\gamma^*},
\end{equation} and $\gamma^*$ is a fixed parameter which influences the variance of $\tilde p_i$. 
Under \eqref{beta.model}, 
\begin{equation}
\label{pt.mean.var}
E(\tp_i|p_i)=p_i \quad\text{and} \quad Var(\tp_i|p_i)=\frac{\gamma^*}{1+\gamma^*}p_i(1-p_i),
\end{equation}
so 
$\tilde p_i$ is an unbiased estimate for $p_i$, 
which becomes more accurate as $\gamma^*$ tends to zero. Figure~\ref{beta.plot} provides an illustration of the density function of $\tp_i$ for three different values of $p_i$. In principle, this model could be extended to incorporate observation specific variance terms $\gamma_i^*$. Unfortunately, in practice $\gamma^*$ needs to be estimated, which would be challenging if we assumed a separate term for each observation. However, in some settings it may be reasonable to model $\gamma_i^*=w_i \gamma^*$, where $w_i$ is a known weighting term, in which case only one parameter needs to be estimated.

Next, we select a loss function for our estimator to minimize. One potential option would be to use a standard squared error loss, $L(\hp_i) = E\left(p_i-\hp_i\right)^2$.
However, this loss function is not the most reasonable approach in this setting. Consider for example the event corresponding to a bond defaulting, or a patient dying during surgery. If the bond has junk status, or the surgery is highly risky, the true probability of default or death might be $p_i=0.26$, in which case an estimate of $\hp_i=0.25$ would be considered very accurate. It is unlikely that an investor or patient would have made a different decision if they had instead been provided with the true probability of $0.26$. 

However, if the bond, or surgery, are considered very safe we might provide an estimated probability of $\hp_i=0.0001$, when the true probability is somewhat higher at $p_i=0.01$. The absolute error in the estimate is actually slightly lower in this case, but the patient or investor might well make a very different decision when given a $1\%$ probability of a negative outcome vs a one in ten thousand chance. 

In this sense, the error between $p_i$ and $\hp_i$ {\it as a percentage of $\hp_i$} 
is a far more meaningful measure of precision. In the first example we have a percentage error of only 4\%, while in the second instance the percentage error is almost 10,000\%,
indicating a far more risky proposition. To capture this concept of relative error we introduce as our measure of accuracy a quantity we call the ``Excess Certainty", which is defined as
\begin{equation}
\label{EC}
    \text{EC}(\hp_i)= \frac{p_i-\hp_i}{\min(\hp_i,1-\hp_i)}.
\end{equation}
In the first example $\text{EC}=0.04$, while in the second example $\text{EC}=99$.  Note, we include $\hat p_i$ in the denominator rather than $p_i$ because we wish to more heavily penalize settings where the estimated risk is far lower than the true risk (irrational exuberance) compared to the alternative where true risk is much lower. 

Ideally, the excess certainty of any probability estimate should be very close to zero. Thus, we adopt the following expected loss function,
\begin{equation}
\label{loss.fn}
    L(\hat p_i, \tp_i)=E_{p_i}\left(\text{EC}(\hp_i)^2|\tp_i\right),
\end{equation}
where the expectation is taken over $p_i$, conditional on $\tp_i$. Our aim is to produce an estimator $\hp_i$ that minimizes \eqref{loss.fn} conditional on the observed value of~$\tp_i$. It is worth noting that if our goal was solely to remove selection bias then we could simply compute $E(p_i|\tp_i)$, which would be equivalent to minimizing $E\left[\left(p_i-\hp_i\right)^2|\tp_i\right]$. Minimizing \eqref{loss.fn} generates a similar shrinkage estimator, which also removes the selection bias, but, as we discuss in the next section, it actually provides additional shrinkage to account for the fact that we wish to minimize the relative, or percentage, error.

\subsection{The Oracle Estimator}

We now derive the oracle estimator, $p_{i0}$, which minimizes the loss function given by \eqref{loss.fn},
\begin{equation}
\label{argmin}
   p_{i0}=\arg\min_a E_{p_i}\left[\text{EC}(a)^2|\tp_i\right].
\end{equation}
Our ECAP estimate aims to approximate the oracle. Theorem~\ref{oracle.thm} below provides a relatively simple closed form expression for $p_{i0}$ and a bound on the minimum reduction in loss from using $p_{i0}$ relative to any other estimator.
\begin{theorem}
\label{oracle.thm}
For any distribution of $\tp_i$,
\begin{equation}
\label{oracle.general}
   p_{i0}=\begin{cases}
   \min\left(E(p_i|\tp_i)+ \frac{Var(p_i|\tp_i)}{E(p_i|\tp_i)}\,,\,0.5\right), & E(p_i|\tp_i)\le 0.5\\
   \max\left(0.5\,,\,E(p_i|\tp_i)- \frac{Var(p_i|\tp_i)}{1-E(p_i|\tp_i)}\right), & E(p_i|\tp_i)>0.5.
   \end{cases}
\end{equation}

Furthermore, for any $p'_i\ne p_{i0}$,
\begin{equation}
\label{L.diff}
    L(p'_i,\tp_i) - L(p_{i0},\tp_i) \ge \begin{cases} E\left(p_i^2|\tp_i\right)\left[\frac{1}{p'_i}-\frac{1}{p_{i0}}\right]^2, &p_{i0}\le 0.5\\
    E\left([1-p_i]^2|\tp_i\right)\left[\frac{1}{1-p'_i}-\frac{1}{1-p_{i0}}\right]^2,&p_{i0}\ge0.5.
    \end{cases}
\end{equation}
\end{theorem}
\begin{remark}
Note that both bounds in \eqref{L.diff} are valid when $p_{i0}=0.5$.
\end{remark}


%
We observe from this result that the oracle estimator starts with the conditional expectation $E(p_i|\tp_i)$ and then shifts the estimate towards $0.5$ by an amount $\frac{Var(p_i|\tp_i)}{\min(E(p_i|\tp_i),1-E(p_i|\tp_i))}$. However, if this would move the estimate past $0.5$ then the estimator simply becomes $0.5$.

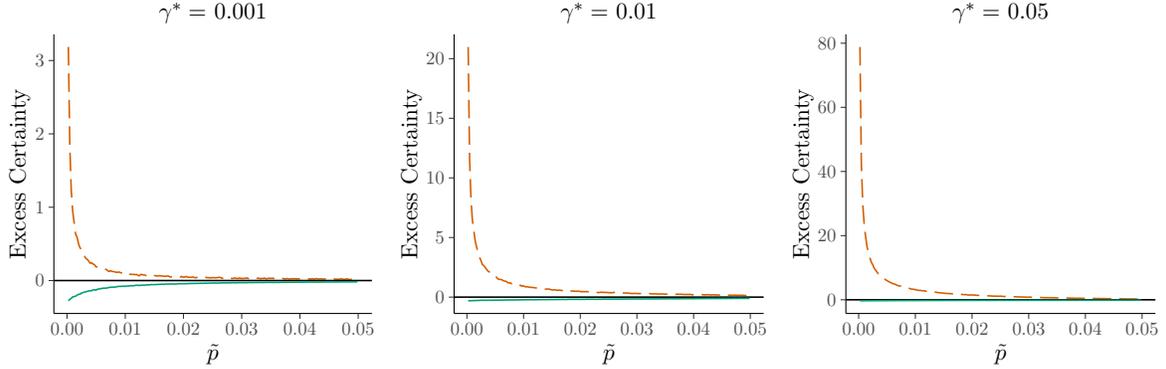
\begin{figure}[t]
    \centering
    \vspace{.5cm}
     \scalebox{1}{%
    \input{Figures/ec_plt1.tex}\hfill
    \input{Figures/ec_plt2.tex}\hfill
    \input{Figures/ec_plt3.tex}}
    	\caption{\label{EC.plot}\small Average excess certainty as a function of $\tp_i$ for three different values of $\gamma^*$ (orange / dashed line). All plots exhibit excess certainty far above zero but the issue grows worse as $\gamma^*$ gets larger, corresponding to more variance in $\tp_i$. The green (solid) line in each plot corresponds to the average excess certainty for the oracle estimator $p_{i0}$.}
\end{figure}

Figure~\ref{EC.plot} plots the average excess certainty \eqref{EC} from using $\tp_i$ to estimate $p_i$ (orange lines) and from using $p_{i0}$ to estimate $p_i$ (green lines), for three different values of $\gamma^*$. Recall that an ideal EC should be zero, but the observed values for $\tp_i$ are far larger, especially for higher values of $\gamma^*$ and lower values of $\tp_i$. Note that, as a consequence of the minimization of the expected squared loss function~\eqref{loss.fn}, the oracle is slightly conservative with a  negative EC, which is due to the variance term in~\eqref{oracle.general}. 

It is worth noting that Theorem~\ref{oracle.thm} applies for any distribution of $\tp_i|p_i$ and does not rely on our model \eqref{beta.model}. If we further assume that \eqref{beta.model} holds, then Theorem~\ref{EandVar}  provides explicit forms for $E(p_i|\tp_i)$ and $Var(p_i|\tp_i)$.

\begin{theorem}
\label{EandVar}
Under~\eqref{beta.model},
\begin{eqnarray}
\label{mui}E(p_i|\tp_i) &=&\mu_i \ \equiv \ \tp_i + \gamma^*\left[g^*(\tp_i)+1-2\tp_i\right]\\
\label{sigmai}
Var(p_i|\tp_i)&=& \sigma^2_i\ \equiv \ \gamma^*\tp_i(1-\tp_i) + {\gamma^*}^2\tp_i(1-\tp_i)\left[{g^*}'(\tp_i)-2\right],
\end{eqnarray}
where $g^*(\tp_i) = \tp_i(1-\tp_i) v^*(\tp_i)$, $v^*(\tilde p_i)=\frac{\partial}{\partial \tilde p_i} \log f^*(\tilde p_i)$ is the score function of $\tilde p_i$ and $f^*(\tilde p_i)$ is the marginal density of $\tilde p_i$. 
%
\end{theorem}
If we also assume that the distribution of $p_i$ is symmetric then further simplifications are possible.
\begin{corollary}
\label{cor.EandVar}
If the prior distribution of $p_i$ is symmetric about $0.5$, then
\begin{equation}
\label{oracle}
   p_{i0}=\begin{cases}
   \min\left(E(p_i|\tp_i)+ \frac{Var(p_i|\tp_i)}{E(p_i|\tp_i)}\,,\,0.5\right), & \tp_i\le 0.5\\
   \max\left(0.5\,,\,E(p_i|\tp_i)- \frac{Var(p_i|\tp_i)}{1-E(p_i|\tp_i)}\right), & \tp_i>0.5,
   \end{cases}
\end{equation}
\begin{equation}
\label{g}
    g^*(0.5)=0,\quad \text{and} \quad g^*(\tp_i)=-g^*(1-\tp_i).
\end{equation}
\end{corollary}




A particularly appealing aspect of Theorem~\ref{EandVar} and its corollary is that $g^*(\tp_i)$ is only a function of the marginal distribution of $\tp_i$, so that it can be estimated directly using the observed probabilities~$\tp_i$. In particular, we do not need to make any assumptions about the distribution of~$p_i$ in order to compute $g^*(\tp_i)$.



%


\subsection{Estimation}
\label{score.sec}

In order to estimate $p_{i0}$ we must form estimates for $g^*(\tp_i)$, its derivative ${g^*}'(t)$, and $\gamma^*$.   
%

\subsubsection{Estimation of $g$}
\label{sec.estimation}
Let $\hat g(\tp)$ represent our estimator of $g^*(\tp)$. Given that $g^*(\tp)$ is a function of the marginal distribution of $\tp_i$, i.e. $f^*(\tp_i)$, then one could estimate $g^*(\tp_i)$ by $\tp_i(1-\tp_i)\hat f'(\tp_i)/\hat f(\tp_i)$, where $\hat f(\tp_i)$ and $\hat f'(\tp_i)$ are respectively estimates for the marginal distribution of $\tp_i$ and its derivative. However, this approach requires dividing by the estimated density function, which can produce a highly unstable estimate in the boundary points, precisely the region we are most interested in.

Instead we directly estimate $g^*(\tp)$ by choosing $\hat g(\tp)$ so as to minimize the risk function, which is defined as $R(g)=E[g(\tp)-g^*(\tp)]^2$ for every candidate function~$g$. The following result provides an explicit form for the risk.
\begin{theorem}
\label{risk.lemma}
Suppose that model~\eqref{beta.model} holds, and the prior for~$p$ has a bounded density.  Then, 
\begin{equation}
    R(g) = E g(\tp)^2+2E\left[ g(\tp)(1-2\tp)+\tp(1-\tp) g'(\tp)\right] +C\label{full.risk}
\end{equation}
for all bounded and differentiable functions~$g$, where~$C$ is a constant that does not depend on~$g$.
\end{theorem}
\begin{remark}
We show in the proof of Theorem~\ref{risk.lemma} that $g^*$ is bounded and differentiable so \eqref{full.risk} holds for $g=g^*$.
\end{remark}

Theorem~\ref{risk.lemma} suggests that we can approximate the risk, up to an irrelevant constant, by
\begin{equation}
\label{emp.risk}
    \hat R(g) = \frac1n \sum_{i=1}^n g(\tp_i)^2 + 2 \frac1n \sum_{i=1}^n  \left[ g(\tp_i)(1-2\tp_i)+\tp_i(1-\tp_i)  g'(\tp_i)\right].
\end{equation}
However, simply minimizing \eqref{emp.risk} would provide a poor estimate for $g^*(\tp)$ because, without any smoothness constraints, $\hat R(g)$ can be trivially minimized. Hence, we place a smoothness penalty on our criterion by minimizing
\begin{equation}
\label{risk.criterion}
    Q(g) = \hat R(g) + \lambda \int g''(\tp)^2d\tp,
\end{equation}
where $\lambda>0$ is a tuning parameter which adjusts the level of smoothness in $g(\tp)$. 
We show in our theoretical analysis in Section~\ref{theory.sec} (see the proof of Theorem~\ref{g.thm}) that, much as with the more standard curve fitting setting, the solution to criteria of the form in \eqref{risk.criterion} can be well approximated using a natural cubic spline, which provides a computationally efficient approach to compute $ g(\tp)$. 


Let $\bb(x)$ represent the vector of basis functions for a natural cubic spline, with knots at $\tp_1, \ldots, \tp_n$, restricted to satisfy $\bb(0.5)={\bf 0}$. Then, in minimizing $Q(g)$ we only need to consider functions of the form $g(\tp) = \bb(\tp)^T\bfeta$, where $\bfeta$ is the basis coefficients. Thus, \eqref{risk.criterion} can be re-expressed as
\begin{equation}
\label{qn.ncs.probs}
   Q_n(\bfeta) = \frac1n \sum_{i=1}^n \bfeta^T\bb(\tp_i)\bb(\tp_i)^T \bfeta + 2\frac1n \sum_{i=1}^n \left[(1-2\tp_i)\bb(\tp_i)^T+\tp_i(1-\tp_i) \bb'(\tp_i)^T\right]\bfeta + \lambda \bfeta^T\Omega\bfeta 
\end{equation}
where $\Omega=\int \bb''(\tp)\bb''(\tp)^Td\tp$. Standard calculations show that \eqref{qn.ncs.probs} is minimized by setting
\begin{equation}
\label{ls.eta.probs}
    \hat \bfeta = -\left(\sum_{i=1}^n\bb(\tp_i)\bb(\tp_i)^T+n\lambda\Omega\right)^{-1} \sum_{i=1}^n \left[(1-2\tp_i)\bb(\tp_i)+\tp_i(1-\tp_i) \bb'(\tp_i)\right].
\end{equation}
If the prior distribution of $p_i$ is not assumed to be symmetric, then $g^*(\tp_i)$ should be directly estimated for $0\le \tp_i\le 1$. However, if the prior is believed to be symmetric this approach is inefficient, because it does not incorporate the identity $g^*(\tp_i)=-g^*(1-\tp_i)$. Hence, a superior approach involves flipping all of the $\tp_i>0.5$ across~$0.5$, thus converting them into $1-\tp_i$, and then using both the flipped and the unflipped~$\tp_i$ to estimate $g(\tp_i)$ between $0$ and $0.5$.  Finally, the identity $\hat g(\tp_i)=-\hat g(1-\tp_i)$ can be used to define~$\hat g$ on $(0.5,1]$. This is the approach we use for the remainder of the paper.

Equation~(\ref{ls.eta.probs}) allows us to compute estimates for $E(p_i|\tp_i)$ and $\text{Var}(p_i|\tp_i)$:
\begin{eqnarray}
\label{mu.hat}\hat \mu_i &=& \tp_i + \hat\gamma(\bb(\tp_i)^T\hat\bfeta+1-2\tp_i) \\
\label{sigma.hat}\hat \sigma^2_i &=& \hat\gamma\tp_i(1-\tp_i)+ {\hat\gamma}^2\tp_i(1-\tp_i)[\bb'(\tp_i)^T\hat\bfeta-2].
\end{eqnarray}
Equations~\eqref{mu.hat} and \eqref{sigma.hat} can then be substituted into \eqref{oracle} to produce the ECAP estimator $\hat p_i$. 


\subsubsection{Estimation of $\lambda$ and $\gamma^*$}
\label{gamma.sec}

In computing \eqref{mu.hat} and \eqref{sigma.hat} we need to provide estimates for $\gamma^*$ and $\lambda$. We choose $\lambda$ so as to minimize a cross-validated version of the estimated risk \eqref{emp.risk}. In particular, we randomly partition the probabilities into $K$ roughly even groups: $G_1,\ldots, G_{K}$. Then, for given values of $\lambda$ and $k$, $\hat \bfeta_{k\lambda}$ is computed via \eqref{ls.eta.probs}, with the probabilities in $G_k$ excluded from the calculation. We then compute the corresponding estimated risk on the probabilities in $G_k$:
$$R_{k\lambda}=\sum_{i\in G_k} \hat h_{ik}^2 + 2\sum_{i\in G_k} \left[(1-2\tp_i)\hat h_{ik}+\tp_i(1-\tp_i) \hat h'_{ik}\right],$$
where $\hat h_{ik}=\bb(\tp_i)^T \hat \bfeta_{k\lambda}$ and $\hat h'_{ik}=\bb'(\tp_i)^T \hat \bfeta_{k\lambda}$. This process is repeated $K$ times for $k=1,\ldots, K$, and 
$$ R_\lambda=\frac1n\sum_{k=1}^K R_{k\lambda}$$
is computed as our cross-validated risk estimate. Finally, we choose $\hat \lambda = \arg\min_\lambda R_\lambda$.

To estimate $\gamma^*$ we need a measure of the accuracy of $\tp_i$ as an estimate of $p_i$. In some cases that information may be available from previous analyses. For example, if the estimates~$\tp_i$ were obtained by fitting a logistic regression model,  we could compute the standard errors on the estimated coefficients and hence form a variance estimate for each~$\tilde p_i$.  We would estimate~$\gamma^*$ by matching the computed variance estimates to the expression~(\ref{pt.mean.var}) for the conditional variance under the ECAP model.  

Alternatively, we can use previously observed outcomes of $A_i$ to estimate $\gamma^*$. Suppose that we observe
$$Z_i=\begin{cases}1 & A_i\text{ occured},\\
0&A_i\text{ did not occur},
\end{cases}$$
for $i= 1,\ldots, n$. Then a natural approach is to compute the (log) likelihood function for $Z_i$. Namely,
\begin{equation}
\label{log.like}
    l_\gamma = \sum_{i:Z_i=1} \log(\hat p^\gamma_i) + \sum_{i:Z_i=0} \log(1-\hat p^\gamma_i),
\end{equation}
where $\hat p^\gamma_i$ is the ECAP estimate generated by substituting in a particular value of $\gamma$ into~\eqref{mu.hat} and~\eqref{sigma.hat}. We then choose the value of~$\gamma$ that maximizes~\eqref{log.like}. 

As an example of this approach, consider the ESPN data recording probabilities of victory for various NCAA football teams throughout each season. To form an estimate for~$\gamma^*$ we can take the observed outcomes of the games from last season (or the first couple of weeks of this season if there are no previous games available), use these results to generate a set of $Z_i$, and then choose the~$\gamma$ that maximizes \eqref{log.like}. One could then form ECAP estimates for future games during the season, possibly updating the~$\gamma$ estimate as new games are played.

\subsection{Large sample results}
\label{theory.sec}

In this section we investigate the large sample behavior of the ECAP estimator. More specifically, we show that, under smoothness assumptions on the function~$g^*$, the ECAP adjusted probabilities are consistent estimators of the corresponding oracle probabilities, defined in~\eqref{argmin}.  We establish an analogous result for the corresponding values of the loss function, defined in~\eqref{loss.fn}.  In addition to demonstrating consistency we also derive the rates of convergence.
Our method of proof takes advantage of the theory of empirical processes,
however, the corresponding arguments go well beyond a simple application of the existing results.

We let~$f^*$ denote the marginal density of the observed $\tp_i$ and define the $L_2(\tilde P)$ norm of a given function~$u(\tp)$ as $\|u\|=[\int_0^1 u^2(\tp)f^*(\tp)d\tp]^{1/2}$.  We denote the corresponding empirical norm, $[(1/n)\sum_{i=1}^n u^2(\tp_i)]^{1/2}$, by $\|u\|_n$.  To simplify the presentation of the results, we define
\begin{equation*}
r_n=n^{-4/7}\lambda_n^{-1}+n^{-2/7}+\lambda_n \qquad \text{and} \qquad s_n=1+n^{-4/7}\lambda_n^{-2}.    
\end{equation*}
We write~$\hat g$ for the minimizer of criterion~(\ref{risk.criterion}) over all natural cubic spline functions~$g$ that correspond to the sequence of~$n$ knots located at the observed~$\tp_i$.  For concreteness, we focus on the case where criterion~(\ref{risk.criterion}) is computed over the entire interval $[0,1]$.  However, all of the results in this section continue to hold if $\hat g$ is determined by only computing the criterion over $[0,0.5]$, according to the estimation approach described in Section~\ref{sec.estimation}.
The following result establishes consistency and rates of convergence for $\hat g$ and $\hat g'$.
\begin{theorem}
\label{g.thm}
If $g^*$ is twice continuously differentiable on $[0,1]$, $f^*$ is bounded away from zero and $n^{-8/21}\ll\lambda_n\ll 1$, then
\begin{equation*}
\|\hat g - g^*\|_n = O_p\big(r_n\big), \qquad \|\hat g' - {g^*}'\|_n= O_p\big(\sqrt{r_ns_n}\big).
\end{equation*}
The above bounds also hold for the $\|\cdot\|$ norm.
\end{theorem}
\begin{remark}
The assumption $n^{-8/21}\ll\lambda_n\ll 1$ implies that the error bounds for $\hat g$ and $\hat g'$ are of order $o_p(1)$.
\end{remark}
When $\lambda_n\asymp n^{-2/7}$, Theorem~\ref{g.thm} yields an $n^{-2/7}$ rate of convergence for $\hat{g}$.  This rate
matches the optimal rate of convergence for estimating the derivative of a density under the corresponding smoothness conditions \citep{stone1980optimal}.

Given a value $\tp$ in the interval $(0,1)$, we define the ECAP estimator, $\hat p=\hat p(\tp)$, by replacing $\tp_i$, $\gamma^*$, and $g$ with~$\tp$,$\hat\gamma$ and $\hat g$, respectively, in the expression for the oracle estimator provided by formulas~(\ref{mui}), (\ref{sigmai}) and~(\ref{oracle}). Thus, we treat~$\hat p$ as a random function of~$\tp$, where the randomness comes from the fact that $\hat p$ depends on the training sample of the observed probabilities~$\tp_i$.  By analogy, we define $p_0$ via~(\ref{oracle}), with~$\tp_i$ replaced by~$\tp$, and view~$p_0$ as a  (deterministic) function of~$\tp$.

We define the function $W_0(\tp)$ as the expected loss for the oracle estimator:
\begin{equation*}
W_0(\tp)=E_p\big[EC\left(p_0(\tp)\right)^2 |\tp\big],
\end{equation*}
where the expected value is taken over the true~$p$ given the corresponding observed probability~$\tp$.
Similarly, we define the random function $\widehat W(\tp)$ as the expected loss for the ECAP estimator:
\begin{equation*}
\widehat W(\tp)=E_p\big[EC\left(\hat p(\tp)\right)^2 |\tp\big],
\end{equation*}
where the expected value is again taken over the true~$p$ given the corresponding~$\tp$. The randomness in the function $\widehat W(\tp)$ is due to the dependence of~$\hat p$ on the training sample $\tp_1,...,\tp_n$.

To state the asymptotic results for $\hat p$ and $\hat W$, we implement a minor technical modification in the estimation of the conditional variance via formula (\ref{sigmai}).  After computing the value of $\hat\sigma^2$, we set it equal to $\max\{\hat\sigma^2,c\sqrt{r_ns_n}\}$, where~$c$ is allowed to be any fixed positive constant.  This ensures that, as the sample size grows, $\hat\sigma^2$ does not approach zero too fast. We note that this technical modification is only used to establish consistency of $\widehat W(\tp)$ in the next theorem; all the other results in this section hold both with and without this modification.
\begin{theorem}
\label{cons.thm}
If $g^*$ is twice continuously differentiable on $[0,1]$, $f^*$ is bounded away from zero, $n^{-8/21}\ll\lambda_n\ll 1$ and $|\hat \gamma-\gamma^*|=o_p(1)$, then 
\begin{equation*}
\|\hat{p} - p_0\|= o_p(1) \qquad\text{and}\qquad  \|\hat{p} - p_0\|_n = o_p(1).
\end{equation*}
If, in addition, $|\hat \gamma-\gamma^*|=O_p(\sqrt{r_ns_n})$, then 
\begin{equation*}
\int\limits_0^1\big|\widehat{W}(\tp) - W_0(\tp)\big|f^*(\tp)d\tp = o_p(1) \qquad\text{and}\qquad 
\frac1n\sum_{i=1}^n\big|\widehat{W}(\tp_i) - W_0(\tp_i)\big| = o_p(1).
\end{equation*}
\end{theorem}

The next result provides the rates of convergence for $\hat p$ and $\hat W$.  
\begin{theorem}
\label{rate.thm}
If $g^*$ is twice continuously differentiable on $[0,1]$, $f^*$ is bounded away from zero, $n^{-8/21}\ll\lambda_n\ll 1$ and $|\hat \gamma-\gamma^*|=O_p\big(\sqrt{r_ns_n}\big)$,  then
\begin{eqnarray*}
\int\limits_{\epsilon}^{1-\epsilon}\big|\hat{p}(\tp) - p_0(\tp)\big|^2f^*(\tp)d\tp =
O_p\big(r_ns_n\big),  &\quad&  \int\limits_{\epsilon}^{1-\epsilon}\big|\widehat{W}(\tp) - W_0(\tp)\big|f^*(\tp)d\tp = O_p\big(r_ns_n\big),\\
\\
\sum_{i:\,  \epsilon\le\tp_i\le 1-\epsilon}\frac1n\big|\hat{p}(\tp_i) - p_0(\tp_i)\big|^2 =
O_p\big(r_ns_n\big)&\,\quad\text{and}\,\quad&  \sum_{i:\,  \epsilon\le\tp_i\le 1-\epsilon}\frac1n\big|\widehat{W}(\tp_i) - W_0(\tp_i)\big| = O_p\big(r_ns_n\big),
\end{eqnarray*}
for each fixed positive~$\epsilon$.
\begin{remark}
The assumption $n^{-8/21}\ll\lambda_n\ll 1$ ensures that all the error bounds are of order $o_p(1)$.  
\end{remark}
\end{theorem}
In Theorem~\ref{rate.thm} we bound the integration limits away from zero and one, because the rate of convergence changes as~$\tp$ approaches those values.  However, we note that~$\epsilon$ can be set to an arbitrarily small value.  The optimal rate of convergence for $\widehat{W}$ is provided in the following result.
\begin{corollary}\label{W.rate.crl}
Suppose that $\lambda_n$ decreases at the rate~$n^{-2/7}$ and $|\hat \gamma-\gamma^*|=O_p(n^{-1/7})$.  If~$f^*$ is bounded away from zero and~$g^*$ is twice continuously differentiable on $[0,1]$, then
\begin{equation*}
\int\limits_{\epsilon}^{1-\epsilon}\big|\widehat{W}(\tp) - W_0(\tp)\big|d\tp = O_p\big(n^{-2/7}\big)\qquad\text{and}\qquad \sum_{i:\,  \epsilon\le\tp_i\le 1-\epsilon}\frac1n\big|\widehat{W}(\tp_i) - W_0(\tp_i)\big|= O_p\big(n^{-2/7}\big),
\end{equation*}
for every positive~$\epsilon$.
\end{corollary}
Corollary~\ref{W.rate.crl} follows directly from Theorem~\ref{rate.thm} by balancing out the components in the expression for~$r_n$.

\section{ECAP Extensions}
\label{extension.sec}

In this section we consider two possible extensions of \eqref{beta.model}, the model for $\tp_i$. In particular, in the next subsection we discuss the setting where~$\tp_i$ can no longer be considered an unbiased estimator for~$p_i$, while in the following subsection we suggest a generalization of the beta model.

\subsection{Incorporating Bias in $\tp_i$}
\label{biased.sec}

So far, we have assumed that $\tp_i$ is an unbiased estimate for $p_i$. In practice probability estimates~$\tp_i$ may exhibit some systematic bias. For example, in Section~\ref{emp.sec} we examine probability predictions from the \href{www.FiveThirtyEight.com}{FiveThirtyEight.com} website on congressional house, senate, and governors races during the 2018 US midterm election.  After comparing the actual election results with the predicted probability of a candidate being elected, there is clear evidence of bias in the estimates \citep{Silver18}. In particular the leading candidate won many more races than would be suggested by the probability estimates. This indicates that the \href{www.FiveThirtyEight.com}{FiveThirtyEight.com} probabilities were overly conservative,  i.e., that in comparison to $p_i$ the estimate~$\tp_i$ was generally closer to $0.5$; for example, $E(\tp_i|p_i)<p_i$ when $p_i>0.5$.

In this section we generalize \eqref{beta.model} to model situations where $E(\tp_i|p_i)\ne p_i$. To achieve this goal we replace \eqref{beta.model} with 
\begin{equation}
\label{bias.alpha}
    \tp_i|p_i \sim Beta(\alpha_i, \beta_i),\quad\text{where} \quad p_i=h_\theta(\alpha_i\gamma^*)=h_\theta(1-\beta_i\gamma^*),
\end{equation}
$h_\theta(\cdot)$ is a prespecified function, and $\theta$ is a parameter which determines the level of bias of $\tp_i$. In particular, \eqref{bias.alpha} implies that for any invertible $h_\theta$,
\begin{equation}
    p_i= h_\theta(E(\tp_i|p_i)),
\end{equation}
so that if $h_\theta(x)=x$, i.e., $h_\theta(\cdot)$ is the identity function, then \eqref{bias.alpha} reduces to \eqref{beta.model}, and $\tp_i$ is an unbiased estimate for $\tp_i$.

\begin{figure}[t!]
	\centering
	 \scalebox{1.5}{%
	\input{Figures/p_expdiff.tex}}
	\caption{\label{ebias.plot}\small Plots of $E(\tp_i|p_i)$ as a function of $p_i$ for different values of $\theta$. When $\theta=0$ (black / solid) the estimates are unbiased. $\theta=2$ (orange / dashed) corresponds to a setting where $\tp_i$ systematically underestimates large values of $p_i$, while $\theta=-3$ (green / dot-dashed) represents a situation where $\tp_i$ is an overestimate for large values of $p_i$.}
\end{figure}
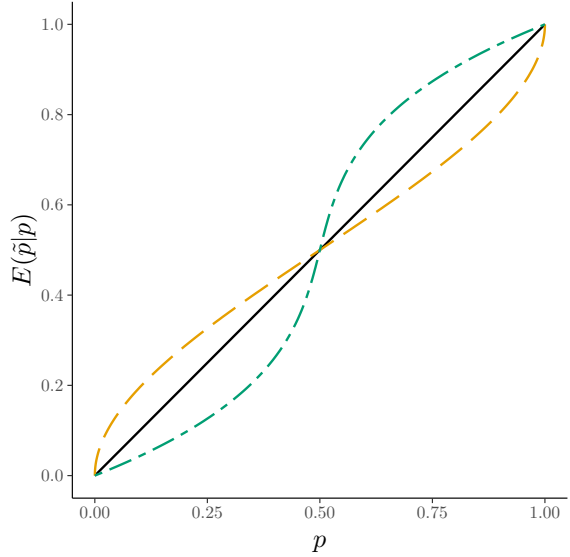

To produce a valid probability model $h_\theta(\cdot)$ needs to satisfy several criteria:
\begin{enumerate}
\item $h_0(x)=x$, so that \eqref{bias.alpha} reduces to \eqref{beta.model} when $\theta=0$.
    \item $h_\theta(1-x)=1-h_\theta(x)$, ensuring that the probabilities of events~$A_i$ and~$A_i^c$ sum to~$1$.
    \item $h_\theta(x)=x$ for $x=0, x=0.5$ and $x=1$.
 \item $h_\theta(\alpha)$ is invertible for values of $\theta$ in a region around zero, so that $E(\tp_i|p_i)$ is unique. 
\end{enumerate}
The simplest polynomial function that satisfies all these constraints is
$$h_\theta(x) = (1-0.5\theta)x - \theta[x^3 - 1.5 x^2],$$ which is invertible for $-4 \le \theta \le 2.$ Note that for $\theta=0$, we have $h_0(x)=x$, which corresponds to the unbiased model~\eqref{beta.model}. However, if $\theta>0$, then $\tp_i$ tends to overestimate small $p_i$ and underestimate large $p_i$, so the probability estimates are overly conservative.  Alternatively, when $\theta<0$, then $\tp_i$ tends to underestimate small $p_i$ and overestimate large $p_i$, so the probability estimates exhibit excess certainty. Figure~\ref{ebias.plot} provides examples of $E(\tp_i|p_i)$ for three different values of $\theta$, with the green line representing probabilities resulting in excess certainty, the orange line overly conservative probabilities, and the black line unbiased probabilities. 

One of the appealing aspects of this model is that the ECAP oracle \eqref{oracle} can still be used to generate an estimator for $p_i$. The only change is in how $E(p_i|\tp_i)$ and $Var(p_i|\tp_i)$ are computed. The following result
allows us to generalize Theorem~\ref{EandVar} to the biased setting to compute $E(p_i|\tp_i)$ and $Var(p_i|\tp_i)$.  
%
\begin{theorem}
\label{bias.thm}
Suppose that model~\eqref{bias.alpha} holds, $p_i$ has a bounded density, and $\mu_i$ and $\sigma_i^2$ are respectively defined as in~\eqref{mui} and~\eqref{sigmai}.  
Then, 
\begin{eqnarray}
   \label{bias.e} E(p_i|\tp_i) &=& \mu_i+0.5\theta\left[3\sigma_i^2-6\mu_i\sigma_i^2+3\mu_i^2-\mu_i-2\mu_i^3\right]+O\big(\theta{\gamma^*}^{3/2}\big)\\
    \label{bias.v}Var(p_i|\tp_i) &=& (1-0.5\theta)^2\sigma_i^2
    +\theta\sigma_i^2\big[3\mu_i(1-\mu_i)(3\theta\mu_i(1-\mu_i)-0.5\theta+1) \big]
    +O\big(\theta{\gamma^*}^{3/2}\big).
\end{eqnarray}
\end{theorem}
%

The remainder terms in the above approximations are of smaller order than the leading terms when~$\gamma^*$ is small, which is typically the case in practice.  As we demonstrate in the proof of Theorem~\ref{bias.thm}, explicit expressions can be provided for the remainder terms.
However, 
the approximation error involved in estimating these expressions is likely to be much higher than any bias from excluding them. Hence, we ignore these terms when estimating $E(p_i|\tp_i)$ and $Var(p_i|\tp_i)$:
\begin{eqnarray}
   \widehat{E(p_i|\tp_i)} &=& \hat\mu_i+0.5\theta\left[3\hat\sigma_i^2-6\hat\mu_i\hat\sigma_i^2+3\hat\mu_i^2-\hat\mu_i-2\hat\mu_i^3\right]\\
    \widehat{Var(p_i|\tp_i)} &=& (1-0.5\theta)^2\hat\sigma_i^2
    +\theta\hat\sigma_i^2\big[3\hat\mu_i(1-\hat\mu_i)(3\theta\hat\mu_i(1-\hat\mu_i)-0.5\theta+1) \big].
\end{eqnarray}

The only remaining issue in implementing this approach involves producing an estimate for~$\theta$. However, this can be achieved using exactly the same maximum likelihood approach as the one used to estimate~$\gamma^*$, which is described in Section~\ref{gamma.sec}. Thus, we now choose both $\theta$ and $\gamma$ to jointly maximize the likelihood function
\begin{equation}
\label{bias.log.like}
    l_{\theta,\gamma} = \sum_{i:Z_i=1} \log(\hat p^{\theta,\gamma}_{i}) + \sum_{i:Z_i=0} \log(1-\hat p^{\theta,\gamma}_{i}),
\end{equation}
where $\hat p^{\theta,\gamma}_{i}$ is the bias corrected ECAP estimate generated by substituting in particular values of~$\gamma$ and~$\theta$. In all other respects, the bias corrected version of ECAP is implemented in an identical fashion to the unbiased version.

%

\subsection{Mixture Distribution}
\label{sec.mixture}
We now consider another possible extension of \eqref{beta.model}, where we believe that~$\tp_i$ is an unbiased estimator for~$p_i$ but find the beta model assumption to be unrealistic. In this setting one could potentially model~$\tp_i$ using a variety of members of the exponential family. However, one appealing alternative is to extend \eqref{beta.model} to a mixture of beta distributions:
\begin{equation}
\label{beta.mixture.model}
    \tp_i|p_i \sim \sum_{k=1}^K w_k Beta(\alpha_{ik}, \beta_{ik}),\quad\text{where} \quad \alpha_{ik}=\frac{c_k p_i}{\gamma^*}, \quad\beta_{ik}=\frac{1-c_k p_i}{\gamma^*},
\end{equation}
and $w_k$ and $c_k$ are predefined weights such that $\sum_k w_k=1$ and $\sum_k w_kc_k=1$. Note that \eqref{beta.model} is a special case of \eqref{beta.mixture.model} with $K=w_1=c_1=1$. 

As $K$ grows, the mixture model can provide as flexible a model as desired, but it also has a number of other appealing characteristics. In particular, under this model it is still the case that $E(\tp_i|p_i)=p_i$. In addition, Theorem~\ref{EandVar.mixture} demonstrates that simple closed form solutions still exist for $E(p_i|\tp_i)$ and $Var(p_i|\tp_i)$, and, hence, also the oracle ECAP estimator $p_{i0}$. 
\begin{theorem}
\label{EandVar.mixture}
Under~\eqref{beta.mixture.model},
\begin{eqnarray}
\label{mui.mixture}E(p_i|\tp_i) &=&\mu_i\sum_{k=1}^K \frac{w_k}{c_k}\\
\label{sigmai.mixture}
Var(p_i|\tp_i)&=& (\sigma_i^2+\mu_i^2)\sum_{k=1}^K \frac{w_k}{c_k^2}-\mu_i^2\left(\sum_{k=1}^K\frac{w_k}{c_k}\right)^2,
\end{eqnarray}
where $\mu_i$ and $\sigma_i^2$ are defined in \eqref{mui} and \eqref{sigmai}.
\end{theorem}
The generalized ECAP estimator can thus be generated by substituting $\hat \mu_i$ and $\hat \sigma^2_i$, given by formulas~(\ref{mu.hat}) and~(\ref{sigma.hat}), into~\eqref{mui.mixture} and~\eqref{sigmai.mixture}. The only additional complication involves computing values for $w_k$ and $c_k$. For settings with a large enough sample size, this could be achieved using a variant of the maximum likelihood approach discussed in Section~\ref{gamma.sec}. However, we do not explore that approach further in this paper.

\section{Simulation Results}
\label{sim.sec}

In Section~\ref{sec.unbiased} we compare ECAP to competing methods under the assumption of unbiasedness in $\tp_i$. We further extend this comparison to the setting where $\tp_i$ represents a potentially biased estimate in Section~\ref{sec.biased}.

\subsection{Unbiased Simulation Results}
\label{sec.unbiased}

\begin{figure}[tp]
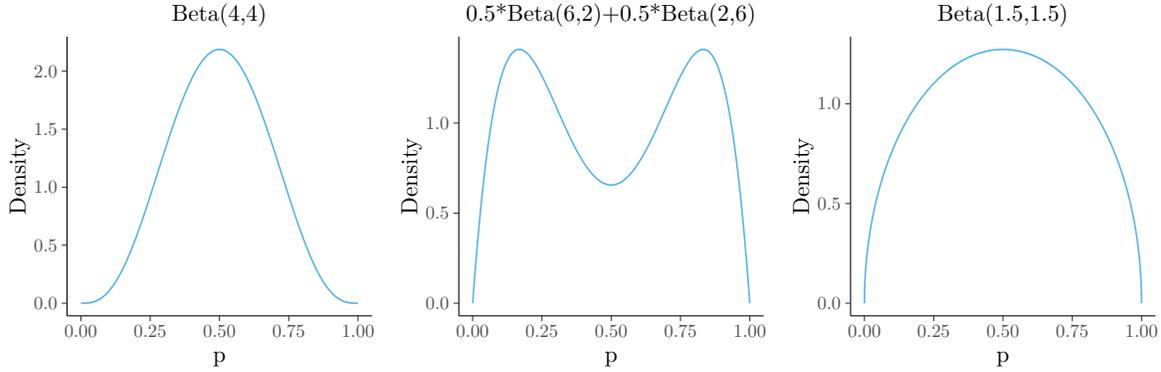

    \centering
    \scalebox{1}{%
    \input{Figures/4-4.tex}\hfill
    \input{Figures/2-6.tex}\hfill
    \input{Figures/15-15.tex}}
    \caption{Distributions of $p$ used in the simulation} 
    \label{fig:figure3}
\end{figure}

In this section our data consists of $n=$ 1,000 triplets $(p_i,\tp_i, Z_i)$ for each simulation. The $p_i$ are generated from one of three possible prior distributions; Beta$(4,4)$, an equal mixture of Beta$(6,2)$ and Beta$(2,6)$, or Beta$(1.5,1.5)$. The corresponding density functions are displayed in Figure~\ref{fig:figure3}. 

Recall that ECAP models $\tp_i$ as coming from a beta distribution, conditional on $p_i$. However, in practice there is no guarantee that the observed data will exactly follow this distribution. Hence, we generate the observed data according to: 
\begin{equation}
\label{sim.model}
    \tp_i = p_i + p^q_i (\tp^{\text{o}}_i-p_i),
\end{equation}
where $\tp_i^{\text{o}}|p_i\sim$ Beta$(\alpha,\beta)$ and $q$ is a tuning parameter. In particular for $q=0$ \eqref{sim.model} generates observations directly from the ECAP model, while larger values of $q$ provide a greater deviation from the beta assumption. In practice we found that setting $q=0$ can result in $\tilde{p}$'s that are so small they are effectively zero ($\tp_i = 10^{-20}$, for example). ECAP is not significantly impacted by these probabilities but, as we show, other approaches can perform extremely poorly in this scenario. Setting $q>0$
prevents pathologic scenarios and allows us to more closely mimic what practitioners will see in real life. 
We found that $q=0.05$ typically gives a reasonable amount of dispersion so we consider settings where either $q=0$ or $q=0.05$.  
We also consider different levels of the conditional variance for~$\tp_i$, by taking~$\gamma^*$ as either~$0.005$ or~$0.03$. Finally, we generate $Z_i$, representing whether event $A_i$ occurs, from a Bernoulli distribution with probability $p_i$. 







We implement the following five approaches: the {\it Unadjusted} method, which simply uses the original probability estimates~$\tp_i$, two implementations of the proposed {\it ECAP} approach (ECAP Opt and ECAP MLE), and two versions of the James Stein approach (JS Opt and JS MLE).
For the proposed ECAP methods, we select~$\lambda$ via the cross-validation procedure in Section~\ref{gamma.sec}. ECAP Opt is an oracle-type implementation of the ECAP methodology, in which we select~$\gamma$ to minimize the average expected loss, defined in \eqref{loss.fn}, over the training data. Alternatively, ECAP MLE makes use of the $Z_i$'s and estimates~$\gamma^*$ using the maximum likelihood approach described in Section~\ref{gamma.sec}.  The James-Stein method we use is similar to its traditional formulation. 
In particular the estimated probability  is computed using
\begin{equation}
\label{js.sim}
    \hat p^{JS}_i = \bar \tp + (1-c)\left(\tp_i - \bar \tp\right),
\end{equation}
where $\bar \tp=\frac{1}{n}\sum_{j=1}^n \tp_j$ and $c$ is a tuning parameter chosen to optimize the estimates.\footnote{To maintain consistency with ECAP we flip all $\tp_i>0.5$ across $0.5$ before forming $\hat p^{JS}_i$ and then flip the estimate back.}
Equation \eqref{js.sim} is a convex combination of $\tp_i$ and the average observed probability $\bar \tp$. 
The JS Opt implementation selects~$c$ to minimize the average expected loss in the same fashion as for ECAP Opt, while
the JS MLE implementation selects~$c$ using the maximum likelihood approach described in Section~\ref{gamma.sec}. Note that ECAP Opt and JS Opt represent optimal situations that can not be implemented in practice because they require knowledge of the true distribution of $p_i$.




\begin{table}[t]
\captionof{table}{Average expected loss for different methods over multiple unbiased simulation scenarios.  Standard errors are provided in parentheses.} \label{tab:title}
\begin{center}
{\small \begin{tabular}{c|c||l|l|l|l}
$\gamma^*$                & \textbf{q}            & \textbf{Method Type} & $\text{Beta}(4, 4)$ & \begin{tabular}[c]{@{}l@{}}$0.5$*Beta($6$,$2$) +\\ $0.5$*Beta($2$,$6$)\end{tabular}          & $\text{Beta}(1.5, 1.5)$           \\ \hline
\multirow{10}{*}{0.005} & \multirow{5}{*}{0}    & Unadjusted                & 0.0116 (0.0001)                  & 44.9824 (43.7241)                           & 3.9$\times 10^{12}$ (3.9$\times 10^{12}$) \\ \cline{3-6} 
                        &                       & ECAP Opt             & 0.0095 (0.0001)                  & 0.0236 (0.0002)                           & 0.0197 (0.0001)                           \\ \cline{3-6} 
                        &                       & JS Opt               & 0.0100 (0.0001)                  & 0.0241 (0.0002)                           & 0.0204 (0.0002)                            \\ \cline{3-6} 
                        &                       & ECAP MLE            & 0.0109 (0.0002)                  & 0.0302 (0.0006)                           & 0.0263 (0.0007)                            \\ \cline{3-6} 
                        &                       & JS MLE               & 0.0121 (0.0003)                  & 1.1590 (0.8569)                           & 4.8941 (4.7526)                            \\ \cline{2-6} 
                        & \multirow{5}{*}{0.05} & Unadjusted                & 0.0100 (0.0001)                  & 0.0308 (0.0006)                           & 0.0273 (0.0006)                            \\ \cline{3-6} 
                        &                       & ECAP Opt             & 0.0085 (0.0000)                  & 0.0196 (0.0001)                           & 0.0166 (0.0001)                            \\ \cline{3-6} 
                        &                       & JS Opt               & 0.0090 (0.0000)                  & 0.0201 (0.0001)                           & 0.0172 (0.0001)                            \\ \cline{3-6} 
                        &                       & ECAP MLE            & 0.0098 (0.0002)                  & 0.0238 (0.0005)                           & 0.0197 (0.0004)                            \\ \cline{3-6} 
                        &                       & JS MLE               & 0.0105 (0.0002)            & 0.0265 (0.0006)                           & 0.0245 (0.0007)                            \\ \hline \hline
\multirow{10}{*}{0.03}  & \multirow{5}{*}{0}    & Unadjusted                & 2.1$\times 10^{8}$ (2.1$\times 10^{8}$)      & 2.4$\times 10^{14}$ (1.6$\times 10^{14}$) & 1.6$\times 10^{15}$ (5.5$\times 10^{14}$) \\ \cline{3-6} 
                        &                       & ECAP Opt             & 0.0391 (0.0002)                  & 0.0854 (0.0004)                           & 0.0740 (0.0004)                            \\ \cline{3-6} 
                        &                       & JS Opt               & 0.0537 (0.0002)                  & 0.0986 (0.0005)                           & 0.0899 (0.0005)                            \\ \cline{3-6} 
                        &                       & ECAP MLE            & 0.0452 (0.0010)                  & 0.1607 (0.0187)                           & 0.1477 (0.0187)                            \\ \cline{3-6} 
                        &                       & JS MLE               & 0.0636 (0.0019)                  & 1.4$\times 10^{13}$ (1.4$\times 10^{13}$) & 1.2$\times 10^{14}$ (1.1$\times 10^{14}$) \\ \cline{2-6} 
                        & \multirow{5}{*}{0.05} & Unadjusted                & 0.0887 (0.0010)                  & 0.3373 (0.0047)                           & 0.2780 (0.0043)                            \\ \cline{3-6} 
                        &                       & ECAP Opt             & 0.0364 (0.0002)                  & 0.0765 (0.0004)                           & 0.0665 (0.0004)                            \\ \cline{3-6} 
                        &                       & JS Opt               & 0.0488 (0.0002)                  & 0.0874 (0.0005)                           & 0.0801 (0.0005)                            \\ \cline{3-6} 
                        &                       & ECAP MLE            & 0.0411 (0.0008)                  & 0.1035 (0.0050)                           & 0.0896 (0.0036)                            \\ \cline{3-6} 
                        &                       & JS MLE               & 0.0558 (0.0011)                  & 0.1213 (0.0066)                           & 0.1235 (0.0071)                            \\ 
\end{tabular}}
\end{center}
\end{table}

In each simulation run we generate both training and test data sets. Each method is fit on the training data. We then calculate $EC(\hat p_i)^2$ for each point in the test data and average over these observations. 
The results for the three prior distributions, two values of $\gamma^*$, and two values of $q$, averaged over 100 simulation runs, are reported in Table~\ref{tab:title}. Since the ECAP Opt and JS Opt approaches both represent oracle type methods, they should be compared with each other.
The ECAP Opt method statistically significantly outperforms its JS counterpart in each of the twelve settings, with larger improvements in the noisy setting where $\gamma^*=0.03$.  The ECAP MLE method is statistically significantly better than the corresponding JS approach in all but four settings.  However, those four settings, correspond to $q=0$ and actually represent situations where JS MLE has failed because it has extremely large excess certainty, which impacts both the mean and standard error. Alternatively, the performance of the ECAP approach remains stable even in the presence of extreme outliers.
Similarly, the ECAP MLE approach statistically significantly outperforms the Unadjusted approach, often by large amounts, except for the five settings with large outliers, which result in extremely bad average performance for the latter method.





\subsection{Biased Simulation}
\label{sec.biased}
In this section we extend the results to the setting where the observed probabilities may be biased, i.e., $E(\tp_i|p_i)\ne p_i$. To do this we generate $\tp_i$ according to  \eqref{bias.alpha} using four different values for $\theta$, $\{-3,-1,0,2\}$. Recall that $\theta<0$ corresponds to anti-conservative data, where~$\tp_i$ tends to be too close to $0$ or $1$, $\theta=0$ represents unbiased observations, and $\theta>0$ corresponds to conservative data, where~$\tp_i$ tends to be too far from $0$ or $1$. In all other respects our data is generated in an identical fashion to that of the unbiased setting.\footnote{Because the observed probabilities are now biased, we replace $p_i$ in \eqref{sim.model} with $E(\tp_i|p_i)$.}

To illustrate the biased setting we opted to focus on the $q=0.05$ with $\gamma^*=0.005$ setting. We also increased the sample size to $n=$ 5,000 because of the increased difficulty of the problem. The two ECAP implementations now require us to estimate three parameters: $\lambda, \gamma$ and $\theta$. We estimate $\lambda$ in the same fashion as previously discussed, while $\gamma$ and $\theta$ are now chosen over a two-dimensional grid of values, with $\theta$ restricted to lie between $-4$ and $2$. The two JS methods remain unchanged.

\begin{table}[t]
\centering
\captionof{table}{Average expected loss for different methods over multiple biased simulation scenarios.} 
\label{table:biasedtable}
{\small \begin{tabular}{l|l||l|l|l}
\textbf{}                    & \textbf{Method Type} & Beta($4$,$4$)       & \begin{tabular}[c]{@{}l@{}}$0.5$*Beta($6$,$2$) +\\ $0.5$*Beta($2$,$6$)\end{tabular}       & Beta($1.5$, $1.5$)  \\ \hline \hline
\multirow{5}{*}{$\theta=-3$} & Unadjusted           & 0.1749 (0.0005) & 0.7837 (0.0025) & 0.6052 (0.0030) \\ \cline{2-5} 
                             & ECAP Opt             & 0.0019 (0.0000) & 0.0109 (0.0000) & 0.0086 (0.0000) \\ \cline{2-5} 
                             & JS Opt               & 0.0609 (0.0002) & 0.2431 (0.0005) & 0.1526 (0.0003) \\ \cline{2-5} 
                             & ECAP MLE             & 0.0026 (0.0001) & 0.0126 (0.0001) & 0.0100 (0.0001) \\ \cline{2-5} 
                             & JS MLE               & 0.0633 (0.0003) & 0.2712 (0.0014) & 0.1707 (0.0011) \\ \hline
\multirow{5}{*}{$\theta=-1$} & Unadjusted           & 0.0319 (0.0001) & 0.1389 (0.0007) & 0.1130 (0.0008) \\ \cline{2-5} 
                             & ECAP Opt             & 0.0051 (0.0000) & 0.0150 (0.0000) & 0.0124 (0.0001) \\ \cline{2-5} 
                             & JS Opt               & 0.0142 (0.0000) & 0.0477 (0.0001) & 0.0361 (0.0001) \\ \cline{2-5} 
                             & ECAP MLE             & 0.0059 (0.0001) & 0.0171 (0.0002) & 0.0149 (0.0003) \\ \cline{2-5} 
                             & JS MLE               & 0.0155 (0.0002) & 0.0541 (0.0008) & 0.0413 (0.0010) \\ \hline
\multirow{5}{*}{$\theta=0$}  & Unadjusted           & 0.0099 (0.0000) & 0.0305 (0.0002) & 0.0275 (0.0003) \\ \cline{2-5} 
                             & ECAP Opt             & 0.0084 (0.0000) & 0.0195 (0.0001) & 0.0164 (0.0001) \\ \cline{2-5} 
                             & JS Opt               & 0.0088 (0.0000) & 0.0199 (0.0001) & 0.0171 (0.0001) \\ \cline{2-5} 
                             & ECAP MLE             & 0.0093 (0.0001) & 0.0224 (0.0003) & 0.0200 (0.0004) \\ \cline{2-5} 
                             & JS MLE               & 0.0094 (0.0001) & 0.0233 (0.0005) & 0.0219 (0.0005) \\ \hline
\multirow{5}{*}{$\theta=2$}  & Unadjusted           & 0.0652 (0.0001) & 0.2419 (0.0003) & 0.1776 (0.0003) \\ \cline{2-5} 
                             & ECAP Opt             & 0.0240 (0.0001) & 0.0614 (0.0002) & 0.0502 (0.0001) \\ \cline{2-5} 
                             & JS Opt               & 0.0652 (0.0001) & 0.2419 (0.0003) & 0.1776 (0.0003) \\ \cline{2-5} 
                             & ECAP MLE             & 0.0255 (0.0002) & 0.0744 (0.0012) & 0.0599 (0.0009) \\ \cline{2-5} 
                             & JS MLE               & 0.0652 (0.0001) & 0.2419 (0.0003) & 0.1776 (0.0003) \\ 
\end{tabular}}
\end{table}

The results, again averaged over $100$ simulation runs, are presented in Table~\ref{table:biasedtable}. In the two settings where $\theta<0$ we note that the unadjusted and JS methods all exhibit significant deterioration in their performance relative to the unbiased $\theta=0$ scenario. By comparison, the two ECAP methods significantly outperform the JS and unadjusted approaches. A similar pattern is observed for $\theta>0$. In this setting all five methods deteriorate, but ECAP is far more robust to the biased setting than unadjusted and JS.

It is perhaps not surprising that the bias corrected version of ECAP outperforms the other methods when the data is indeed biased. However, just as interestingly, even in the unbiased setting ($\theta=0$) we still observe that ECAP outperforms its JS counterpart, despite the fact that ECAP must estimate $\theta$. This is likely a result of the fact that ECAP is able to accurately estimate $\theta$. Over all simulation runs and settings, ECAP Opt and ECAP MLE respectively averaged absolute errors of only~$0.0681$ and $0.2666$ in estimating $\theta$.






\section{Empirical Results}
\label{emp.sec}

In this section we illustrate ECAP on two real world data sets. Section~\ref{sec.espn} contains our results analyzing ESPN's probability estimates from NCAA football games, while Section~\ref{sec.538} examines probability estimates from the 2018 US midterm elections. Given that for real data $p_i$ is never observed, we need to compute an estimate of $EC(\hat p_i)$. Hence, we choose a small window $\delta$, for example $\delta =[0,0.02]$, and consider all observations for which $\tp_i$ falls within $\delta$.\footnote{In this section, for simplicity of notation, we have flipped all probabilities greater than $0.5$, and the associated $Z_i$ around $0.5$ so $\delta=[0,0.02]$ also includes probabilities between $0.98$ and $1$.} We then estimate $p_i$ via $\bar p_\delta=\frac{1}{n_\delta}\sum_{i=1}^n Z_i \delta_i$, where $\delta_i=I(\tp_i\in \delta)$ and $n_\delta=\sum_{i=1}^n \delta_i$. Hence we can estimate EC using
\begin{equation}
\label{se.ec}
    \widehat{EC}_\delta(\bar{\hat{p_\delta}}) = \frac{\bar p_\delta - \bar{\hat{p_\delta}}}{\bar{\hat{p_\delta}}},
\end{equation}
where $\bar{\hat{p_\delta}}=\frac{1}{n_\delta}\sum_{i=1}^n \hat p_i \delta_i$. 












\subsection{ESPN NCAA Football Data}
\label{sec.espn}

\begin{figure}[t]
	\centering
	\vspace{-3cm}
   \includegraphics[angle=270,origin=c,width=0.7\textwidth]{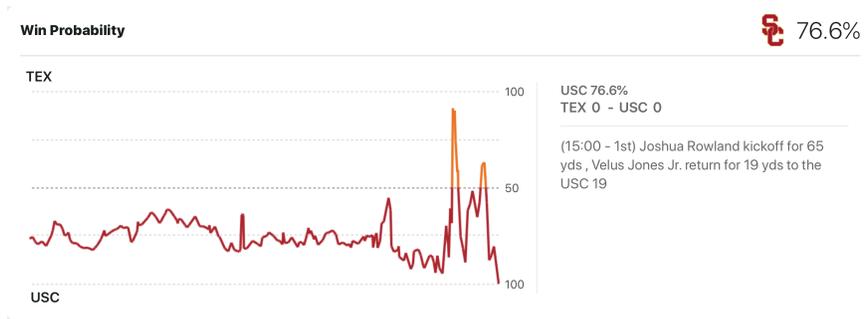}
	\vspace{-3cm}
	\caption{\label{espn_web.plot}\small A screenshot of the NCAA football win probabilities publicly available on ESPN's website. USC vs. Texas (2017)}
\end{figure}

Each year there are approximately 1,200 Division 1 NCAA football games played within the US. For the last several seasons ESPN has been producing automatic win probability estimates for every game. These probabilities update in real time after every play. 
Figure~\ref{espn_web.plot} provides an example of a fully realized game between the University of Southern California (USC) and the University of Texas at Austin (TEX) during the 2017 season. For most of the game the probability of a USC win hovers around 75\% but towards the end of the game the probability starts to oscillate wildly, with both teams having high win probabilities, before USC ultimately wins.\footnote{The game was not chosen at random.} These gyrations are quite common and occasionally result in a team with a high win probability ultimately losing. Of course even a team with a 99\% win probability will end up losing 1\% of the time so these unusual outcomes do not necessarily indicate an error, or selection bias issue, with the probability estimates.



To assess the accuracy of ESPN's estimation procedure we collected data from the 2016 and 2017 NCAA football seasons. We obtained this unique data set by scrapping the win probabilities, and ultimate winning team, for a total of 1,722 games (about 860 per season), involving an average of approximately 180 probabilities per game. Each game runs for 60 minutes, although the clock is often stopped. For any particular time point $t$ during these 60 minutes, we took the probability estimate closest to $t$ in each of the individual games. We used the entire data set, 2016 and 2017, to compute $\bar{p}_\delta$, which represents the ideal gold standard. However, this estimator is impractical in practice because we would need to collect data over two full years to implement it. By comparison, we used only the 2016 season to fit ECAP and ultimately to compute $\bar{\hat{p}}_\delta$. We then calculated $\widehat{EC}_\delta(\bar{\hat{p_\delta}},t)$ for both the raw ESPN probabilities and the adjusted ECAP estimates. The intuition here is that $\widehat{EC}_\delta(\bar{\hat{p_\delta}},t)$ provides a comparison of these estimates to the ideal, but unrealistic, $\bar{p}_\delta$.

\begin{figure}[t]
	    \centering
	     \scalebox{1}{%
		\input{Figures/ESPN_Revision_Plot.tex}}
		\caption{\label{espn.plot} Empirical EC in both the unadjusted and ECAP setting with $\delta=[0,0.02]$.}
		\label{plot:espn_ecap}
\end{figure}
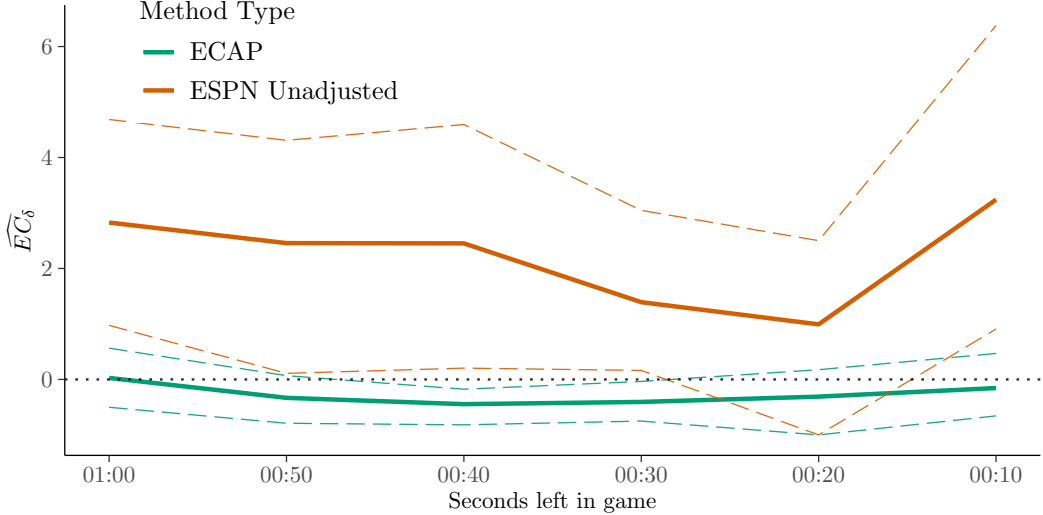

In general we found that $\widehat{EC}_\delta(\bar{\hat{p_\delta}},t)$ computed on the ESPN probabilities was not systematically different from zero, suggesting ESPN's probabilities were reasonably accurate. However, we observed that, for extreme values of $\delta$, $\widehat{EC}_\delta(\bar{\hat{p_\delta}},t)$ was well above zero towards the end of the games. Consider, for example, the solid orange line in Figure~\ref{espn.plot}, which plots $\widehat{EC_\delta}(\bar{\hat{p_\delta}},t)$ using $\delta=[0,0.02]$ at six different time points during the final minute of these games. We observe that excess certainty is consistently well above zero. 
The $90\%$ bootstrap confidence intervals (dashed lines), generated by sampling with replacement from the probabilities that landed inside $\delta_i$, demonstrate that the difference from zero is statistically significant for most time points.
This suggests that towards the end of the game ESPN's probabilities are too extreme i.e. there are more upsets then would be predicted by their estimates.

Next we applied the unbiased implementation of ECAP, i.e. with $\theta=0$, separately to each of these six time points and computed $\widehat{EC_\delta}(t)$ for the associated ECAP probability estimates. To estimate the out of sample performance of our method, we randomly picked half of the 2016 games to estimate $\gamma^*$, and then used ECAP to produce probability estimates on the other half. We repeated this process 100 times and averaged the resulting $\widehat{EC}_\delta(\bar{\hat{p_\delta}},t)$ independently for each time point.  The solid green line in Figure~\ref{espn.plot} provides the estimated excess certainty. ECAP appears to work well on this data, with excess certainty estimates close to zero and confidence intervals that contain zero at most time points. Notice also that ECAP is consistently producing a slightly negative excess certainty, which is actually necessary to minimize the expected loss function \eqref{loss.fn}, as demonstrated in Figure~\ref{EC.plot}. Interestingly this excess certainty pattern in the ESPN probabilities is no longer apparent in data for the 2018 season, suggesting that ESPN also identified this as an issue and applied a correction to their estimation procedure.

\subsection{Election Data}
\label{sec.538}

Probabilities have increasingly been used to predict election results. For example, news organizations, political campaigns, and others, often attempt to predict the probability of a given candidate winning a governors race, or a seat in the house, or senate. Among other uses, political parties can use these estimates to optimize their funding allocations across hundreds of different races. 
In this section we illustrate ECAP using probability estimates produced by the  \href{www.FiveThirtyEight.com}{FiveThirtyEight.com} website during the 2018 US midterm election cycle. FiveThrityEight used three different methods, {\it Classic, Deluxe}, and {\it Lite}, to generate probability estimates for every governor, house, and senate seat up for election, resulting in 506 probability estimates for each of the three methods. 


Interestingly a previous analysis of this data \citep{Silver18} showed that the FiveThirtyEight probability estimates appeared to be overly conservative i.e. the leading candidate won more often than would have been predicted by their probabilities. Hence, we should be able to improve the probability estimates using the bias corrected version of ECAP from Section~\ref{biased.sec}. We first computed $\widehat{EC}_\delta(\bar{\hat{p_\delta}})$ on the unadjusted FiveThirtyEight probability estimates using two different values for $\delta$ i.e. $\delta_1=[0,0.1]$ and $\delta_2=[0.1,0.2]$. We used wider windows for $\delta$ in comparison to the ESPN data because we only had one third as many observations.  The results for the three methods used by FiveThirtyEight are shown in Table~\ref{table:538table}. Notice that for all three methods and both values of $\delta$ the unadjusted estimates are far below zero and several are close to $-1$, the minimum possible value. These results validate the previous analysis suggesting the FiveThirtyEight estimates are systematically conservatively biased. 


\begin{table}[t]
\centering
\captionof{table}{Bias corrected ECAP adjustment of FiveThirtyEight's 2018 election probabilities. Reported average $\widehat{EC}_\delta$.} \label{table:538table}
{\small \begin{tabular}{ll||rr}
Method               & Adjustment & \textbf{$\delta_1$} & \textbf{$\delta_2$} \\ \hline \hline
\multirow{2}{*}{\textbf{Classic}} & Unadjusted          & -0.6910          & -0.8361          \\ \cline{2-4} 
                                  & ECAP                & -0.2880          & -0.0734          \\ \hline
\multirow{2}{*}{\textbf{Deluxe}}  & Unadjusted          & -0.4276          & -0.8137          \\ \cline{2-4} 
                                  & ECAP                & -0.0364          & 0.1802           \\ \hline
\multirow{2}{*}{\textbf{Lite}}    & Unadjusted          & -0.8037          & -0.8302          \\ \cline{2-4} 
                                  & ECAP                & -0.3876         & -0.1118         \\ 
\end{tabular}}
\end{table}

\begin{figure}[t]
	    \centering
	     \scalebox{1}{%
		\input{Figures/cap_tilde.tex}}
		\caption{ECAP bias corrected probabilities vs original FiveThirtyEight probability from classic method.}
		\label{plot:ecap_v_538unadj}
\end{figure}
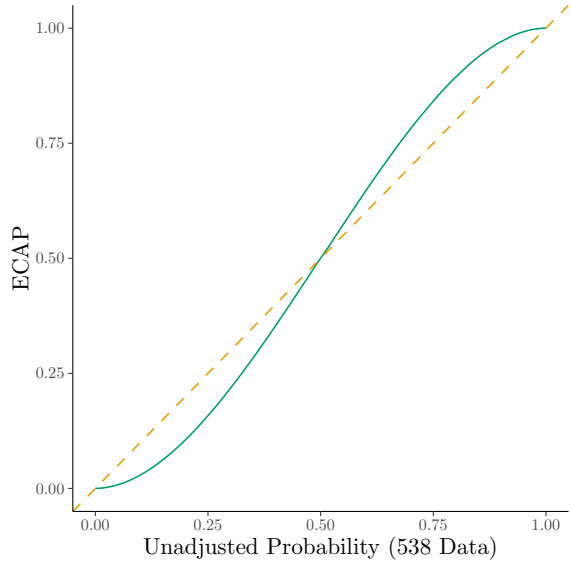

Next we applied ECAP separately to each of the three sets of probability estimates, with the value of $\theta$ chosen using the MLE approach previously described. Again the results are provided in Table~\ref{table:538table}. ECAP appears to have significantly reduced the level of bias, with most values of $\widehat{EC}_\delta(\bar{\hat{p_\delta}})$ close to zero, and in one case actually slightly above zero. For the Deluxe method with $\delta_1$, ECAP has an almost perfect level of excess certainty. 
For the Classic and Lite methods, $\theta=2$ was chosen by ECAP for both values of~$\delta$, representing the largest possible level of bias correction.  For the Deluxe method, ECAP selected $\theta=1.9$. Figure~\ref{plot:ecap_v_538unadj} demonstrates the significant level of correction that ECAP applies to the classic method FiveThirtyEight estimates. For example, ECAP adjusts probability estimates of $0.8$ to $0.89$ and estimates of $0.9$ to $0.97$.

\section{Discussion}
\label{discussion.sec}

In this article, we have convincingly demonstrated both theoretically and empirically that probability estimates are subject to selection bias, even when the individual estimates are unbiased. Our proposed ECAP method applies a novel non-parametric empirical Bayes approach to adjust both biased and unbiased probabilities, and hence produce more accurate estimates. The results in both the simulation study and on real data sets demonstrate that ECAP can successfully correct for selection bias, allowing us to use the probabilities with a higher level of confidence when selecting extreme values.

There are a number of possible areas for future work. For example, the ESPN data contains an interesting time series structure to the probabilities, with each game consisting of a probability function measured over 60 minutes. Our current method treats each time point independently and adjusts the probabilities accordingly. However, one may be able to leverage more power by incorporating all time points simultaneously using some form of functional data analysis. Another potential area of exploration involves the type of data on which ECAP is implemented. For example, consider a setting involving a large number of hypothesis tests and associated p-values, $\tp_1,\ldots, \tp_n$. There has been much discussion recently of the limitations around using p-values. A superior approach would involve thresholding based on the posterior probability of the null hypothesis being true i.e. $p_i=P(H_{0i}|X_i)$. Of course, in general, $p_i$ is difficult to compute which is why we use the p-value $\tp_i$. However, if we were to treat $\tp_i$ as a, possibly biased, estimate of $p_i$, then it may be possible to use a modified version of ECAP to estimate $p_i$. If such an approach could be implemented it would likely have a significant impact in the area of multiple hypothesis testing.



\appendix

\section{Proof of Theorem~\ref{oracle.thm}}

We begin by computing the derivative of the loss function,
$$L(x)=\begin{cases}
\frac{1}{x^2}E(p_i^2|\tp_i)-\frac2x E(p_i|\tp_i)+1&x\le 0.5\\
\frac{1}{(1-x)^2}E(p_i^2|\tp_i)-\frac{2x}{(1-x)^2} E(p_i|\tp_i)+\left(\frac{x}{1-x}\right)^2&x> 0.5.
\end{cases}$$
We have
\begin{eqnarray*}
\frac{\partial L}{\partial x} &=& \begin{cases}
-\frac{2}{x^3}E(p_i^2|\tp_i)+\frac2{x^2} E(p_i|\tp_i)&x< 0.5\\
\frac{2}{(1-x)^3}E(p_i^2|\tp_i)-\frac{2(1+x)}{(1-x)^3} E(p_i|\tp_i)+\frac{2x}{(1-x)^3}&x> 0.5
\end{cases}\\
&\propto& \begin{cases}
-E(p_i^2|\tp_i)+x E(p_i|\tp_i)&x< 0.5\\
E(p_i^2|\tp_i)-E(p_i|\tp_i)+x(1-E(p_i|\tp_i)) &x> 0.5.
\end{cases}
\end{eqnarray*}
Note that~$L$ is a continuous function.  If $E(p_i|\tp_i)\le 0.5$ and $x^*=E(p_i^2|\tp_i)/E(p_i|\tp_i)\le0.5$ then algebraic manipulations show that $\partial L/\partial x$ is negative for all $x<x^*$ and positive for $x>x^*$. Hence, $p_{i0}=x^*=E(p_i|\tp_i) + Var(p_i|\tp_i)/E(p_i|\tp_i)$ minimizes~$L$. Alternatively, if $E(p_i|\tp_i)\le0.5$ and $x^*=E(p_i^2|\tp_i)/E(p_i|\tp_i)\ge0.5$ then $\partial L/\partial x$ is negative for all $x<0.5$ and positive for all $x>0.5$, so $L$ is minimized by $p_{i0}=0.5$.

Analogous arguments show that if $E(p_i|\tp_i)>0.5$ and $x^*=E(p_i^2|\tp_i)/(1-E(p_i|\tp_i))>0.5$, then $\partial L/\partial x$ is negative for all $x<x^*$, zero at $x=x^*$ and positive for $x>x^*$. Hence, $p_{i0}=x^*=E(p_i|\tp_i) + Var(p_i|\tp_i)/(1-E(p_i|\tp_i))$ will minimize $L$. Alternatively, if $E(p_i|\tp_i)>0.5$ and $x^*=E(p_i^2|\tp_i)/(1-E(p_i|\tp_i))<0.5$ then $\partial L/\partial x$ is negative for all $x<0.5$ and positive for all $x>0.5$, so $L$ is minimized by $p_{i0}=0.5$.

To prove the second result, first suppose $E(p_i|\tp_i)\le0.5$ and $p_{i0}<0.5$, in which case $L(p_{i0}) = 1 - E(p_i^2|\tp_i)/p_{i0}^2$. Now 
let $\tilde L(p_i') = E\left(\left(\frac{p_i-p_i'}{p_i'}\right)^2|\tp_i\right)= \frac{1}{{p_i'}^2}E(p_i^2|\tp_i) - \frac{2}{p_i'}E(p_i|\tp_i)+1$. Note that $\tilde L(p_i')\le L(p_i')$ with equality for $p_i'\le 0.5$.  Hence,
$$L(p_i')-L(p_{i0})\ge \tilde L(p_i')-L(p_{i0}) = E(p_i^2|\tp_i)\left(\frac{1}{{p_i'}^2} + \frac{1}{p_{i0}^2}\right)- \frac{2}{p_i'}E(p_i|\tp_i)  = E(p_i^2|\tp_i)\left( \frac{1}{p_i'}-\frac{1}{p_{i0}}\right)^2.$$

Now consider the case $E(p_i|\tp_i)\le0.5$ and $p_{i0}=0.5$.  Note that this implies $2E(p_i^2|\tp_i)>E(p_i|\tp_i)$.  If~$p_i'\le0.5$, then
$$L(p_i')-L(p_{i0})= E(p_i^2|\tp_i)\left(\frac{1}{{p_i'}^2} -4\right)- 2E(p_i|\tp_i)\left(\frac{1}{{p_i'}} -2\right)  \ge E(p_i^2|\tp_i)\left( \frac{1}{p_i'}-\frac{1}{0.5}\right)^2.$$
Also note that $$\left( \frac{1}{p_i'}-\frac{1}{0.5}\right)^2\ge
\left( \frac{1}{1-p_i'}-\frac{1}{0.5}\right)^2.
$$
Alternatively, if~$p_i'>0.5$, then
$$L(p_i')-L(p_{i0})\ge \tilde{L}(1-p_i')-L(p_{i0})\ge E(p_i^2|\tp_i)\left( \frac{1}{1-p_i'}-\frac{1}{0.5}\right)^2.$$
Observe that $$\left( \frac{1}{1-p_i'}-\frac{1}{0.5}\right)^2\ge
\left( \frac{1}{p_i'}-\frac{1}{0.5}\right)^2.
$$
Consequently, we have shown that 
$$L(p_i')-L(p_{i0})\ge  E(p_i^2|\tp_i)
\max\left(\left[ \frac{1}{p_i'}-\frac{1}{0.5}\right]^2,
\left[\frac{1}{1-p_i'}-\frac{1}{0.5}\right]^2\right)
$$
when $E(p_i|\tp_i)\le0.5$ and $p_{i0}=0.5$.

Thus, we have established the result for the case $E(p_i|\tp_i)\le0.5$  Finally, consider the case $E(p_i|\tp_i)\ge0.5$.  The result follows by repeating the argument from the case $E(p_i|\tp_i)<0.5$ while replacing all of the probabilities with their complements, i.e., by replacing~$p_i$, $p_{i0}$ and $p_i'$ with $1-p_i$, $1-p_{i0}$ and $1-p_i'$, respectively.


\section{Proof of Theorem~\ref{EandVar} and Corollary~\ref{cor.EandVar}}
Throughout the proof, we omit the subscript~$i$, for the simplicity of notation.  We let $f_{\tp}(\tp|p)$ denote the conditional density of $\tp$ given that the corresponding true probability equals~$p$, and define $f_{X}(x|p)$ by analogy for the random variable~$X=\log(\tp/[1-\tp])$.  We will slightly abuse the notation and not distinguish between the random variable and its value in the case of $\tp$, $p$ and~$\eta$.  

According to model~\eqref{beta.model}, we have $f_{\tp}(\tp|p)=B(p/{\gamma^*},(1-p)/{\gamma^*})^{-1}\tp^{p/{\gamma^*}-1}(1-\tp)^{(1-p)/{\gamma^*}-1}$ where~$B(\cdot)$ denotes the beta function. Hence, writing~$B$ for $B(p/{\gamma^*},(1-p)/{\gamma^*})$, we derive
\begin{eqnarray}
\log(f_{\tp}(\tp|p)) &=& -\log B + \left(\frac{p}{{\gamma^*}}-1\right)[\log \tp - \log(1-\tp)] + (1/{\gamma^*}-2)\log(1-\tp)\nonumber\\
\label{eq.exp.fam}
&=& -\log B + \eta x + (1/{\gamma^*}-2)\log(1-\tp) 
\end{eqnarray}
where $\eta=\frac{p}{{\gamma^*}}-1$ and $x=\log \frac{\tp}{1-\tp}$.
Standard calculations show that 
\begin{equation}
\label{ptilde.dist}
    f_X(x|p)= f_{\tp}(\tp|p) \frac{e^x}{(1+e^x)^2} = f_{\tp}(\tp|p) \tp(1-\tp).
\end{equation}
Note that $\log(1-\tp)=-\log(1+e^x)$, and hence 
\begin{eqnarray*}
\log(f_{X}(x|p))&=&-\log B + \eta x - (1/{\gamma^*}-2)\log(1+e^x)+x - 2\log(1+e^x)\\
&=& -\log B + \eta x +x- 1/{\gamma^*}\log(1+e^x)\\
&=& -\log B + \eta x +l_h(x),
\end{eqnarray*}
where $l_h(x)=x- 1/{\gamma^*}\log(1+e^x)$.

Consequently, we can apply Tweedie's formula \citep{efron2011} to derive
$$E(p/{\gamma^*}-1|\tp)=E(\eta|x)=v_X(x)-l_h'(x)=v_X(x)-1 + \frac{1}{{\gamma^*}}\frac{e^x}{1+e^x}= v_X(x)+\frac{\tp}{{\gamma^*}}-1,$$
where $v_{X}(x)=(d f_{X}(x)/dx)/f_{X}(x)$ and $f_X$ is the density of~$X$.  This implies
$$E(p|\tp) = \tp+{\gamma^*} v_X(x).$$
In addition, we have
\begin{eqnarray*}
\frac{d f_X(x)}{dx}&=& \frac{d f_{\tp}(\tp)}{d\tp}\frac{d\tp}{dx}\frac{e^x}{(1+e^x)^2}+ f_{\tp}(\tp) \frac{e^x(1+e^x)^2-2e^{2x}(1+e^x)}{(1+e^x)^4}\\
&=& \frac{d f_{\tp}(\tp)}{d\tp}\left(\frac{e^x}{(1+e^x)^2}\right)^2+ f_{\tp}(\tp)\frac{e^x}{(1+e^x)^2}\frac{1-e^x}{1+e^x}.
\end{eqnarray*}
Using the unconditional analog of formula~\eqref{ptilde.dist}, we derive
\begin{eqnarray*}
v_X(x)&=&\frac{d f_X(x)/dx}{f_X(x)}\\
&=& \frac{d f_{\tp}(\tp)/d\tp}{f_{\tp}(\tp)}\frac{e^x}{(1+e^x)^2}+\frac{1-e^x}{1+e^x}\\
&=& v_{\tp}(\tp) \tp(1-\tp) +1-2\tp,
\end{eqnarray*}
where $v_{\tp}(\tp)=(d f_{\tp}(\tp)/d\tp)/f_{\tp}(\tp).$
Thus,
$$E(p|\tp) = \tp + {\gamma^*}(\tp(1-\tp)v_{\tp}(\tp)+ 1-2\tp).$$
Similarly, again by Tweedie's formula,
$$Var(p/{\gamma^*}-1|\tp)=Var(\eta|x)=v'_X(x)-l_h''(x)=v'_X(x)+\frac{1}{{\gamma^*}}\frac{e^x}{(1+e^x)^2}= v'_X(x)+\frac{\tp(1-\tp)}{{\gamma^*}},$$
which implies
$$Var(p|\tp) = {\gamma^*}\tp(1-\tp)+{\gamma^*}^2 v'_X(x).$$
Noting that $$v'_X(x) =\tp(1-\tp)[v'_{\tp}(\tp) \tp(1-\tp) + v_{\tp}(\tp)(1-2\tp)-2],$$
we derive
$$Var(p|\tp) = {\gamma^*}^2\tp(1-\tp)[v'_{\tp}(\tp) \tp(1-\tp) + v_{\tp}(\tp)(1-2\tp)-2]+ {\gamma^*}\tp(1-\tp).$$
If we define $g^*(\tp)=\tp(1-\tp)v_{\tp}(\tp)$, then 
\begin{eqnarray*}
E(p|\tp) &=& \tp + {\gamma^*}(g(\tp)+1-2\tp)\\
Var(p|\tp) &=& {\gamma^*}\tp(1-\tp)+{\gamma^*}^2\tp(1-\tp)[g'(\tp)-2].
\end{eqnarray*}
This  completes the proof of Theorem~\ref{EandVar}.

Finally, we establish some properties of $g^*(\tp)$ and prove Corollary~\ref{cor.EandVar}. We denote the marginal density of~$\tp$ by~$f$. First note that $g(1-\tp)=-\tp(1-\tp)f'(1-\tp|p)/f(1-\tp|p)$. If $h(p)$ represents the prior density for~$p$, then
\begin{equation}
\label{f.marginal}
    f(\tp)=\int_0^{1} B(\alpha,\beta)^{-1}\tp^{\alpha-1}(1-\tp)^{\beta-1}h(p)dp.
\end{equation}
Because function~$h$ is bounded, differentiation under the integral sign is justified, and hence
\begin{equation}
\label{f.prime}
    f'(\tp)=\int_0^{1}  B(\alpha,\beta)^{-1}\left\{(\alpha-1)\tp^{\alpha-2}(1-\tp)^{\beta-1}-(\beta-1)\tp^{\alpha-1}(1-\tp)^{\beta-2}\right\}h(p)dp, 
    \end{equation}
where $\alpha=p/{\gamma^*}$ and $\beta=(1-p)/{\gamma^*}$. Substituting $p^*=1-p$ we get 
$$f(1-\tp)=\int_0^{1} B(\beta,\alpha)^{-1}\tp^{\alpha-1}(1-\tp)^{\beta-1}h(1-p^*)dp^*=f(\tp)$$
and 
$$f'(1-\tp)=\int_0^{1}  B(\beta,\alpha)^{-1}\left\{(\alpha-1)\tp^{\alpha-2}(1-\tp)^{\beta-1}-(\beta-1)\tp^{\alpha-1}(1-\tp)^{\beta-2}\right\}h(1-p^*)dp^*=f'(\tp),$$
provided $h(p)=h(1-p)$. Hence, $g^*(1-\tp)=-\tp(1-\tp)f'(\tp)/f(\tp)=-g^*(\tp)$. By continuity of $g^*(\tp)$ this result also implies $g^*(0.5)=0$. 

To complete the proof of Corollary~\ref{cor.EandVar}, we note that under the assumption that the distribution of~$p_i$ is symmetric, the conditional expected value $E(p_i|\tp_i)$ lies on the same side of~$0.5$ as~$\tp_i$.

\section{Proof of Theorem~\ref{risk.lemma}}

As before, we denote the marginal density of~$\tp$ by~$f$.
First, we derive a bound for~${g^*}$. Note that $-1\le \alpha-1\le \frac{1}{{\gamma^*}}$ and, similarly, $-1\le \beta-1\le \frac{1}{{\gamma^*}}$. Hence, by \eqref{f.marginal} and \eqref{f.prime},
\begin{equation*}
  \nonumber  -\left(1-\tp+\frac{\tp}{{\gamma^*}}\right)f(\tp)\le \tp(1-\tp) f'(\tp)\le \left((1-\tp)\frac{1}{{\gamma^*}}+\tp\right)f(\tp),
\end{equation*}
which implies
\begin{equation}
\label{g.bound}
|{g^*}(\tp)|\le \frac{1}{{\gamma^*}}.
\end{equation}
Next, note that
\begin{eqnarray}
\lim_{\tp\rightarrow 0} \tp(1-\tp) f(\tp) &=& 0\label{lim1}\qquad\text{and}\\
\lim_{\tp\rightarrow 1} \tp(1-\tp) f(\tp) &=& 0.\label{lim2}
\end{eqnarray}
Observe that
\begin{eqnarray}
R( g(\tp)) &=& E(\left( g(\tp)- {g^*}(\tp)\right)^2\nonumber\\
&=& E g(\tp)^2-2E\left\{ g(\tp) {g^*}(\tp)\right\}+C\nonumber\\
&=& E g(\tp)^2-2\int_0^1 \left\{ g(\tp) \tp(1-\tp) \frac{f'(\tp)}{f(\tp)}\right\} f(\tp)d\tp+C\nonumber\\
&=& E g(\tp)^2-2\left[ g(\tp)\tp(1-\tp) f(\tp)\right]^1_0+2\int_0^1 \left[ g(\tp)(1-2\tp)+\tp(1-\tp)  g'(\tp)\right]f(\tp)d\tp +C \nonumber\\
&=& E g(\tp)^2+2\int_0^1 \left[ g(\tp)(1-2\tp)+\tp(1-\tp)  g'(\tp)\right]f(\tp)d\tp +C\label{full.risk.prf}
\end{eqnarray}
where $C$ is a constant that does not depend on~$g$, and the second to last line follows via integration by parts. Note the last line holds when~$ g$ is bounded, because by \eqref{lim1},
$$\lim_{\tp\rightarrow 0}  g(\tp)\tp(1-\tp) f(\tp) = 0,$$
and by \eqref{lim2},
$$\lim_{\tp\rightarrow 1}  g(\tp)\tp(1-\tp) f(\tp) = 0.$$  
In particular, due to the inequality~(\ref{g.bound}), the relationship~(\ref{full.risk.prf}) holds when~$ g$ is the true function~${g^*}$.

\section{Proof of Theorem~\ref{g.thm}}
\label{sec:proof.asympt}
We write $\mathcal{G}_N$ for the class of all natural cubic spline functions $g$ on $[0,1]$ that correspond to the sequence of~$n$ knots located at the observed~$\tp_i$.  Given a function~$g$, we define $s_g(\tp)=2[g(\tp)(1-2\tp)+\tp(1-\tp)g'(\tp)]$ and $I^2(g) = \int_0^1 [g''(\tp)]^2 d\tp$.   We also denote $(1/n)\sum_{i=1}^n g^2(\tp_i)$ and $\int_0^1 g(\tp)f^*(\tp)d\tp$ by~$\|g\|^2_n$ and~$\|g\|^2$, respectively.

By Lemma~\ref{lem1} in Appendix~\ref{append.sup.res}, there exists $g^*_N\in\mathcal{G}_N$, such that $\|g^*_N-g^*\|^2=O_p(\lambda_n^2)$ and
\begin{equation*}
\|\hat g - g^*_N\|^2 + \lambda_n^2 I^2(\hat g)\le
O_p\Big(n^{-2/7}\|\hat g - g^*_N\|\Big) + O_p\Big(n^{-4/7} I(\hat g) \Big)+O_p\Big(n^{-4/7} +\lambda_n^2\Big).
\end{equation*}


We consider two possible cases (a) $n^{-4/7} I(\hat g)\le n^{-2/7}\|\hat g - g^*_N\|+n^{-4/7}+\lambda_n^2$ and (b) $n^{-4/7} I(\hat g)> n^{-2/7}\|\hat g - g^*_N\|+n^{-4/7}+\lambda_n^2$.

Under (a) we have
\begin{equation}
\|\hat g - g^*_N\|^2 +\lambda^2_n I^2(\hat g)\le  O_p\Big(n^{-2/7}\|\hat g - g^*_N\|\Big) +O_p\Big(n^{-4/7}+\lambda^2_n\Big).
\end{equation}
It follows that $\|\hat g - g^*_N\|=O_p(n^{-2/7}+\lambda_n)$ and $I^2(\hat g )=O_p(n^{-4/7}\lambda_n^{-2}+1)$.  However, taking into account the case (a) condition, we also have $I^2(\hat g )=O_p(n^{4/7}\lambda^2_n+1)$, thus leading to $I(\hat g )=O_p(1)$.

Under (b) we have
\begin{equation}
\label{case.b.ineq}
\|\hat g - g^*_N\|^2 + \lambda_n^2 I^2(\hat g)\le O_p\Big(n^{-4/7}I(\hat g )\Big).
\end{equation}
It follows that $I(\hat g )=O_p(n^{-4/7}\lambda_n^{-2})$ and $\|\hat g - g^*_N\|=O_p(n^{-4/7}\lambda_n^{-1})$.

Collecting all the stochastic bounds we derived, and using the fact that~$f^*$ is bounded away from zero, we deduce
\begin{equation*}
\|\hat g - g^*_N\| = O_p(n^{-4/7}\lambda_n^{-1}+n^{-2/7}+\lambda_n) \qquad\text{and}\qquad
I(\hat g)=O_p(1+n^{-4/7}\lambda_n^{-2})
\end{equation*}
Using the bound $\|g^*_N-g^*\|^2=O_p(\lambda_n^2)$, together with the definitions of $r_n$ and~$s_n$, we derive
\begin{equation}
\label{g.bounds}
\|\hat g - g^*\| = O_p(r_n)   \qquad\text{and}\qquad I(\hat g - g^*)=O_p(1+n^{-4/7}\lambda_n^{-2}).
\end{equation}
Applying Lemma 10.9 in \cite{van2000applications}, which builds on the interpolation inequality of \cite{agmon1965lectures}, we derive $\|\hat{g}' - {g^*}'\| = O_p(\sqrt{r_n s_n})$.
This establishes the error bounds for $\hat g$ and $\hat g'$ with respect to the $\|\cdot\|$ norm.  

To derive the corresponding results with respect to the $\|\cdot\|_n$ norm, we first apply bound~(\ref{nu_n_dg2}), in which we replace~$g^*_N$ with~$g^*$.  It follows that
\begin{equation*}
 \|\hat g - g^*\|^2_n-\|\hat g - g^*\|^2 = 
 ({\tilde P}_n-{\tilde P})[\hat g-{g^*}]^2=
 o_p\Big(\|\hat g - g^*\|^2\Big) + O_p\Big(n^{-1}I^2(\hat g - g^*)\Big),
\end{equation*}
where we use the notation from the proof of Lemma~\ref{lem1}.
Because bounds~(\ref{g.bounds}) together with the assumption $\lambda_n\gg n^{-8/21}$ imply
\begin{equation*}
I(\hat g - g^*)=O\Big(n^{-4/7}n^{16/21}\Big)=O\Big(n^{4/21}\Big),
\end{equation*}
we can then derive
\begin{equation*}
 \|\hat g - g^*\|^2_n=O\Big(\|\hat g - g^*\|^2\Big) + O_p\Big(n^{-13/21}\Big).
\end{equation*}
Because $r_n\ge n^{-2/7}$, we have $r_n^2\ge n^{-13/21}$.  Consequently, $ \|\hat g - g^*\|^2_n=O(r_n^2)$,
which establishes the analog of the first bound in~(\ref{g.bounds}) for the $\|\cdot\|_n$ norm. 

It is only left to derive $\|\hat{g}' - {g^*}'\|_n=O_p(\sqrt{r_ns_n})$.  Applying Lemma~17 in \cite{meier2009high}, in conjunction with Corollary~5 from the same paper, in which we take ${\gamma^*}=2/3$ and $\lambda=n^{-3/14}$, we derive
\begin{equation*}
({\tilde P}_n-{\tilde P})[\hat{g}' - {g^*}']^2 =O_p\Big(n^{-5/14}\Big[ \|\hat{g}' - {g^*}'\|^2 + n^{-2/7} I^2(\hat{g}' - {g^*}')\Big] \Big).
\end{equation*}
Consequently,
\begin{equation*}
\|\hat{g}' - {g^*}'\|^2_n =O_p\Big(\|\hat{g}' - {g^*}'\|^2\Big)+O_p\Big(n^{-9/14}I^2(\hat{g}' - {g^*}') \Big).
\end{equation*}
Taking into account bound~(\ref{g.bounds}), the definition of~$s_n$, the assumption $\lambda_n\gg n^{-8/21}$ and the inequality $r_n\ge n^{-2/7}$,we derive
\begin{equation*}
n^{-9/14}I^2(\hat{g}' - {g^*}')=O_p\Big(n^{-9/14}s_n^2\Big)=O_p\Big(n^{-19/42}s_n\Big)=O_p\Big(r_ns_n\Big).
\end{equation*}
Thus, $\|\hat{g}' - {g^*}'\|^2_n=O_p(\|\hat{g}' - {g^*}'\|^2+r_ns_n)=O_p(r_ns_n)$, which completes the proof.

\section{Proof of Theorem~\ref{cons.thm}}
We will take advantage of the results in Theorem~\ref{rate.thm}, which are established independently from Theorem~\ref{cons.thm}.  We will focus on proving the results involving integrals, because the results for the averages follow by an analogous argument with minimal modifications.

We start by establishing consistency of $\hat p$.  Fixing an arbitrary positive~$\tilde\epsilon$, identifying a positive~$\epsilon$ for which $\tilde P(0,\epsilon)+\tilde P(1-\epsilon,1)\le \tilde\epsilon/2$, and noting that $\hat p$ and $p_0$ fall in $[0,1]$ for every~$\tp$, we derive
\begin{equation*}
\|\hat p - p_0 \|^2 \le \tilde\epsilon/2 + \int_{\epsilon}^{1-\epsilon}|\hat p(\tp)-p_0(\tp)|^2f^*(\tp)dp.
\end{equation*}
By Theorem~\ref{rate.thm}, the second term on the right-hand side of the above display is $o_p(1)$.  Consequently, 
\begin{equation*}
P\Big(\|\hat p - p_0 \|^2 >\tilde\epsilon\Big)\rightarrow0 \quad\text{as}\quad n\rightarrow\infty.
\end{equation*}
As the above statement holds for every fixed positive~$\tilde\epsilon$, we have established that $\|\hat p - p_0 \|=o_p(1)$.

We now focus on showing consistency for~$\hat W$.  Note that
\begin{equation*}
[\hat\mu^2(\tp)+\hat\sigma^2(\tp)]^2/\hat\mu^2(\tp) \ge 
[2\hat\mu(\tp)\hat\sigma(\tp)]^2/[\hat\mu^2(\tp)]=4\hat\sigma^2(\tp).
\end{equation*}
Thus, the definition of~$\hat p$ implies $\hat p^2(\tp)\ge \hat\sigma^2(\tp) \wedge 0.25$, and also $\hat p(\tp)\ge \hat\mu(\tp)\wedge 0.5$, for every~$\tp\in(0,1)$.
Writing~$p$ for the true probability corresponding to the observed~$\tp$, we then derive
\begin{eqnarray}
    \widehat W(\tp)=\frac{E_p\Big[(p-\hat p(\tp))^2|\tp\Big]}{\hat p^2(\tp)}&\le& \frac{\sigma^2(\tp)}{\hat p^2(\tp)}+
    \frac{[\hat p(\tp) -\mu(\tp)]^2}{\hat p^2(\tp)}\nonumber\\
    \nonumber\\
        &\le&
    \frac{|\hat\sigma^2(\tp)-\sigma^2(\tp)|}{4\hat\sigma^2(\tp)}+
    \frac{[\hat \mu(\tp) -\mu(\tp)]^2}{\hat\sigma^2(\tp)}+7. \label{W.hat.bnd}
\end{eqnarray}
By Theorem~\ref{rate.thm}, we have  $\|\hat\sigma^2-\sigma^2\|=O_p(\sqrt{r_ns_n})=o_p(1)$ and $\|\hat\mu-\mu\|=O_p(\sqrt{r_ns_n})=o_p(1)$. Fix an arbitrary positive~$\epsilon$ and define $A_{\epsilon}=(0,\epsilon)\cup(1-\epsilon,1)$.  Applying the Cauchy-Schwarz inequality, and using the imposed technical modification of the ECAP approach to bound $\hat\sigma^2$ below, we derive
\begin{equation*}
    \int_{A_{\epsilon}} \frac{|\hat\sigma^2(\tp)-\sigma^2(\tp)|}{\hat\sigma^2(\tp)}f^*(\tp)d\tp\le
    [\tilde PA_{\epsilon}]^{1/2} \frac{\|\hat\sigma^2-\sigma^2\|}{c\sqrt{r_ns_n}}=
    [\tilde PA_{\epsilon}]^{1/2}O_p(1)=O_p(\epsilon^{1/2}).
\end{equation*}
Similarly, we derive
\begin{equation*}
    \int_{A_{\epsilon}} \frac{|\hat\mu(\tp)-\mu(\tp)|^2}{\hat\sigma^2(\tp)}f^*(\tp)d\tp\le
    \int_{A_{\epsilon}} \frac{|\hat\mu(\tp)-\mu(\tp)|}{\hat\sigma^2(\tp)}f^*(\tp)d\tp=
    O_p(\epsilon^{1/2}).
\end{equation*}
Note that $|W_0(\tp)|\le 1$ for every~$\tp$. Thus, combining the bounds for the terms in~(\ref{W.hat.bnd}) with the corresponding bound for $|\hat W - W_0|$ in Theorem~\ref{rate.thm}, we derive
\begin{equation*}
\int_0^1 \big|\widehat W(\tp) - W_0(\tp)\big|f^*(\tp)d\tp=O_p(\epsilon^{1/2})+o_p(1).
\end{equation*}
As this bound holds for every positive~$\epsilon$, we deduce that $\int_0^1 \big|\widehat W(\tp) - W_0(\tp)\big|f^*(\tp)d\tp=o_p(1)$.

\section{Proof of Theorem~\ref{rate.thm}}

We build on the results of Theorem~\ref{g.thm} to derive the rate of convergence for~$\hat\mu$ and~$\widehat{W}$ for a fixed positive~$\epsilon$.  Continuity and positivity of $\mu(\tp)$ and $p_0(\tp)$ imply that both functions are bounded away from zero on the interval $[\epsilon,1-\epsilon]$. Applying Lemma 10.9 in \cite{van2000applications}, we derive  $\|\hat g - g^*\|_{\infty} = O_p(r_n^{3/4} s_n^{1/4})$.
Because $n^{-8/21}\ll\lambda_n\ll 1$, we have $\|\hat g - g^*\|_{\infty} = o_p(1)$, which implies $\sup_{[\epsilon,1-\epsilon]}|\hat \mu(\tp) - \mu(\tp)| = o_p(1)$.  Also note that $\hat p(\tp)\ge \hat \mu(\tp)$ for all~$\tp$.  Consequently, there exists an event with probability tending to one, on which random functions~$\hat p(\tp)$ and $\hat\mu(\tp)$ are bounded away from zero on the interval $[\epsilon,1-\epsilon]$.  The stated error bounds for~$\hat p$ then follow directly from this observation and the error bounds for~$\hat g$ and~$\hat g'$ in Theorem~\ref{g.thm}.

For the remainder of the proof we restrict our attention to the event~A (whose probability tends to one), on which functions~$p_0$ and~$\hat p$ are both bounded away from zero on $[\epsilon,1-\epsilon]$.  We write~$p$ for the true probability corresponding to the observed~$\tp$, define $G(q)=E[(p-q)^2/q^2|\tp]$
and note that $G'(q)={2(q-p^*)E(p|\tp)}/{q^3}$.  
Let $p^*$ be the minimizer of $G$, given by $p^*=E[p|\tp]+{Var[p|\tp]}/{E[p|\tp]}$.
Denote by $\hat p^*$ our estimator of~$p^*$, which is obtained by replacing the conditional expected value and variance in the above formula by their ECAP estimators.  While~$p^*$ and $\hat p^*$ depend on~$\tp$, we will generally suppress this dependence in the notation for simplicity.  Note that for~$\tp\in[\epsilon,1-\epsilon]$, functions~$p^*$ and~$\hat p^*$ are both bounded away from zero on the set~$A$.

Fix an arbitrary~$\tp\le0.5$.  Define events $A_1=A\cap\{p^*\le0.5,\hat p^*\le0.5\}$, $A_2=A\cap\{p^*>0.5,\hat p^*\le0.5\}$, $A_3=A\cap\{p^*\le0.5,\hat p^*>0.5\}$ and $A_4=A\cap\{p^*>0.5,\hat p^*>0.5\}$.  Note that $A_4$ implies $\hat p=p_0=0.5$.  Writing Taylor expansions for function~$G$ near~$p^*$ and~$0.5$, we derive the following bounds, which hold for some universal constant~$c$ that depends only on~$\epsilon$:
\begin{eqnarray*}
\big|W_0(\tp)-\widehat{W}(\tp)\big|1_{\{A\}}&=&\big|G(p^*)-G(\hat p^*)\big|1_{\{A_1\}}+\big|G(0.5)-G(\hat p^*)\big|1_{\{A_2\}}+\big|G(p^*)-G(0.5)\big|1_{\{A_3\}}\\
&\le& c\big(p^*-\hat p^*)^21_{\{A_1\}}+c\big(0.5-\hat p^*)^21_{\{A_2\}}+c\big(p^*-0.5)^21_{\{A_3\}}\\
&\le& c\big(p^*-\hat p^*)^2.
 \end{eqnarray*}
Analogous arguments derive the above bound for $\tp>0.5$.  The rate of convergence for~$\widehat{W}$ then follows directly from the error bounds for~$\hat g$ and~$\hat g'$ in Theorem~\ref{g.thm}.

\section{Proof of Theorem~\ref{bias.thm}}
\label{prf.bias.thm}

Throughout the proof we drop the subscript~$i$ for the simplicity of notation. First note that the derivations in the proof of Theorem~\ref{EandVar} also give $E({\gamma^*}\alpha|\tp)=\mu$ and $Var({\gamma^*}\alpha|\tp)=\sigma^2$,
where $\mu$ and $\sigma^2$ are respectively defined in \eqref{mui} and \eqref{sigmai}. These identities hold for both the unbiased and biased versions of the model. The only difference is in how ${\gamma^*}\alpha$ relates to $p$.
Note that
\begin{eqnarray}
E(p|\tp) &=& E(h({\gamma^*}\alpha)|\tp)\nonumber=(1-0.5\theta) E({\gamma^*}\alpha|\tp) - \theta[E({\gamma^*}^3\alpha^3|\tp)-1.5E({\gamma^*}^2\alpha^2|\tp)]\nonumber\\
\label{bias.e.proof}&=& (1-0.5\theta)\mu -\theta[s_3 + 3\mu\sigma^2 + \mu^3 - 1.5\sigma^2-1.5\mu^2],
\end{eqnarray}
where  we use~$s_k$ to denote the $k$-th conditional central moment of ${\gamma^*}\alpha$ given~$\tp$. By Lemma~\ref{lem.mom.appr} in Appendix~\ref{append.sup.res}, the~$s_3$ term in~\eqref{bias.e.proof} is $O({\gamma^*}^{3/2})$, which leads to the stated approximation for $E(p|\tp)$.
We also have
$$Var(p|\tp) = Var(h({\gamma^*}\alpha)|\tp) = (1-0.5\theta)^2\sigma^2 + \theta a,$$
where $a = \theta Var({\gamma^*}^3 \alpha^3 - 1.5 {\gamma^*}^2\alpha^2|\tp) - (1-0.5\theta)Cov({\gamma^*}\alpha,{\gamma^*}^3\alpha^3 -1.5{\gamma^*}^2\alpha^2|\tp)$.
It is only left to show that $a=O({\gamma^*}^{3/2})$.  A routine calculation yields
\begin{equation*}
    a=\sigma_i^2\big[3\mu(1-\mu)(3\theta\mu(1-\mu)-0.5\theta+1) \big]+O\Big(\sum_{k=3}^6[\sigma^k+s_k]\Big).
\end{equation*}
By Lemma~\ref{lem.mom.appr}, the remainder term is $O({\gamma^*}^{3/2})$, which completes the proof.

\section{Proof of Theorem~\ref{EandVar.mixture}}

We use the notation from the proof of Theorem \ref{EandVar}.  In particular, we omit the subscript~$i$ throughout most of the proof, for the simplicity of the exposition.
We represent $\tp$ as $\sum_{k=1}^KI_{\{\mathcal{I}=k\}}\xi_k$, where $\xi_k|p\sim Beta(\alpha_k,\beta_k)$,$\alpha_k=c_kp/\gamma^*$, $\beta_k=(1-c_kp)/\gamma^*$, and~$\mathcal{I}$ is a discrete random variable independent of~$p$ and $\xi_k$, whose probability distribution is given by $P(\mathcal{I}=k)=w_k$ for $k=1,...,K$.

Note that
\begin{equation*} 
f_{\xi_k}(\tp|p)=B\Big(\frac{c_kp}{\gamma^*}, \frac{1-c_kp}{\gamma^*}\Big)^{-1}\tp^{\frac{c_kp}{\gamma^*} -1 }(1-\tp)^{\frac{1-c_kp}{\gamma^*}-1}. 
\end{equation*}
Hence, writing $B$ for $B(\frac{c_kp}{\gamma^*}, \frac{1-c_kp}{\gamma^*})$, we derive
\begin{equation*} \begin{split}
\log(f_{\xi_k}(\tp|p) )&=-\log B + \Big( \frac{c_kp}{\gamma^*} -1 \Big) \log\tp + \Big( \frac{1-c_kp}{\gamma^*} -1 \Big) \log(1-\tp) \\
&= -\log B +\frac{c_kp}{\gamma^*}\log\tp -\log\tp+\frac{1-c_kp}{\gamma^*} \log(1-\tp)-\log(1-\tp)\\
&= -\log B + p\frac{c_k}{\gamma^*} \log\Big( \frac{\tp}{1-\tp}\Big) -\log\tp +\frac{1}{\gamma^*} \log(1-\tp)-\log(1-\tp) \\
&= -\log B +\eta x - \log\tp +\Big(\frac{1-\gamma^*}{\gamma^*}\Big)\log(1-\tp),
\end{split}
\end{equation*}
where we've defined $\eta=p\frac{c_1}{\gamma^*}$ and $x=\log \big({\tp}/[{1-\tp}]\big)$. 
Repeating the derivations in the proof of Theorem \ref{EandVar} directly below display~(\ref{eq.exp.fam}), we derive 
\begin{equation*} \begin{split}
E(p|\xi_k=\tp)&=\frac{1}{c_k} \Big[ \gamma^*\Big(g^*(\tp)+1-2\tp\Big)+\tp\Big] \\
Var(p|\xi_k=\tp) &= \frac{1}{c_k^2} \Big[ \gamma^{*2} \Big( \tp(1-\tp) ({g^*}'(\tp)-2) +\gamma^*\tp(1-\tp)\Big) \Big].
\end{split}
\end{equation*}
Consequently, 
\begin{equation} \begin{split}
E(p|\tp, \mathcal{I}=k)&=E(p|\xi_k=\tp)=\frac{1}{c_k} \Big[ \gamma^*\Big(g^*(\tp)+1-2\tp\Big)+\tp\Big] \\
Var(p|\tp, \mathcal{I}=k) &= Var(p|\xi_k=\tp)=\frac{1}{c_k^2} \Big[ \gamma^{*2} \Big( \tp(1-\tp) ({g^*}'(\tp)-2) +\gamma^*\tp(1-\tp)\Big) \Big]
\end{split}
\end{equation}
Applying the law of total probability and using the fact that~$\mathcal{I}$ and~$\tp$ are independent, we derive
$$E(p|\tp)= \sum_{k=1}w_kE(p|\tp,\mathcal{I}=k)= \sum_{k=1}^K \frac{w_k}{c_k} \Big[ \gamma\Big( g^{*}(\tp)+1-2\tp  \Big) +\tp \Big]. $$ 
By the law of total variance, we also have 
\begin{eqnarray*}
Var(p|\tp)&=&
\sum_{k=1}w_k Var(p|\tp,\mathcal{I}=k)+\sum_{k=1}w_kE^2(p|\tp,\mathcal{I}=k)-\big[\sum_{k=1}w_kE(p|\tp,\mathcal{I}=k)\big]^2\\
&=&  \sum_{k=1}^K  \frac{w_k}{c_k^2} \big[ \gamma^{*2}  \tp(1-\tp) ({g^*}'(\tp)-2) +\gamma^*\tp(1-\tp) \big] +  
\big[ \gamma^*(g^*(\tp)+1-2\tp)+\tp\big] ^2 \big[ \sum_{k=1}^K\frac{w_k}{c_k^2}-  \big( \sum_{k=1}^K\frac{w_k}{c_k} \big)^2\big]. 
\end{eqnarray*}
To complete the proof, we use formulas
$$\mu_i=\tp_i+\gamma^*[g^*(\tp_i)+1-2\tp_i]\qquad\text{and}\qquad \sigma_i^2=\gamma^*\tp_i(1-\tp_i)+\gamma^{*2}\tp_i(1-\tp)[g^{*'}(\tp_i)-2]$$
to rewrite the above expressions as
$E(p_i|\tp_i) = \mu_i \sum_{k=1}^K {w_k}/{c_k}$
and 
\begin{equation*} 
Var(p_i|\tp_i)=\sum_{k=1}^K \frac{w_k}{c_k^2} \sigma_i^2 + \mu_i^2\sum_{k=1}^K \frac{w_k}{c_k^2}  - \mu_i^2\Big( \sum_{k=1}^K \frac{w_k}{c_k} \Big)^2 = (\sigma_i^2+\mu_i^2)\sum_{k=1}^K \frac{w_k}{c_k^2} -\mu_i^2 \Big( \sum_{k=1}^K \frac{w_k}{c_k} \Big)^2.
\end{equation*}


\section{Supplementary Results}
\label{append.sup.res}

\begin{lemma}
\label{lem1}
Under the conditions of Theorem~\ref{rate.thm}, there exists a function $g^*_N\in\mathcal{G}_N$, such that $\|g^*_N-g^*\|^2=O_p(\lambda_n^{2})$ and
\begin{equation*}
\|\hat g - g^*_N\|^2 + \lambda_n^2 I^2(\hat g)\le 
O_p\Big(n^{-2/7}\|\hat g - g^*_N\|\Big) + O_p\Big(n^{-4/7} I(\hat g) \Big)+O_p\Big(n^{-4/7} +\lambda_n^2\Big).
\end{equation*}
\end{lemma}
\noindent\textbf{Proof of Lemma~\ref{lem1}}.
We will use the empirical process theory notation and write ${\tilde P}_n g$ and~${\tilde P}g$ for $(1/n)\sum_{i=1}^n g(\tp_i)$ and $\int_0^1 g(\tp)f^*(\tp)d\tp$, respectively. 
Using the new notation, criterion~(\ref{risk.criterion}) can be written as follows:
\begin{equation*}
Q_n(g) = {\tilde P}_n g^2 + {\tilde P}_n s_g + \lambda_n^2 I^2(g).
\end{equation*}
 As we showed in the proof of Theorem~\ref{risk.lemma}, equality ${\tilde P}g^2 + {\tilde P} s_g = \|g-g^*\|^2$ holds for every candidate function~$g\in\mathcal{G}_N$.  Consequently,
\begin{equation*}
Q_n(g) = \|g-g^*\|^2 + ({\tilde P}_n-{\tilde P})g^2 + ({\tilde P}_n-{\tilde P}) s_g + \lambda_n^2 I^2(g).
\end{equation*}

Let $g^*_N$ be a function in $\mathcal{G}_N$ that interpolates~$g^*$ at points $\{0,\tp_1,...,\tp_n,1\}$, with two additional constraints: ${g^*_N}'(0)={g^*}'(0)$ and ${g^*_N}'(1)={g^*}'(1)$.  
A standard partial integration argument \citep[similar to that in ][for example]{green1993nonparametric} shows that $I(g^*_N)\le I(g^*)$, which also implies that ${g^*_N}'$ is uniformly bounded.  Furthermore, we have $\|g^*_N-g^*\|_{\infty}=O_p(\log(n)/n)$ by the maximum spacing results for the uniform distribution \citep[for example]{shorack2009empirical}, the boundedness away from zero  assumption on~$f^*$ and the boundedness of ${g^*_N}'$.  Consequently, $\|g^*_N-g^*\|^2=O_p(\lambda_n^2)$.

Because $Q_n(\hat g)\le Q_n(g^*_N)$, we then have
\begin{equation}
\label{basic_ineq}
\|\hat g - g^*_N\|^2 + \lambda_n^2 I^2(\hat g)\le ({\tilde P}_n-{\tilde P})[{g^*_N}^2-\hat g^2] + ({\tilde P}_n-{\tilde P})[s_{g^*_N}-s_{\hat g}] + \lambda_n^2 [I^2(g^*_N)+1].
\end{equation}
Note that
\begin{equation}
\label{g-sq.ineq}
({\tilde P}_n-{\tilde P})[{g^*_N}^2-\hat g^2] =-({\tilde P}_n-{\tilde P})[\hat g-{g^*_N}]^2 - ({\tilde P}_n-{\tilde P}){g^*_N}[\hat g-{g^*_N}].
\end{equation}
Applying Lemma~17 in \cite{meier2009high}, in conjunction with Corollary~5 from the same paper, in which we take $\gamma=2/5$ and $\lambda=n^{-1/2}$, we derive
\begin{equation}
\label{nu_n_dg2}
({\tilde P}_n-{\tilde P})[{g^*_N}-\hat g]^2 = O_p\Big(n^{-1/5}\|\hat g - g^*_N\|^2\Big) + O_p\Big(n^{-1} I^2(\hat g - g^*_N) \Big).
\end{equation}
Applying Corollary 5 in \cite{meier2009high} with the same~$\gamma$ and~$\lambda$ yields
$$({\tilde P}_n-{\tilde P}){g^*_N}[\hat g-{g^*_N}] = O_p\Big(n^{-2/5}\sqrt{\|{g^*_N}[\hat g - g^*_N]\|^2 + n^{-4/5} I^2({g^*_N}[\hat g - g^*_N])}\Big).$$
Using Lemma 10.9 in \cite{van2000applications} to express the $L_2$ norm of the first derivative in terms of the norms of the second derivative and the original function, we derive
\begin{equation}
\label{nu_n_dg}
({\tilde P}_n-{\tilde P}){g^*_N}[\hat g-{g^*_N}] = O_p\Big(n^{-2/5}\|\hat g - g^*_N\|\Big) + O_p\Big(n^{-4/5} I(\hat g - g^*_N) \Big).
\end{equation}
Applying Corollary 5 in \cite{meier2009high} with $\gamma=2/3, \lambda=n^{-3/14}$ and using Lemma 10.9 in \cite{van2000applications} again, we derive
\begin{equation*}
({\tilde P}_n-{\tilde P})[s_{g^*_N}-s_{\hat g}] = O_p\Big(n^{-3/7}\|s_{g^*_N}-s_{\hat g}\|\Big) + O_p\Big(n^{-4/7} I(\hat g - g^*_N) \Big).
\end{equation*}
Hence, by Lemma 10.9 in \cite{van2000applications},
\begin{equation*}
({\tilde P}_n-{\tilde P})[s_{g^*_N}-s_{\hat g}] = O_p\Big(n^{-3/7}\|\hat g - g^*_N\|^{1/2}I^{1/2}(\hat g - g^*_N)\Big) + O_p\Big(n^{-4/7} I(\hat g - g^*_N) \Big),
\end{equation*}
which leads to
\begin{equation}
\label{nu_n_ds}({\tilde P}_n-{\tilde P})[s_{g^*_N}-s_{\hat g}] = O_p\Big(n^{-2/7}\|\hat g - g^*_N\|\Big) + O_p\Big(n^{-4/7} I(\hat g) \Big)+O_p\Big([n^{-4/7}I(\hat g^*_N)\Big)
\end{equation}

Combining (\ref{basic_ineq})-(\ref{nu_n_ds}), and noting the imposed assumptions on~$\lambda_n$, we arrive at
\begin{equation*}
\|\hat g - g^*_N\|^2 + \lambda_n^2 I^2(\hat g)\le 
O_p\Big(n^{-2/7}\|\hat g - g^*_N\|\Big) + O_p\Big(n^{-4/7} I(\hat g) \Big)+O_p\Big(n^{-4/7} +\lambda_n^2\Big),
\end{equation*}
which completes the proof of Lemma~\ref{lem1}.

\bigskip

\begin{lemma}
\label{lem.mom.appr}
Under the conditions of Theorem~\ref{bias.thm},
\begin{equation*}
\sigma^2=O({\gamma^*})
\quad\text{and}\quad  s_k=O({\gamma^*}^{3/2}),\;\;\text{for}\;\; k\ge3.
\end{equation*}
\end{lemma}
\noindent \textbf{Proof of Lemma~\ref{lem.mom.appr}}.
We first show that $E(t-\tp|\tp)=O(\sqrt{\gamma^*})$ as $\gamma^*$ tends to zero, where we write~$t$ for the quantity $\alpha\gamma^*=h_{\theta}^{-1}(p)$. This result will useful for establishing the stated bound for~$s_k$.

Throughout the proof we use expression $\gtrsim$ to denote inequality $\ge$ up to a multiplicative factor equal to a positive constant that does not depend on~$\gamma^*$.  We use an analogous agreement for the $\lesssim$ expression.  We write $f_c(\tp)$ for the conditional density of $\tp$ given $t=c$, write $f(\tp)$ for the marginal density of $\tp$,  and write $m_{\theta}(t)$ for the marginal density of $t$.  In the new notation, we have
\begin{equation*}
E\big(|t-\tp|\big|\tp\big)={\int\limits_{0}^{1}|t-\tp|f_{t}(\tp)m_{\theta}(t)[f(\tp)]^{-1}dt}.
\end{equation*}

Using Stirling's approximation for the Gamma function,
$\Gamma(x)=e^{-x}x^{x-1/2}(2\pi)^{1/2}[1+O(1/x)]$, and applying the bound $x\Gamma(x)=O(1)$ when~$t$ is close to zero or one, we derive the following bounds as $\tau$ tends to infinity: 
\begin{eqnarray*}
    \sqrt{\tau}E\big(|t-\tp|\big|\tp\big)
    &=&\int\limits_0^1 \sqrt{\tau}|t-\tp|\frac{\Gamma(\tau)}{\Gamma(t\tau)\Gamma(q\tau)}\tp^{t\tau-1}\tilde q^{q\tau-1}m_{\theta}(t)[f(\tp)]^{-1}dt\\
&\lesssim&\int\limits_0^1 \sqrt{\tau}|t-\tp|\frac1{\sqrt{2\pi}}[\tp/t]^{t\tau}(\tilde q/q)^{q\tau}\sqrt{tq\tau}m_{\theta}(t)[f(\tp)]^{-1}[\tp\tilde q]^{-1}dt\\
    &\lesssim& \int\limits_{0}^{1} \sqrt{\tau}|t-\tp|e^{-\frac{\tau(t-\tp)^2}{18}}\sqrt{\tau}dt.
\end{eqnarray*}
Implementing a change of variable, $v=\sqrt{\tau}(t-\tp)$, we derive
\begin{equation*}
    \sqrt{\tau}\big|E\big(t-\tp\big|\tp\big)\big|\lesssim  \int\limits_{\mathbb{R}} |v|e^{-v^2/18}dv=O(1).
\end{equation*}
Consequently, $E(t-\tp|\tp)=O(1/\sqrt{\tau})=O(\sqrt{\gamma^*})$.

We now bound $E\big([t-\tp]^2\big|\tp\big)$ using a similar argument.  Following the arguments in the derivations above, we arrive at
\begin{eqnarray*}
        \tau E\big([t-\tp]^2\big|\tp\big)&\lesssim& \int\limits_{0}^{1} \tau(t-\tp)^2\frac1{\sqrt{2\pi}}e^{-\frac{\tau(t-\tp)^2}{18}}\sqrt{\tau}m_{\theta}(t)dt.
\end{eqnarray*}
Implementing a change of variable, $v=\sqrt{\tau}(t-\tp)$, we conclude that
\begin{equation*}
    \tau E\big([t-\tp]^2\big|\tp\big)\lesssim  \int\limits_{\mathbb{R}} v^2e^{-v^2/18}dv = O(1).
\end{equation*}

Thus, we have established
\begin{equation}
\label{s_k.bnds.12}
   E\big([t-\tp]^k\big|\tp\big)=O({\gamma^*}^{k/2}), \quad\text{for}\quad k\in\{1,2\}.
\end{equation}
Analogous arguments lead to bounds $E\big([t-\tp]^k\big|\tp\big)=O({\gamma^*}^{3/2})$ for $k\ge3$. We complete the proof of the lemma by noting that 
\begin{equation*}
   s_k=O\Big(E\Big([t-\tp]^k\Big|\tp\Big)\Big)+O\big({\gamma^*}^{3/2}\big), \quad\text{for}\quad k\ge 2.
\end{equation*}
When $k=2$ the above approximation follows from $\sigma^2\le E\big([t-\tp]^2\big|\tp\big)$, and when $k=3$ it follows from~(\ref{s_k.bnds.12}) and
\begin{eqnarray*}
    s_3&=&E\Big([t-\tp]^3\Big|\tp\Big)+3E\Big([t-\tp]^2\Big|\tp\Big)E\Big(\tp-t\Big|\tp\Big)
    +3E\Big(t-\tp\Big|\tp\Big)E^2\Big(\tp-t\Big|\tp\Big)+E^3\Big(\tp-t\Big|\tp\Big)\\
    &=&E\Big([t-\tp]^3\Big|\tp\Big)+O\big({\gamma^*}^{3/2}\big).
\end{eqnarray*}
The derivations for~$k\ge4$ are analogous.

\bibliographystyle{chicago}
\bibliography{bibs.bib}

\end{document}

%% file: Figures/intro_plt1.tex
\begin{tikzpicture}[x=1pt,y=1pt]
\definecolor{fillColor}{RGB}{255,255,255}
\path[use as bounding box,fill=fillColor,fill opacity=0.00] (0,0) rectangle (144.54,144.54);
\begin{scope}
\path[clip] (  0.00,  0.00) rectangle (144.54,144.54);
\definecolor{drawColor}{RGB}{255,255,255}
\definecolor{fillColor}{RGB}{255,255,255}

\path[draw=drawColor,line width= 0.4pt,line join=round,line cap=round,fill=fillColor] (  0.00,  0.00) rectangle (144.54,144.54);
\end{scope}
\begin{scope}
\path[clip] ( 25.12, 20.97) rectangle (141.04,141.04);
\definecolor{fillColor}{RGB}{255,255,255}

\path[fill=fillColor] ( 25.12, 20.97) rectangle (141.04,141.04);
\definecolor{drawColor}{RGB}{135,206,235}

\path[draw=drawColor,line width= 0.4pt,line join=round,line cap=round] ( 47.46, 45.03) circle (  1.43);

\path[draw=drawColor,line width= 0.4pt,line join=round,line cap=round] (115.90,118.83) circle (  1.43);

\path[draw=drawColor,line width= 0.4pt,line join=round,line cap=round] ( 70.67, 61.68) circle (  1.43);

\path[draw=drawColor,line width= 0.4pt,line join=round,line cap=round] ( 64.55, 49.26) circle (  1.43);

\path[draw=drawColor,line width= 0.4pt,line join=round,line cap=round] ( 93.91,101.01) circle (  1.43);

\path[draw=drawColor,line width= 0.4pt,line join=round,line cap=round] ( 94.16, 95.18) circle (  1.43);

\path[draw=drawColor,line width= 0.4pt,line join=round,line cap=round] ( 42.81, 33.35) circle (  1.43);

\path[draw=drawColor,line width= 0.4pt,line join=round,line cap=round] ( 61.00, 61.35) circle (  1.43);

\path[draw=drawColor,line width= 0.4pt,line join=round,line cap=round] ( 91.29, 69.30) circle (  1.43);

\path[draw=drawColor,line width= 0.4pt,line join=round,line cap=round] ( 97.00, 81.48) circle (  1.43);

\path[draw=drawColor,line width= 0.4pt,line join=round,line cap=round] ( 84.27, 64.69) circle (  1.43);

\path[draw=drawColor,line width= 0.4pt,line join=round,line cap=round] ( 83.52, 55.28) circle (  1.43);

\path[draw=drawColor,line width= 0.4pt,line join=round,line cap=round] ( 86.63, 95.44) circle (  1.43);

\path[draw=drawColor,line width= 0.4pt,line join=round,line cap=round] ( 89.11, 96.94) circle (  1.43);

\path[draw=drawColor,line width= 0.4pt,line join=round,line cap=round] (122.36,127.17) circle (  1.43);

\path[draw=drawColor,line width= 0.4pt,line join=round,line cap=round] (118.27, 96.71) circle (  1.43);

\path[draw=drawColor,line width= 0.4pt,line join=round,line cap=round] ( 41.40, 31.37) circle (  1.43);

\path[draw=drawColor,line width= 0.4pt,line join=round,line cap=round] (104.78,108.15) circle (  1.43);

\path[draw=drawColor,line width= 0.4pt,line join=round,line cap=round] (125.52, 77.64) circle (  1.43);

\path[draw=drawColor,line width= 0.4pt,line join=round,line cap=round] ( 59.41, 62.83) circle (  1.43);

\path[draw=drawColor,line width= 0.4pt,line join=round,line cap=round] ( 53.89, 48.35) circle (  1.43);

\path[draw=drawColor,line width= 0.4pt,line join=round,line cap=round] ( 31.11, 26.44) circle (  1.43);

\path[draw=drawColor,line width= 0.4pt,line join=round,line cap=round] ( 43.27, 47.69) circle (  1.43);

\path[draw=drawColor,line width= 0.4pt,line join=round,line cap=round] ( 39.46, 36.61) circle (  1.43);

\path[draw=drawColor,line width= 0.4pt,line join=round,line cap=round] ( 54.82, 57.42) circle (  1.43);

\path[draw=drawColor,line width= 0.4pt,line join=round,line cap=round] (114.14, 89.40) circle (  1.43);

\path[draw=drawColor,line width= 0.4pt,line join=round,line cap=round] ( 93.66, 67.50) circle (  1.43);

\path[draw=drawColor,line width= 0.4pt,line join=round,line cap=round] (126.88,109.81) circle (  1.43);

\path[draw=drawColor,line width= 0.4pt,line join=round,line cap=round] ( 89.45, 60.28) circle (  1.43);

\path[draw=drawColor,line width= 0.4pt,line join=round,line cap=round] (110.35, 74.41) circle (  1.43);

\path[draw=drawColor,line width= 0.4pt,line join=round,line cap=round] ( 70.05, 51.87) circle (  1.43);

\path[draw=drawColor,line width= 0.4pt,line join=round,line cap=round] ( 69.42, 50.69) circle (  1.43);

\path[draw=drawColor,line width= 0.4pt,line join=round,line cap=round] ( 47.70, 41.36) circle (  1.43);

\path[draw=drawColor,line width= 0.4pt,line join=round,line cap=round] ( 77.99, 54.91) circle (  1.43);

\path[draw=drawColor,line width= 0.4pt,line join=round,line cap=round] ( 57.13, 59.67) circle (  1.43);

\path[draw=drawColor,line width= 0.4pt,line join=round,line cap=round] ( 65.46, 57.54) circle (  1.43);

\path[draw=drawColor,line width= 0.4pt,line join=round,line cap=round] (124.68, 79.35) circle (  1.43);

\path[draw=drawColor,line width= 0.4pt,line join=round,line cap=round] ( 51.08, 40.08) circle (  1.43);

\path[draw=drawColor,line width= 0.4pt,line join=round,line cap=round] ( 91.46, 76.56) circle (  1.43);

\path[draw=drawColor,line width= 0.4pt,line join=round,line cap=round] ( 51.69, 42.52) circle (  1.43);

\path[draw=drawColor,line width= 0.4pt,line join=round,line cap=round] ( 59.59, 40.90) circle (  1.43);

\path[draw=drawColor,line width= 0.4pt,line join=round,line cap=round] (113.62,123.78) circle (  1.43);

\path[draw=drawColor,line width= 0.4pt,line join=round,line cap=round] ( 47.99, 32.45) circle (  1.43);

\path[draw=drawColor,line width= 0.4pt,line join=round,line cap=round] ( 90.56, 72.73) circle (  1.43);

\path[draw=drawColor,line width= 0.4pt,line join=round,line cap=round] ( 74.34, 63.86) circle (  1.43);

\path[draw=drawColor,line width= 0.4pt,line join=round,line cap=round] ( 58.11, 40.60) circle (  1.43);

\path[draw=drawColor,line width= 0.4pt,line join=round,line cap=round] ( 34.59, 28.44) circle (  1.43);

\path[draw=drawColor,line width= 0.4pt,line join=round,line cap=round] ( 40.55, 29.98) circle (  1.43);

\path[draw=drawColor,line width= 0.4pt,line join=round,line cap=round] ( 63.08, 48.29) circle (  1.43);

\path[draw=drawColor,line width= 0.4pt,line join=round,line cap=round] (115.16,101.13) circle (  1.43);

\path[draw=drawColor,line width= 0.4pt,line join=round,line cap=round] ( 54.01, 52.22) circle (  1.43);

\path[draw=drawColor,line width= 0.4pt,line join=round,line cap=round] ( 52.27, 56.73) circle (  1.43);

\path[draw=drawColor,line width= 0.4pt,line join=round,line cap=round] (123.34, 82.53) circle (  1.43);

\path[draw=drawColor,line width= 0.4pt,line join=round,line cap=round] (135.77,109.23) circle (  1.43);

\path[draw=drawColor,line width= 0.4pt,line join=round,line cap=round] (119.83,104.74) circle (  1.43);

\path[draw=drawColor,line width= 0.4pt,line join=round,line cap=round] (126.91,113.34) circle (  1.43);

\path[draw=drawColor,line width= 0.4pt,line join=round,line cap=round] ( 79.91, 82.75) circle (  1.43);

\path[draw=drawColor,line width= 0.4pt,line join=round,line cap=round] ( 53.49, 39.44) circle (  1.43);

\path[draw=drawColor,line width= 0.4pt,line join=round,line cap=round] ( 43.15, 40.55) circle (  1.43);

\path[draw=drawColor,line width= 0.4pt,line join=round,line cap=round] ( 59.40, 39.02) circle (  1.43);

\path[draw=drawColor,line width= 0.4pt,line join=round,line cap=round] (116.81, 78.52) circle (  1.43);

\path[draw=drawColor,line width= 0.4pt,line join=round,line cap=round] ( 35.64, 26.48) circle (  1.43);

\path[draw=drawColor,line width= 0.4pt,line join=round,line cap=round] (115.39, 73.37) circle (  1.43);

\path[draw=drawColor,line width= 0.4pt,line join=round,line cap=round] ( 40.64, 38.05) circle (  1.43);

\path[draw=drawColor,line width= 0.4pt,line join=round,line cap=round] (111.52, 68.90) circle (  1.43);

\path[draw=drawColor,line width= 0.4pt,line join=round,line cap=round] ( 62.09, 46.79) circle (  1.43);

\path[draw=drawColor,line width= 0.4pt,line join=round,line cap=round] (111.80,126.08) circle (  1.43);

\path[draw=drawColor,line width= 0.4pt,line join=round,line cap=round] ( 87.33, 67.70) circle (  1.43);

\path[draw=drawColor,line width= 0.4pt,line join=round,line cap=round] ( 68.25, 69.35) circle (  1.43);

\path[draw=drawColor,line width= 0.4pt,line join=round,line cap=round] ( 39.38, 36.24) circle (  1.43);

\path[draw=drawColor,line width= 0.4pt,line join=round,line cap=round] (110.78,113.34) circle (  1.43);

\path[draw=drawColor,line width= 0.4pt,line join=round,line cap=round] (110.90, 83.42) circle (  1.43);

\path[draw=drawColor,line width= 0.4pt,line join=round,line cap=round] (126.14,100.20) circle (  1.43);

\path[draw=drawColor,line width= 0.4pt,line join=round,line cap=round] (132.89, 99.96) circle (  1.43);

\path[draw=drawColor,line width= 0.4pt,line join=round,line cap=round] ( 84.62, 57.78) circle (  1.43);

\path[draw=drawColor,line width= 0.4pt,line join=round,line cap=round] ( 88.28, 62.52) circle (  1.43);

\path[draw=drawColor,line width= 0.4pt,line join=round,line cap=round] ( 46.99, 48.05) circle (  1.43);

\path[draw=drawColor,line width= 0.4pt,line join=round,line cap=round] ( 47.09, 44.70) circle (  1.43);

\path[draw=drawColor,line width= 0.4pt,line join=round,line cap=round] (113.63, 94.19) circle (  1.43);

\path[draw=drawColor,line width= 0.4pt,line join=round,line cap=round] (109.86, 97.43) circle (  1.43);

\path[draw=drawColor,line width= 0.4pt,line join=round,line cap=round] (113.40,112.70) circle (  1.43);

\path[draw=drawColor,line width= 0.4pt,line join=round,line cap=round] ( 99.50, 96.88) circle (  1.43);

\path[draw=drawColor,line width= 0.4pt,line join=round,line cap=round] ( 69.94, 49.37) circle (  1.43);

\path[draw=drawColor,line width= 0.4pt,line join=round,line cap=round] ( 30.39, 26.42) circle (  1.43);

\path[draw=drawColor,line width= 0.4pt,line join=round,line cap=round] (131.72, 95.39) circle (  1.43);

\path[draw=drawColor,line width= 0.4pt,line join=round,line cap=round] (119.22, 91.94) circle (  1.43);

\path[draw=drawColor,line width= 0.4pt,line join=round,line cap=round] ( 52.31, 54.72) circle (  1.43);

\path[draw=drawColor,line width= 0.4pt,line join=round,line cap=round] ( 82.42, 50.32) circle (  1.43);

\path[draw=drawColor,line width= 0.4pt,line join=round,line cap=round] ( 97.56,103.69) circle (  1.43);

\path[draw=drawColor,line width= 0.4pt,line join=round,line cap=round] (128.05,135.58) circle (  1.43);

\path[draw=drawColor,line width= 0.4pt,line join=round,line cap=round] ( 30.73, 26.61) circle (  1.43);

\path[draw=drawColor,line width= 0.4pt,line join=round,line cap=round] ( 58.09, 45.32) circle (  1.43);

\path[draw=drawColor,line width= 0.4pt,line join=round,line cap=round] ( 76.09, 52.16) circle (  1.43);

\path[draw=drawColor,line width= 0.4pt,line join=round,line cap=round] (118.24, 95.07) circle (  1.43);

\path[draw=drawColor,line width= 0.4pt,line join=round,line cap=round] (122.68, 75.43) circle (  1.43);

\path[draw=drawColor,line width= 0.4pt,line join=round,line cap=round] ( 56.34, 36.89) circle (  1.43);

\path[draw=drawColor,line width= 0.4pt,line join=round,line cap=round] ( 64.19, 58.24) circle (  1.43);

\path[draw=drawColor,line width= 0.4pt,line join=round,line cap=round] ( 62.25, 48.72) circle (  1.43);

\path[draw=drawColor,line width= 0.4pt,line join=round,line cap=round] ( 49.19, 37.34) circle (  1.43);

\path[draw=drawColor,line width= 0.4pt,line join=round,line cap=round] (102.25, 57.63) circle (  1.43);
\definecolor{drawColor}{RGB}{0,0,0}

\path[draw=drawColor,line width= 0.6pt,dash pattern=on 4pt off 4pt ,line join=round] ( 25.12, 22.95) -- (141.04,115.52);
\end{scope}
\begin{scope}
\path[clip] (  0.00,  0.00) rectangle (144.54,144.54);
\definecolor{drawColor}{RGB}{0,0,0}

\path[draw=drawColor,line width= 0.4pt,line join=round] ( 25.12, 20.97) --
	( 25.12,141.04);
\end{scope}
\begin{scope}
\path[clip] (  0.00,  0.00) rectangle (144.54,144.54);
\definecolor{drawColor}{gray}{0.30}

\node[text=drawColor,anchor=base east,inner sep=0pt, outer sep=0pt, scale=  0.56] at ( 21.97, 24.50) {0.00};

\node[text=drawColor,anchor=base east,inner sep=0pt, outer sep=0pt, scale=  0.56] at ( 21.97, 67.23) {0.05};

\node[text=drawColor,anchor=base east,inner sep=0pt, outer sep=0pt, scale=  0.56] at ( 21.97,109.97) {0.10};
\end{scope}
\begin{scope}
\path[clip] (  0.00,  0.00) rectangle (144.54,144.54);
\definecolor{drawColor}{gray}{0.20}

\path[draw=drawColor,line width= 0.4pt,line join=round] ( 23.37, 26.42) --
	( 25.12, 26.42);

\path[draw=drawColor,line width= 0.4pt,line join=round] ( 23.37, 69.16) --
	( 25.12, 69.16);

\path[draw=drawColor,line width= 0.4pt,line join=round] ( 23.37,111.90) --
	( 25.12,111.90);
\end{scope}
\begin{scope}
\path[clip] (  0.00,  0.00) rectangle (144.54,144.54);
\definecolor{drawColor}{RGB}{0,0,0}

\path[draw=drawColor,line width= 0.4pt,line join=round] ( 25.12, 20.97) --
	(141.04, 20.97);
\end{scope}
\begin{scope}
\path[clip] (  0.00,  0.00) rectangle (144.54,144.54);
\definecolor{drawColor}{gray}{0.20}

\path[draw=drawColor,line width= 0.4pt,line join=round] ( 29.47, 19.22) --
	( 29.47, 20.97);

\path[draw=drawColor,line width= 0.4pt,line join=round] ( 56.23, 19.22) --
	( 56.23, 20.97);

\path[draw=drawColor,line width= 0.4pt,line join=round] ( 82.98, 19.22) --
	( 82.98, 20.97);

\path[draw=drawColor,line width= 0.4pt,line join=round] (109.74, 19.22) --
	(109.74, 20.97);

\path[draw=drawColor,line width= 0.4pt,line join=round] (136.50, 19.22) --
	(136.50, 20.97);
\end{scope}
\begin{scope}
\path[clip] (  0.00,  0.00) rectangle (144.54,144.54);
\definecolor{drawColor}{gray}{0.30}

\node[text=drawColor,anchor=base,inner sep=0pt, outer sep=0pt, scale=  0.56] at ( 29.47, 13.96) {0.000};

\node[text=drawColor,anchor=base,inner sep=0pt, outer sep=0pt, scale=  0.56] at ( 56.23, 13.96) {0.025};

\node[text=drawColor,anchor=base,inner sep=0pt, outer sep=0pt, scale=  0.56] at ( 82.98, 13.96) {0.050};

\node[text=drawColor,anchor=base,inner sep=0pt, outer sep=0pt, scale=  0.56] at (109.74, 13.96) {0.075};

\node[text=drawColor,anchor=base,inner sep=0pt, outer sep=0pt, scale=  0.56] at (136.50, 13.96) {0.100};
\end{scope}
\begin{scope}
\path[clip] (  0.00,  0.00) rectangle (144.54,144.54);
\definecolor{drawColor}{RGB}{0,0,0}

\node[text=drawColor,anchor=base,inner sep=0pt, outer sep=0pt, scale=  0.70] at ( 83.08,  5.44) {$p$};
\end{scope}
\begin{scope}
\path[clip] (  0.00,  0.00) rectangle (144.54,144.54);
\definecolor{drawColor}{RGB}{0,0,0}

\node[text=drawColor,rotate= 90.00,anchor=base,inner sep=0pt, outer sep=0pt, scale=  0.70] at (  8.32, 81.00) {$\tilde{p}$};
\end{scope}
\end{tikzpicture}

%% file: Figures/intro_plt2.tex
\begin{tikzpicture}[x=1pt,y=1pt]
\definecolor{fillColor}{RGB}{255,255,255}
\path[use as bounding box,fill=fillColor,fill opacity=0.00] (0,0) rectangle (144.54,144.54);
\begin{scope}
\path[clip] (  0.00,  0.00) rectangle (144.54,144.54);
\definecolor{drawColor}{RGB}{255,255,255}
\definecolor{fillColor}{RGB}{255,255,255}

\path[draw=drawColor,line width= 0.4pt,line join=round,line cap=round,fill=fillColor] (  0.00,  0.00) rectangle (144.54,144.54);
\end{scope}
\begin{scope}
\path[clip] ( 25.46, 21.31) rectangle (141.04,141.04);
\definecolor{fillColor}{RGB}{255,255,255}

\path[fill=fillColor] ( 25.46, 21.31) rectangle (141.04,141.04);
\definecolor{drawColor}{RGB}{230,159,0}

\path[draw=drawColor,line width= 0.6pt,line join=round] ( 30.72, 26.75) --
	( 31.24, 28.39) --
	( 31.77, 29.56) --
	( 32.30, 30.44) --
	( 32.83, 31.31) --
	( 33.36, 32.27) --
	( 33.88, 33.04) --
	( 34.41, 33.53) --
	( 34.94, 34.10) --
	( 35.47, 35.17) --
	( 36.00, 35.53) --
	( 36.52, 36.29) --
	( 37.05, 36.81) --
	( 37.58, 37.51) --
	( 38.11, 38.21) --
	( 38.64, 38.74) --
	( 39.16, 39.38) --
	( 39.69, 39.82) --
	( 40.22, 40.18) --
	( 40.75, 41.15) --
	( 41.28, 41.50) --
	( 41.80, 42.04) --
	( 42.33, 42.62) --
	( 42.86, 43.48) --
	( 43.39, 43.52) --
	( 43.92, 44.34) --
	( 44.44, 45.00) --
	( 44.97, 45.30) --
	( 45.50, 46.02) --
	( 46.03, 47.10) --
	( 46.56, 47.13) --
	( 47.08, 48.32) --
	( 47.61, 48.16) --
	( 48.14, 48.45) --
	( 48.67, 48.88) --
	( 49.20, 49.85) --
	( 49.72, 50.07) --
	( 50.25, 50.88) --
	( 50.78, 51.59) --
	( 51.31, 51.56) --
	( 51.84, 52.72) --
	( 52.36, 52.76) --
	( 52.89, 53.36) --
	( 53.42, 54.29) --
	( 53.95, 54.58) --
	( 54.48, 54.73) --
	( 55.00, 55.41) --
	( 55.53, 56.75) --
	( 56.06, 56.87) --
	( 56.59, 56.62) --
	( 57.12, 58.47) --
	( 57.64, 58.02) --
	( 58.17, 58.74) --
	( 58.70, 59.60) --
	( 59.23, 59.65) --
	( 59.76, 60.63) --
	( 60.28, 60.66) --
	( 60.81, 61.07) --
	( 61.34, 62.65) --
	( 61.87, 62.67) --
	( 62.40, 62.84) --
	( 62.92, 63.41) --
	( 63.45, 64.34) --
	( 63.98, 64.96) --
	( 64.51, 65.26) --
	( 65.04, 65.73) --
	( 65.56, 66.62) --
	( 66.09, 66.88) --
	( 66.62, 67.87) --
	( 67.15, 67.60) --
	( 67.68, 68.44) --
	( 68.20, 68.66) --
	( 68.73, 69.35) --
	( 69.26, 70.52) --
	( 69.79, 71.20) --
	( 70.32, 70.87) --
	( 70.84, 72.27) --
	( 71.37, 71.67) --
	( 71.90, 71.84) --
	( 72.43, 73.58) --
	( 72.96, 74.35) --
	( 73.48, 73.71) --
	( 74.01, 74.42) --
	( 74.54, 74.98) --
	( 75.07, 75.88) --
	( 75.60, 76.35) --
	( 76.12, 76.23) --
	( 76.65, 77.72) --
	( 77.18, 78.12) --
	( 77.71, 78.52) --
	( 78.24, 79.15) --
	( 78.76, 79.15) --
	( 79.29, 79.48) --
	( 79.82, 80.69) --
	( 80.35, 81.52) --
	( 80.88, 81.50) --
	( 81.40, 82.32) --
	( 81.93, 82.11) --
	( 82.46, 82.94) --
	( 82.99, 84.33) --
	( 83.52, 83.62) --
	( 84.04, 84.49) --
	( 84.57, 85.16) --
	( 85.10, 85.88) --
	( 85.63, 86.03) --
	( 86.16, 87.01) --
	( 86.68, 88.16) --
	( 87.21, 87.91) --
	( 87.74, 88.31) --
	( 88.27, 89.01) --
	( 88.80, 89.32) --
	( 89.32, 90.07) --
	( 89.85, 90.10) --
	( 90.38, 91.33) --
	( 90.91, 91.64) --
	( 91.44, 91.40) --
	( 91.96, 93.35) --
	( 92.49, 93.21) --
	( 93.02, 93.06) --
	( 93.55, 94.13) --
	( 94.08, 95.80) --
	( 94.60, 94.88) --
	( 95.13, 95.99) --
	( 95.66, 95.93) --
	( 96.19, 96.48) --
	( 96.72, 97.35) --
	( 97.24, 97.39) --
	( 97.77, 98.31) --
	( 98.30, 98.60) --
	( 98.83, 98.69) --
	( 99.35,100.07) --
	( 99.88,100.37) --
	(100.41,101.61) --
	(100.94,101.34) --
	(101.47,102.94) --
	(101.99,102.98) --
	(102.52,103.23) --
	(103.05,103.67) --
	(103.58,104.30) --
	(104.11,105.20) --
	(104.63,104.59) --
	(105.16,106.09) --
	(105.69,106.93) --
	(106.22,107.13) --
	(106.75,107.00) --
	(107.27,106.25) --
	(107.80,107.83) --
	(108.33,108.80) --
	(108.86,110.07) --
	(109.39,110.55) --
	(109.91,110.69) --
	(110.44,111.60) --
	(110.97,111.75) --
	(111.50,112.15) --
	(112.03,112.15) --
	(112.55,113.10) --
	(113.08,114.07) --
	(113.61,114.05) --
	(114.14,115.33) --
	(114.67,115.65) --
	(115.19,115.70) --
	(115.72,116.49) --
	(116.25,115.70) --
	(116.78,117.05) --
	(117.31,118.26) --
	(117.83,119.52) --
	(118.36,120.23) --
	(118.89,119.09) --
	(119.42,119.97) --
	(119.95,121.28) --
	(120.47,120.76) --
	(121.00,120.30) --
	(121.53,122.37) --
	(122.06,123.00) --
	(122.59,122.70) --
	(123.11,123.99) --
	(123.64,124.20) --
	(124.17,124.99) --
	(124.70,124.91) --
	(125.23,125.12) --
	(125.75,128.01) --
	(126.28,127.05) --
	(126.81,127.98) --
	(127.34,127.30) --
	(127.87,128.16) --
	(128.39,127.69) --
	(128.92,131.15) --
	(129.45,129.67) --
	(129.98,130.94) --
	(130.51,130.40) --
	(131.03,131.39) --
	(131.56,131.94) --
	(132.09,132.02) --
	(132.62,132.58) --
	(133.15,133.13) --
	(133.67,133.75) --
	(134.20,134.25) --
	(134.73,135.55) --
	(135.26,135.26) --
	(135.79,135.60);
\definecolor{drawColor}{RGB}{0,0,0}

\path[draw=drawColor,line width= 0.6pt,dash pattern=on 4pt off 4pt ,line join=round] ( 25.46, 15.73) -- (141.04,132.97);
\end{scope}
\begin{scope}
\path[clip] (  0.00,  0.00) rectangle (144.54,144.54);
\definecolor{drawColor}{RGB}{0,0,0}

\path[draw=drawColor,line width= 0.4pt,line join=round] ( 25.46, 21.31) --
	( 25.46,141.04);
\end{scope}
\begin{scope}
\path[clip] (  0.00,  0.00) rectangle (144.54,144.54);
\definecolor{drawColor}{gray}{0.30}

\node[text=drawColor,anchor=base east,inner sep=0pt, outer sep=0pt, scale=  0.56] at ( 22.31, 40.12) {0.01};

\node[text=drawColor,anchor=base east,inner sep=0pt, outer sep=0pt, scale=  0.56] at ( 22.31, 61.58) {0.02};

\node[text=drawColor,anchor=base east,inner sep=0pt, outer sep=0pt, scale=  0.56] at ( 22.31, 83.05) {0.03};

\node[text=drawColor,anchor=base east,inner sep=0pt, outer sep=0pt, scale=  0.56] at ( 22.31,104.51) {0.04};

\node[text=drawColor,anchor=base east,inner sep=0pt, outer sep=0pt, scale=  0.56] at ( 22.31,125.98) {0.05};
\end{scope}
\begin{scope}
\path[clip] (  0.00,  0.00) rectangle (144.54,144.54);
\definecolor{drawColor}{gray}{0.20}

\path[draw=drawColor,line width= 0.4pt,line join=round] ( 23.71, 42.04) --
	( 25.46, 42.04);

\path[draw=drawColor,line width= 0.4pt,line join=round] ( 23.71, 63.51) --
	( 25.46, 63.51);

\path[draw=drawColor,line width= 0.4pt,line join=round] ( 23.71, 84.98) --
	( 25.46, 84.98);

\path[draw=drawColor,line width= 0.4pt,line join=round] ( 23.71,106.44) --
	( 25.46,106.44);

\path[draw=drawColor,line width= 0.4pt,line join=round] ( 23.71,127.91) --
	( 25.46,127.91);
\end{scope}
\begin{scope}
\path[clip] (  0.00,  0.00) rectangle (144.54,144.54);
\definecolor{drawColor}{RGB}{0,0,0}

\path[draw=drawColor,line width= 0.4pt,line join=round] ( 25.46, 21.31) --
	(141.04, 21.31);
\end{scope}
\begin{scope}
\path[clip] (  0.00,  0.00) rectangle (144.54,144.54);
\definecolor{drawColor}{gray}{0.20}

\path[draw=drawColor,line width= 0.4pt,line join=round] ( 30.24, 19.56) --
	( 30.24, 21.31);

\path[draw=drawColor,line width= 0.4pt,line join=round] ( 51.40, 19.56) --
	( 51.40, 21.31);

\path[draw=drawColor,line width= 0.4pt,line join=round] ( 72.56, 19.56) --
	( 72.56, 21.31);

\path[draw=drawColor,line width= 0.4pt,line join=round] ( 93.73, 19.56) --
	( 93.73, 21.31);

\path[draw=drawColor,line width= 0.4pt,line join=round] (114.89, 19.56) --
	(114.89, 21.31);

\path[draw=drawColor,line width= 0.4pt,line join=round] (136.05, 19.56) --
	(136.05, 21.31);
\end{scope}
\begin{scope}
\path[clip] (  0.00,  0.00) rectangle (144.54,144.54);
\definecolor{drawColor}{gray}{0.30}

\node[text=drawColor,anchor=base,inner sep=0pt, outer sep=0pt, scale=  0.56] at ( 30.24, 14.30) {0.00};

\node[text=drawColor,anchor=base,inner sep=0pt, outer sep=0pt, scale=  0.56] at ( 51.40, 14.30) {0.01};

\node[text=drawColor,anchor=base,inner sep=0pt, outer sep=0pt, scale=  0.56] at ( 72.56, 14.30) {0.02};

\node[text=drawColor,anchor=base,inner sep=0pt, outer sep=0pt, scale=  0.56] at ( 93.73, 14.30) {0.03};

\node[text=drawColor,anchor=base,inner sep=0pt, outer sep=0pt, scale=  0.56] at (114.89, 14.30) {0.04};

\node[text=drawColor,anchor=base,inner sep=0pt, outer sep=0pt, scale=  0.56] at (136.05, 14.30) {0.05};
\end{scope}
\begin{scope}
\path[clip] (  0.00,  0.00) rectangle (144.54,144.54);
\definecolor{drawColor}{RGB}{0,0,0}

\node[text=drawColor,anchor=base,inner sep=0pt, outer sep=0pt, scale=  0.75] at ( 83.25,  5.44) {$\tilde{p}$};
\end{scope}
\begin{scope}
\path[clip] (  0.00,  0.00) rectangle (144.54,144.54);
\definecolor{drawColor}{RGB}{0,0,0}

\node[text=drawColor,rotate= 90.00,anchor=base,inner sep=0pt, outer sep=0pt, scale=  0.75] at (  8.67, 81.18) {$E(p|\tilde{p})$};
\end{scope}
\end{tikzpicture}

%% file: Figures/intro_plt3.tex
\begin{tikzpicture}[x=1pt,y=1pt]
\definecolor{fillColor}{RGB}{255,255,255}
\path[use as bounding box,fill=fillColor,fill opacity=0.00] (0,0) rectangle (144.54,144.54);
\begin{scope}
\path[clip] (  0.00,  0.00) rectangle (144.54,144.54);
\definecolor{drawColor}{RGB}{255,255,255}
\definecolor{fillColor}{RGB}{255,255,255}

\path[draw=drawColor,line width= 0.4pt,line join=round,line cap=round,fill=fillColor] (  0.00,  0.00) rectangle (144.54,144.54);
\end{scope}
\begin{scope}
\path[clip] ( 21.11, 21.31) rectangle (141.04,141.04);
\definecolor{fillColor}{RGB}{255,255,255}

\path[fill=fillColor] ( 21.11, 21.31) rectangle (141.04,141.04);
\definecolor{drawColor}{RGB}{34,139,34}

\path[draw=drawColor,line width= 0.6pt,line join=round] ( 26.56,135.60) --
	( 27.11, 88.32) --
	( 27.66, 70.86) --
	( 28.20, 61.08) --
	( 28.75, 55.24) --
	( 29.30, 51.65) --
	( 29.85, 48.64) --
	( 30.39, 45.78) --
	( 30.94, 43.69) --
	( 31.49, 42.91) --
	( 32.04, 41.15) --
	( 32.59, 40.26) --
	( 33.13, 39.20) --
	( 33.68, 38.51) --
	( 34.23, 37.91) --
	( 34.78, 37.20) --
	( 35.33, 36.69) --
	( 35.87, 36.04) --
	( 36.42, 35.39) --
	( 36.97, 35.34) --
	( 37.52, 34.77) --
	( 38.07, 34.41) --
	( 38.61, 34.11) --
	( 39.16, 34.03) --
	( 39.71, 33.40) --
	( 40.26, 33.33) --
	( 40.80, 33.16) --
	( 41.35, 32.79) --
	( 41.90, 32.69) --
	( 42.45, 32.81) --
	( 43.00, 32.33) --
	( 43.54, 32.51) --
	( 44.09, 31.97) --
	( 44.64, 31.69) --
	( 45.19, 31.49) --
	( 45.74, 31.57) --
	( 46.28, 31.29) --
	( 46.83, 31.30) --
	( 47.38, 31.26) --
	( 47.93, 30.90) --
	( 48.47, 31.06) --
	( 49.02, 30.76) --
	( 49.57, 30.69) --
	( 50.12, 30.75) --
	( 50.67, 30.57) --
	( 51.21, 30.34) --
	( 51.76, 30.32) --
	( 52.31, 30.53) --
	( 52.86, 30.31) --
	( 53.41, 29.97) --
	( 53.95, 30.35) --
	( 54.50, 29.95) --
	( 55.05, 29.95) --
	( 55.60, 30.00) --
	( 56.15, 29.79) --
	( 56.69, 29.87) --
	( 57.24, 29.66) --
	( 57.79, 29.57) --
	( 58.34, 29.83) --
	( 58.88, 29.63) --
	( 59.43, 29.48) --
	( 59.98, 29.45) --
	( 60.53, 29.51) --
	( 61.08, 29.49) --
	( 61.62, 29.39) --
	( 62.17, 29.33) --
	( 62.72, 29.38) --
	( 63.27, 29.27) --
	( 63.82, 29.35) --
	( 64.36, 29.11) --
	( 64.91, 29.15) --
	( 65.46, 29.05) --
	( 66.01, 29.05) --
	( 66.56, 29.17) --
	( 67.10, 29.17) --
	( 67.65, 28.94) --
	( 68.20, 29.10) --
	( 68.75, 28.83) --
	( 69.29, 28.72) --
	( 69.84, 28.95) --
	( 70.39, 28.98) --
	( 70.94, 28.70) --
	( 71.49, 28.71) --
	( 72.03, 28.70) --
	( 72.58, 28.75) --
	( 73.13, 28.71) --
	( 73.68, 28.56) --
	( 74.23, 28.72) --
	( 74.77, 28.68) --
	( 75.32, 28.63) --
	( 75.87, 28.63) --
	( 76.42, 28.51) --
	( 76.96, 28.45) --
	( 77.51, 28.55) --
	( 78.06, 28.59) --
	( 78.61, 28.47) --
	( 79.16, 28.50) --
	( 79.70, 28.35) --
	( 80.25, 28.39) --
	( 80.80, 28.52) --
	( 81.35, 28.29) --
	( 81.90, 28.33) --
	( 82.44, 28.34) --
	( 82.99, 28.35) --
	( 83.54, 28.27) --
	( 84.09, 28.33) --
	( 84.64, 28.42) --
	( 85.18, 28.28) --
	( 85.73, 28.24) --
	( 86.28, 28.25) --
	( 86.83, 28.20) --
	( 87.37, 28.22) --
	( 87.92, 28.13) --
	( 88.47, 28.23) --
	( 89.02, 28.18) --
	( 89.57, 28.05) --
	( 90.11, 28.25) --
	( 90.66, 28.14) --
	( 91.21, 28.03) --
	( 91.76, 28.09) --
	( 92.31, 28.24) --
	( 92.85, 28.03) --
	( 93.40, 28.10) --
	( 93.95, 28.00) --
	( 94.50, 28.00) --
	( 95.05, 28.03) --
	( 95.59, 27.95) --
	( 96.14, 28.00) --
	( 96.69, 27.95) --
	( 97.24, 27.88) --
	( 97.78, 27.99) --
	( 98.33, 27.95) --
	( 98.88, 28.03) --
	( 99.43, 27.92) --
	( 99.98, 28.04) --
	(100.52, 27.97) --
	(101.07, 27.93) --
	(101.62, 27.91) --
	(102.17, 27.91) --
	(102.72, 27.95) --
	(103.26, 27.80) --
	(103.81, 27.91) --
	(104.36, 27.94) --
	(104.91, 27.89) --
	(105.45, 27.80) --
	(106.00, 27.64) --
	(106.55, 27.76) --
	(107.10, 27.80) --
	(107.65, 27.88) --
	(108.19, 27.87) --
	(108.74, 27.81) --
	(109.29, 27.85) --
	(109.84, 27.80) --
	(110.39, 27.78) --
	(110.93, 27.71) --
	(111.48, 27.75) --
	(112.03, 27.79) --
	(112.58, 27.72) --
	(113.13, 27.80) --
	(113.67, 27.77) --
	(114.22, 27.71) --
	(114.77, 27.73) --
	(115.32, 27.59) --
	(115.86, 27.67) --
	(116.41, 27.73) --
	(116.96, 27.80) --
	(117.51, 27.81) --
	(118.06, 27.63) --
	(118.60, 27.66) --
	(119.15, 27.74) --
	(119.70, 27.62) --
	(120.25, 27.52) --
	(120.80, 27.67) --
	(121.34, 27.67) --
	(121.89, 27.58) --
	(122.44, 27.65) --
	(122.99, 27.62) --
	(123.54, 27.64) --
	(124.08, 27.57) --
	(124.63, 27.54) --
	(125.18, 27.76) --
	(125.73, 27.61) --
	(126.27, 27.64) --
	(126.82, 27.52) --
	(127.37, 27.55) --
	(127.92, 27.45) --
	(128.47, 27.72) --
	(129.01, 27.53) --
	(129.56, 27.59) --
	(130.11, 27.49) --
	(130.66, 27.53) --
	(131.21, 27.52) --
	(131.75, 27.48) --
	(132.30, 27.48) --
	(132.85, 27.47) --
	(133.40, 27.48) --
	(133.94, 27.47) --
	(134.49, 27.53) --
	(135.04, 27.46) --
	(135.59, 27.44);
\definecolor{drawColor}{RGB}{0,0,0}

\path[draw=drawColor,line width= 0.6pt,dash pattern=on 4pt off 4pt ,line join=round] ( 21.11, 26.75) -- (141.04, 26.75);
\end{scope}
\begin{scope}
\path[clip] (  0.00,  0.00) rectangle (144.54,144.54);
\definecolor{drawColor}{RGB}{0,0,0}

\path[draw=drawColor,line width= 0.4pt,line join=round] ( 21.11, 21.31) --
	( 21.11,141.04);
\end{scope}
\begin{scope}
\path[clip] (  0.00,  0.00) rectangle (144.54,144.54);
\definecolor{drawColor}{gray}{0.30}

\node[text=drawColor,anchor=base east,inner sep=0pt, outer sep=0pt, scale=  0.56] at ( 17.96, 34.05) {2};

\node[text=drawColor,anchor=base east,inner sep=0pt, outer sep=0pt, scale=  0.56] at ( 17.96, 52.50) {4};

\node[text=drawColor,anchor=base east,inner sep=0pt, outer sep=0pt, scale=  0.56] at ( 17.96, 70.95) {6};

\node[text=drawColor,anchor=base east,inner sep=0pt, outer sep=0pt, scale=  0.56] at ( 17.96, 89.41) {8};

\node[text=drawColor,anchor=base east,inner sep=0pt, outer sep=0pt, scale=  0.56] at ( 17.96,107.86) {10};

\node[text=drawColor,anchor=base east,inner sep=0pt, outer sep=0pt, scale=  0.56] at ( 17.96,126.31) {12};
\end{scope}
\begin{scope}
\path[clip] (  0.00,  0.00) rectangle (144.54,144.54);
\definecolor{drawColor}{gray}{0.20}

\path[draw=drawColor,line width= 0.4pt,line join=round] ( 19.36, 35.98) --
	( 21.11, 35.98);

\path[draw=drawColor,line width= 0.4pt,line join=round] ( 19.36, 54.43) --
	( 21.11, 54.43);

\path[draw=drawColor,line width= 0.4pt,line join=round] ( 19.36, 72.88) --
	( 21.11, 72.88);

\path[draw=drawColor,line width= 0.4pt,line join=round] ( 19.36, 91.34) --
	( 21.11, 91.34);

\path[draw=drawColor,line width= 0.4pt,line join=round] ( 19.36,109.79) --
	( 21.11,109.79);

\path[draw=drawColor,line width= 0.4pt,line join=round] ( 19.36,128.24) --
	( 21.11,128.24);
\end{scope}
\begin{scope}
\path[clip] (  0.00,  0.00) rectangle (144.54,144.54);
\definecolor{drawColor}{RGB}{0,0,0}

\path[draw=drawColor,line width= 0.4pt,line join=round] ( 21.11, 21.31) --
	(141.04, 21.31);
\end{scope}
\begin{scope}
\path[clip] (  0.00,  0.00) rectangle (144.54,144.54);
\definecolor{drawColor}{gray}{0.20}

\path[draw=drawColor,line width= 0.4pt,line join=round] ( 26.07, 19.56) --
	( 26.07, 21.31);

\path[draw=drawColor,line width= 0.4pt,line join=round] ( 48.03, 19.56) --
	( 48.03, 21.31);

\path[draw=drawColor,line width= 0.4pt,line join=round] ( 69.98, 19.56) --
	( 69.98, 21.31);

\path[draw=drawColor,line width= 0.4pt,line join=round] ( 91.94, 19.56) --
	( 91.94, 21.31);

\path[draw=drawColor,line width= 0.4pt,line join=round] (113.90, 19.56) --
	(113.90, 21.31);

\path[draw=drawColor,line width= 0.4pt,line join=round] (135.86, 19.56) --
	(135.86, 21.31);
\end{scope}
\begin{scope}
\path[clip] (  0.00,  0.00) rectangle (144.54,144.54);
\definecolor{drawColor}{gray}{0.30}

\node[text=drawColor,anchor=base,inner sep=0pt, outer sep=0pt, scale=  0.56] at ( 26.07, 14.30) {0.00};

\node[text=drawColor,anchor=base,inner sep=0pt, outer sep=0pt, scale=  0.56] at ( 48.03, 14.30) {0.01};

\node[text=drawColor,anchor=base,inner sep=0pt, outer sep=0pt, scale=  0.56] at ( 69.98, 14.30) {0.02};

\node[text=drawColor,anchor=base,inner sep=0pt, outer sep=0pt, scale=  0.56] at ( 91.94, 14.30) {0.03};

\node[text=drawColor,anchor=base,inner sep=0pt, outer sep=0pt, scale=  0.56] at (113.90, 14.30) {0.04};

\node[text=drawColor,anchor=base,inner sep=0pt, outer sep=0pt, scale=  0.56] at (135.86, 14.30) {0.05};
\end{scope}
\begin{scope}
\path[clip] (  0.00,  0.00) rectangle (144.54,144.54);
\definecolor{drawColor}{RGB}{0,0,0}

\node[text=drawColor,anchor=base,inner sep=0pt, outer sep=0pt, scale=  0.75] at ( 81.07,  5.44) {$\tilde{p}$};
\end{scope}
\begin{scope}
\path[clip] (  0.00,  0.00) rectangle (144.54,144.54);
\definecolor{drawColor}{RGB}{0,0,0}

\node[text=drawColor,rotate= 90.00,anchor=base,inner sep=0pt, outer sep=0pt, scale=  0.75] at (  8.67, 81.18) {$E(p|\tilde{p}) / \tilde{p}$};
\end{scope}
\end{tikzpicture}

%% file: Figures/density_plt.tex
\begin{tikzpicture}[x=1pt,y=1pt]
\definecolor{fillColor}{RGB}{255,255,255}
\path[use as bounding box,fill=fillColor,fill opacity=0.00] (0,0) rectangle (361.35,144.54);
\begin{scope}
\path[clip] (  0.00,  0.00) rectangle (361.35,144.54);
\definecolor{drawColor}{RGB}{255,255,255}
\definecolor{fillColor}{RGB}{255,255,255}

\path[draw=drawColor,line width= 0.4pt,line join=round,line cap=round,fill=fillColor] (  0.00,  0.00) rectangle (361.35,144.54);
\end{scope}
\begin{scope}
\path[clip] ( 24.25, 21.65) rectangle (357.85,141.04);
\definecolor{fillColor}{RGB}{255,255,255}

\path[fill=fillColor] ( 24.25, 21.65) rectangle (357.85,141.04);
\definecolor{drawColor}{RGB}{0,114,178}

\path[draw=drawColor,line width= 0.6pt,line join=round] ( 39.42, 27.08) --
	( 40.02, 53.52) --
	( 40.63, 75.06) --
	( 41.24, 92.32) --
	( 41.85,105.88) --
	( 42.45,116.29) --
	( 43.06,124.02) --
	( 43.67,129.48) --
	( 44.28,133.03) --
	( 44.89,134.99) --
	( 45.49,135.61) --
	( 46.10,135.15) --
	( 46.71,133.79) --
	( 47.32,131.71) --
	( 47.92,129.06) --
	( 48.53,125.97) --
	( 49.14,122.55) --
	( 49.75,118.88) --
	( 50.35,115.04) --
	( 50.96,111.10) --
	( 51.57,107.12) --
	( 52.18,103.13) --
	( 52.79, 99.17) --
	( 53.39, 95.28) --
	( 54.00, 91.48) --
	( 54.61, 87.78) --
	( 55.22, 84.20) --
	( 55.82, 80.75) --
	( 56.43, 77.44) --
	( 57.04, 74.27) --
	( 57.65, 71.25) --
	( 58.26, 68.38) --
	( 58.86, 65.65) --
	( 59.47, 63.07) --
	( 60.08, 60.62) --
	( 60.69, 58.32) --
	( 61.29, 56.15) --
	( 61.90, 54.11) --
	( 62.51, 52.20) --
	( 63.12, 50.40) --
	( 63.73, 48.72) --
	( 64.33, 47.14) --
	( 64.94, 45.67) --
	( 65.55, 44.30) --
	( 66.16, 43.02) --
	( 66.76, 41.83) --
	( 67.37, 40.72) --
	( 67.98, 39.68) --
	( 68.59, 38.72) --
	( 69.20, 37.83) --
	( 69.80, 37.00) --
	( 70.41, 36.24) --
	( 71.02, 35.53) --
	( 71.63, 34.87) --
	( 72.23, 34.26) --
	( 72.84, 33.69) --
	( 73.45, 33.17) --
	( 74.06, 32.68) --
	( 74.67, 32.24) --
	( 75.27, 31.83) --
	( 75.88, 31.45) --
	( 76.49, 31.09) --
	( 77.10, 30.77) --
	( 77.70, 30.47) --
	( 78.31, 30.20) --
	( 78.92, 29.94) --
	( 79.53, 29.71) --
	( 80.13, 29.49) --
	( 80.74, 29.29) --
	( 81.35, 29.11) --
	( 81.96, 28.94) --
	( 82.57, 28.79) --
	( 83.17, 28.65) --
	( 83.78, 28.52) --
	( 84.39, 28.40) --
	( 85.00, 28.29) --
	( 85.60, 28.19) --
	( 86.21, 28.10) --
	( 86.82, 28.01) --
	( 87.43, 27.93) --
	( 88.04, 27.86) --
	( 88.64, 27.79) --
	( 89.25, 27.73) --
	( 89.86, 27.68) --
	( 90.47, 27.63) --
	( 91.07, 27.58) --
	( 91.68, 27.54) --
	( 92.29, 27.50) --
	( 92.90, 27.46) --
	( 93.51, 27.43) --
	( 94.11, 27.40) --
	( 94.72, 27.37) --
	( 95.33, 27.35) --
	( 95.94, 27.33) --
	( 96.54, 27.30) --
	( 97.15, 27.29) --
	( 97.76, 27.27) --
	( 98.37, 27.25) --
	( 98.98, 27.24) --
	( 99.58, 27.22) --
	(100.19, 27.21) --
	(100.80, 27.20) --
	(101.41, 27.19) --
	(102.01, 27.18) --
	(102.62, 27.17) --
	(103.23, 27.16) --
	(103.84, 27.16) --
	(104.45, 27.15) --
	(105.05, 27.14) --
	(105.66, 27.14) --
	(106.27, 27.13) --
	(106.88, 27.13) --
	(107.48, 27.12) --
	(108.09, 27.12) --
	(108.70, 27.12) --
	(109.31, 27.11) --
	(109.92, 27.11) --
	(110.52, 27.11) --
	(111.13, 27.11) --
	(111.74, 27.10) --
	(112.35, 27.10) --
	(112.95, 27.10) --
	(113.56, 27.10) --
	(114.17, 27.10) --
	(114.78, 27.10) --
	(115.38, 27.09) --
	(115.99, 27.09) --
	(116.60, 27.09) --
	(117.21, 27.09) --
	(117.82, 27.09) --
	(118.42, 27.09) --
	(119.03, 27.09) --
	(119.64, 27.09) --
	(120.25, 27.09) --
	(120.85, 27.09) --
	(121.46, 27.09) --
	(122.07, 27.09) --
	(122.68, 27.09) --
	(123.29, 27.08) --
	(123.89, 27.08) --
	(124.50, 27.08) --
	(125.11, 27.08) --
	(125.72, 27.08) --
	(126.32, 27.08) --
	(126.93, 27.08) --
	(127.54, 27.08) --
	(128.15, 27.08) --
	(128.76, 27.08) --
	(129.36, 27.08) --
	(129.97, 27.08) --
	(130.58, 27.08) --
	(131.19, 27.08) --
	(131.79, 27.08) --
	(132.40, 27.08) --
	(133.01, 27.08) --
	(133.62, 27.08) --
	(134.23, 27.08) --
	(134.83, 27.08) --
	(135.44, 27.08) --
	(136.05, 27.08) --
	(136.66, 27.08) --
	(137.26, 27.08) --
	(137.87, 27.08) --
	(138.48, 27.08) --
	(139.09, 27.08) --
	(139.70, 27.08) --
	(140.30, 27.08) --
	(140.91, 27.08) --
	(141.52, 27.08) --
	(142.13, 27.08) --
	(142.73, 27.08) --
	(143.34, 27.08) --
	(143.95, 27.08) --
	(144.56, 27.08) --
	(145.17, 27.08) --
	(145.77, 27.08) --
	(146.38, 27.08) --
	(146.99, 27.08) --
	(147.60, 27.08) --
	(148.20, 27.08) --
	(148.81, 27.08) --
	(149.42, 27.08) --
	(150.03, 27.08) --
	(150.63, 27.08) --
	(151.24, 27.08) --
	(151.85, 27.08) --
	(152.46, 27.08) --
	(153.07, 27.08) --
	(153.67, 27.08) --
	(154.28, 27.08) --
	(154.89, 27.08) --
	(155.50, 27.08) --
	(156.10, 27.08) --
	(156.71, 27.08) --
	(157.32, 27.08) --
	(157.93, 27.08) --
	(158.54, 27.08) --
	(159.14, 27.08) --
	(159.75, 27.08) --
	(160.36, 27.08) --
	(160.97, 27.08) --
	(161.57, 27.08) --
	(162.18, 27.08) --
	(162.79, 27.08) --
	(163.40, 27.08) --
	(164.01, 27.08) --
	(164.61, 27.08) --
	(165.22, 27.08) --
	(165.83, 27.08) --
	(166.44, 27.08) --
	(167.04, 27.08) --
	(167.65, 27.08) --
	(168.26, 27.08) --
	(168.87, 27.08) --
	(169.48, 27.08) --
	(170.08, 27.08) --
	(170.69, 27.08) --
	(171.30, 27.08) --
	(171.91, 27.08) --
	(172.51, 27.08) --
	(173.12, 27.08) --
	(173.73, 27.08) --
	(174.34, 27.08) --
	(174.95, 27.08) --
	(175.55, 27.08) --
	(176.16, 27.08) --
	(176.77, 27.08) --
	(177.38, 27.08) --
	(177.98, 27.08) --
	(178.59, 27.08) --
	(179.20, 27.08) --
	(179.81, 27.08) --
	(180.42, 27.08) --
	(181.02, 27.08) --
	(181.63, 27.08) --
	(182.24, 27.08) --
	(182.85, 27.08) --
	(183.45, 27.08) --
	(184.06, 27.08) --
	(184.67, 27.08) --
	(185.28, 27.08) --
	(185.88, 27.08) --
	(186.49, 27.08) --
	(187.10, 27.08) --
	(187.71, 27.08) --
	(188.32, 27.08) --
	(188.92, 27.08) --
	(189.53, 27.08) --
	(190.14, 27.08) --
	(190.75, 27.08) --
	(191.35, 27.08) --
	(191.96, 27.08) --
	(192.57, 27.08) --
	(193.18, 27.08) --
	(193.79, 27.08) --
	(194.39, 27.08) --
	(195.00, 27.08) --
	(195.61, 27.08) --
	(196.22, 27.08) --
	(196.82, 27.08) --
	(197.43, 27.08) --
	(198.04, 27.08) --
	(198.65, 27.08) --
	(199.26, 27.08) --
	(199.86, 27.08) --
	(200.47, 27.08) --
	(201.08, 27.08) --
	(201.69, 27.08) --
	(202.29, 27.08) --
	(202.90, 27.08) --
	(203.51, 27.08) --
	(204.12, 27.08) --
	(204.73, 27.08) --
	(205.33, 27.08) --
	(205.94, 27.08) --
	(206.55, 27.08) --
	(207.16, 27.08) --
	(207.76, 27.08) --
	(208.37, 27.08) --
	(208.98, 27.08) --
	(209.59, 27.08) --
	(210.20, 27.08) --
	(210.80, 27.08) --
	(211.41, 27.08) --
	(212.02, 27.08) --
	(212.63, 27.08) --
	(213.23, 27.08) --
	(213.84, 27.08) --
	(214.45, 27.08) --
	(215.06, 27.08) --
	(215.67, 27.08) --
	(216.27, 27.08) --
	(216.88, 27.08) --
	(217.49, 27.08) --
	(218.10, 27.08) --
	(218.70, 27.08) --
	(219.31, 27.08) --
	(219.92, 27.08) --
	(220.53, 27.08) --
	(221.13, 27.08) --
	(221.74, 27.08) --
	(222.35, 27.08) --
	(222.96, 27.08) --
	(223.57, 27.08) --
	(224.17, 27.08) --
	(224.78, 27.08) --
	(225.39, 27.08) --
	(226.00, 27.08) --
	(226.60, 27.08) --
	(227.21, 27.08) --
	(227.82, 27.08) --
	(228.43, 27.08) --
	(229.04, 27.08) --
	(229.64, 27.08) --
	(230.25, 27.08) --
	(230.86, 27.08) --
	(231.47, 27.08) --
	(232.07, 27.08) --
	(232.68, 27.08) --
	(233.29, 27.08) --
	(233.90, 27.08) --
	(234.51, 27.08) --
	(235.11, 27.08) --
	(235.72, 27.08) --
	(236.33, 27.08) --
	(236.94, 27.08) --
	(237.54, 27.08) --
	(238.15, 27.08) --
	(238.76, 27.08) --
	(239.37, 27.08) --
	(239.98, 27.08) --
	(240.58, 27.08) --
	(241.19, 27.08) --
	(241.80, 27.08) --
	(242.41, 27.08) --
	(243.01, 27.08) --
	(243.62, 27.08) --
	(244.23, 27.08) --
	(244.84, 27.08) --
	(245.45, 27.08) --
	(246.05, 27.08) --
	(246.66, 27.08) --
	(247.27, 27.08) --
	(247.88, 27.08) --
	(248.48, 27.08) --
	(249.09, 27.08) --
	(249.70, 27.08) --
	(250.31, 27.08) --
	(250.91, 27.08) --
	(251.52, 27.08) --
	(252.13, 27.08) --
	(252.74, 27.08) --
	(253.35, 27.08) --
	(253.95, 27.08) --
	(254.56, 27.08) --
	(255.17, 27.08) --
	(255.78, 27.08) --
	(256.38, 27.08) --
	(256.99, 27.08) --
	(257.60, 27.08) --
	(258.21, 27.08) --
	(258.82, 27.08) --
	(259.42, 27.08) --
	(260.03, 27.08) --
	(260.64, 27.08) --
	(261.25, 27.08) --
	(261.85, 27.08) --
	(262.46, 27.08) --
	(263.07, 27.08) --
	(263.68, 27.08) --
	(264.29, 27.08) --
	(264.89, 27.08) --
	(265.50, 27.08) --
	(266.11, 27.08) --
	(266.72, 27.08) --
	(267.32, 27.08) --
	(267.93, 27.08) --
	(268.54, 27.08) --
	(269.15, 27.08) --
	(269.76, 27.08) --
	(270.36, 27.08) --
	(270.97, 27.08) --
	(271.58, 27.08) --
	(272.19, 27.08) --
	(272.79, 27.08) --
	(273.40, 27.08) --
	(274.01, 27.08) --
	(274.62, 27.08) --
	(275.23, 27.08) --
	(275.83, 27.08) --
	(276.44, 27.08) --
	(277.05, 27.08) --
	(277.66, 27.08) --
	(278.26, 27.08) --
	(278.87, 27.08) --
	(279.48, 27.08) --
	(280.09, 27.08) --
	(280.70, 27.08) --
	(281.30, 27.08) --
	(281.91, 27.08) --
	(282.52, 27.08) --
	(283.13, 27.08) --
	(283.73, 27.08) --
	(284.34, 27.08) --
	(284.95, 27.08) --
	(285.56, 27.08) --
	(286.16, 27.08) --
	(286.77, 27.08) --
	(287.38, 27.08) --
	(287.99, 27.08) --
	(288.60, 27.08) --
	(289.20, 27.08) --
	(289.81, 27.08) --
	(290.42, 27.08) --
	(291.03, 27.08) --
	(291.63, 27.08) --
	(292.24, 27.08) --
	(292.85, 27.08) --
	(293.46, 27.08) --
	(294.07, 27.08) --
	(294.67, 27.08) --
	(295.28, 27.08) --
	(295.89, 27.08) --
	(296.50, 27.08) --
	(297.10, 27.08) --
	(297.71, 27.08) --
	(298.32, 27.08) --
	(298.93, 27.08) --
	(299.54, 27.08) --
	(300.14, 27.08) --
	(300.75, 27.08) --
	(301.36, 27.08) --
	(301.97, 27.08) --
	(302.57, 27.08) --
	(303.18, 27.08) --
	(303.79, 27.08) --
	(304.40, 27.08) --
	(305.01, 27.08) --
	(305.61, 27.08) --
	(306.22, 27.08) --
	(306.83, 27.08) --
	(307.44, 27.08) --
	(308.04, 27.08) --
	(308.65, 27.08) --
	(309.26, 27.08) --
	(309.87, 27.08) --
	(310.48, 27.08) --
	(311.08, 27.08) --
	(311.69, 27.08) --
	(312.30, 27.08) --
	(312.91, 27.08) --
	(313.51, 27.08) --
	(314.12, 27.08) --
	(314.73, 27.08) --
	(315.34, 27.08) --
	(315.95, 27.08) --
	(316.55, 27.08) --
	(317.16, 27.08) --
	(317.77, 27.08) --
	(318.38, 27.08) --
	(318.98, 27.08) --
	(319.59, 27.08) --
	(320.20, 27.08) --
	(320.81, 27.08) --
	(321.41, 27.08) --
	(322.02, 27.08) --
	(322.63, 27.08) --
	(323.24, 27.08) --
	(323.85, 27.08) --
	(324.45, 27.08) --
	(325.06, 27.08) --
	(325.67, 27.08) --
	(326.28, 27.08) --
	(326.88, 27.08) --
	(327.49, 27.08) --
	(328.10, 27.08) --
	(328.71, 27.08) --
	(329.32, 27.08) --
	(329.92, 27.08) --
	(330.53, 27.08) --
	(331.14, 27.08) --
	(331.75, 27.08) --
	(332.35, 27.08) --
	(332.96, 27.08) --
	(333.57, 27.08) --
	(334.18, 27.08) --
	(334.79, 27.08) --
	(335.39, 27.08) --
	(336.00, 27.08) --
	(336.61, 27.08) --
	(337.22, 27.08) --
	(337.82, 27.08) --
	(338.43, 27.08) --
	(339.04, 27.08) --
	(339.65, 27.08) --
	(340.26, 27.08) --
	(340.86, 27.08) --
	(341.47, 27.08) --
	(342.08, 27.08) --
	(342.69, 27.08);
\definecolor{drawColor}{RGB}{213,94,0}

\path[draw=drawColor,line width= 0.6pt,dash pattern=on 2pt off 2pt on 6pt off 2pt ,line join=round] ( 39.42, 27.08) --
	( 40.02, 27.08) --
	( 40.63, 27.08) --
	( 41.24, 27.08) --
	( 41.85, 27.08) --
	( 42.45, 27.08) --
	( 43.06, 27.08) --
	( 43.67, 27.08) --
	( 44.28, 27.08) --
	( 44.89, 27.08) --
	( 45.49, 27.08) --
	( 46.10, 27.08) --
	( 46.71, 27.08) --
	( 47.32, 27.08) --
	( 47.92, 27.08) --
	( 48.53, 27.09) --
	( 49.14, 27.09) --
	( 49.75, 27.10) --
	( 50.35, 27.10) --
	( 50.96, 27.11) --
	( 51.57, 27.13) --
	( 52.18, 27.15) --
	( 52.79, 27.17) --
	( 53.39, 27.21) --
	( 54.00, 27.25) --
	( 54.61, 27.30) --
	( 55.22, 27.36) --
	( 55.82, 27.44) --
	( 56.43, 27.54) --
	( 57.04, 27.65) --
	( 57.65, 27.78) --
	( 58.26, 27.94) --
	( 58.86, 28.11) --
	( 59.47, 28.32) --
	( 60.08, 28.55) --
	( 60.69, 28.81) --
	( 61.29, 29.11) --
	( 61.90, 29.44) --
	( 62.51, 29.80) --
	( 63.12, 30.20) --
	( 63.73, 30.63) --
	( 64.33, 31.11) --
	( 64.94, 31.62) --
	( 65.55, 32.17) --
	( 66.16, 32.76) --
	( 66.76, 33.39) --
	( 67.37, 34.05) --
	( 67.98, 34.76) --
	( 68.59, 35.50) --
	( 69.20, 36.27) --
	( 69.80, 37.08) --
	( 70.41, 37.92) --
	( 71.02, 38.79) --
	( 71.63, 39.69) --
	( 72.23, 40.61) --
	( 72.84, 41.55) --
	( 73.45, 42.51) --
	( 74.06, 43.49) --
	( 74.67, 44.48) --
	( 75.27, 45.48) --
	( 75.88, 46.48) --
	( 76.49, 47.49) --
	( 77.10, 48.50) --
	( 77.70, 49.51) --
	( 78.31, 50.51) --
	( 78.92, 51.50) --
	( 79.53, 52.47) --
	( 80.13, 53.43) --
	( 80.74, 54.37) --
	( 81.35, 55.29) --
	( 81.96, 56.18) --
	( 82.57, 57.05) --
	( 83.17, 57.88) --
	( 83.78, 58.69) --
	( 84.39, 59.45) --
	( 85.00, 60.18) --
	( 85.60, 60.88) --
	( 86.21, 61.53) --
	( 86.82, 62.14) --
	( 87.43, 62.71) --
	( 88.04, 63.23) --
	( 88.64, 63.71) --
	( 89.25, 64.14) --
	( 89.86, 64.53) --
	( 90.47, 64.87) --
	( 91.07, 65.16) --
	( 91.68, 65.41) --
	( 92.29, 65.61) --
	( 92.90, 65.77) --
	( 93.51, 65.88) --
	( 94.11, 65.95) --
	( 94.72, 65.97) --
	( 95.33, 65.95) --
	( 95.94, 65.88) --
	( 96.54, 65.78) --
	( 97.15, 65.63) --
	( 97.76, 65.45) --
	( 98.37, 65.23) --
	( 98.98, 64.97) --
	( 99.58, 64.68) --
	(100.19, 64.36) --
	(100.80, 64.01) --
	(101.41, 63.62) --
	(102.01, 63.21) --
	(102.62, 62.77) --
	(103.23, 62.31) --
	(103.84, 61.82) --
	(104.45, 61.32) --
	(105.05, 60.79) --
	(105.66, 60.24) --
	(106.27, 59.68) --
	(106.88, 59.11) --
	(107.48, 58.52) --
	(108.09, 57.91) --
	(108.70, 57.30) --
	(109.31, 56.68) --
	(109.92, 56.05) --
	(110.52, 55.42) --
	(111.13, 54.78) --
	(111.74, 54.14) --
	(112.35, 53.49) --
	(112.95, 52.84) --
	(113.56, 52.20) --
	(114.17, 51.55) --
	(114.78, 50.91) --
	(115.38, 50.27) --
	(115.99, 49.63) --
	(116.60, 49.00) --
	(117.21, 48.37) --
	(117.82, 47.75) --
	(118.42, 47.14) --
	(119.03, 46.53) --
	(119.64, 45.94) --
	(120.25, 45.35) --
	(120.85, 44.77) --
	(121.46, 44.20) --
	(122.07, 43.64) --
	(122.68, 43.09) --
	(123.29, 42.55) --
	(123.89, 42.02) --
	(124.50, 41.50) --
	(125.11, 40.99) --
	(125.72, 40.50) --
	(126.32, 40.02) --
	(126.93, 39.54) --
	(127.54, 39.08) --
	(128.15, 38.64) --
	(128.76, 38.20) --
	(129.36, 37.78) --
	(129.97, 37.36) --
	(130.58, 36.96) --
	(131.19, 36.58) --
	(131.79, 36.20) --
	(132.40, 35.83) --
	(133.01, 35.48) --
	(133.62, 35.13) --
	(134.23, 34.80) --
	(134.83, 34.48) --
	(135.44, 34.17) --
	(136.05, 33.87) --
	(136.66, 33.58) --
	(137.26, 33.30) --
	(137.87, 33.03) --
	(138.48, 32.77) --
	(139.09, 32.52) --
	(139.70, 32.28) --
	(140.30, 32.04) --
	(140.91, 31.82) --
	(141.52, 31.61) --
	(142.13, 31.40) --
	(142.73, 31.20) --
	(143.34, 31.01) --
	(143.95, 30.83) --
	(144.56, 30.65) --
	(145.17, 30.48) --
	(145.77, 30.32) --
	(146.38, 30.17) --
	(146.99, 30.02) --
	(147.60, 29.87) --
	(148.20, 29.74) --
	(148.81, 29.61) --
	(149.42, 29.48) --
	(150.03, 29.37) --
	(150.63, 29.25) --
	(151.24, 29.14) --
	(151.85, 29.04) --
	(152.46, 28.94) --
	(153.07, 28.85) --
	(153.67, 28.76) --
	(154.28, 28.67) --
	(154.89, 28.59) --
	(155.50, 28.51) --
	(156.10, 28.43) --
	(156.71, 28.36) --
	(157.32, 28.30) --
	(157.93, 28.23) --
	(158.54, 28.17) --
	(159.14, 28.11) --
	(159.75, 28.06) --
	(160.36, 28.01) --
	(160.97, 27.96) --
	(161.57, 27.91) --
	(162.18, 27.86) --
	(162.79, 27.82) --
	(163.40, 27.78) --
	(164.01, 27.74) --
	(164.61, 27.70) --
	(165.22, 27.67) --
	(165.83, 27.64) --
	(166.44, 27.61) --
	(167.04, 27.58) --
	(167.65, 27.55) --
	(168.26, 27.52) --
	(168.87, 27.50) --
	(169.48, 27.47) --
	(170.08, 27.45) --
	(170.69, 27.43) --
	(171.30, 27.41) --
	(171.91, 27.39) --
	(172.51, 27.37) --
	(173.12, 27.36) --
	(173.73, 27.34) --
	(174.34, 27.32) --
	(174.95, 27.31) --
	(175.55, 27.30) --
	(176.16, 27.28) --
	(176.77, 27.27) --
	(177.38, 27.26) --
	(177.98, 27.25) --
	(178.59, 27.24) --
	(179.20, 27.23) --
	(179.81, 27.22) --
	(180.42, 27.21) --
	(181.02, 27.21) --
	(181.63, 27.20) --
	(182.24, 27.19) --
	(182.85, 27.18) --
	(183.45, 27.18) --
	(184.06, 27.17) --
	(184.67, 27.17) --
	(185.28, 27.16) --
	(185.88, 27.16) --
	(186.49, 27.15) --
	(187.10, 27.15) --
	(187.71, 27.14) --
	(188.32, 27.14) --
	(188.92, 27.14) --
	(189.53, 27.13) --
	(190.14, 27.13) --
	(190.75, 27.13) --
	(191.35, 27.12) --
	(191.96, 27.12) --
	(192.57, 27.12) --
	(193.18, 27.12) --
	(193.79, 27.11) --
	(194.39, 27.11) --
	(195.00, 27.11) --
	(195.61, 27.11) --
	(196.22, 27.11) --
	(196.82, 27.10) --
	(197.43, 27.10) --
	(198.04, 27.10) --
	(198.65, 27.10) --
	(199.26, 27.10) --
	(199.86, 27.10) --
	(200.47, 27.10) --
	(201.08, 27.10) --
	(201.69, 27.09) --
	(202.29, 27.09) --
	(202.90, 27.09) --
	(203.51, 27.09) --
	(204.12, 27.09) --
	(204.73, 27.09) --
	(205.33, 27.09) --
	(205.94, 27.09) --
	(206.55, 27.09) --
	(207.16, 27.09) --
	(207.76, 27.09) --
	(208.37, 27.09) --
	(208.98, 27.09) --
	(209.59, 27.09) --
	(210.20, 27.09) --
	(210.80, 27.09) --
	(211.41, 27.09) --
	(212.02, 27.09) --
	(212.63, 27.09) --
	(213.23, 27.08) --
	(213.84, 27.08) --
	(214.45, 27.08) --
	(215.06, 27.08) --
	(215.67, 27.08) --
	(216.27, 27.08) --
	(216.88, 27.08) --
	(217.49, 27.08) --
	(218.10, 27.08) --
	(218.70, 27.08) --
	(219.31, 27.08) --
	(219.92, 27.08) --
	(220.53, 27.08) --
	(221.13, 27.08) --
	(221.74, 27.08) --
	(222.35, 27.08) --
	(222.96, 27.08) --
	(223.57, 27.08) --
	(224.17, 27.08) --
	(224.78, 27.08) --
	(225.39, 27.08) --
	(226.00, 27.08) --
	(226.60, 27.08) --
	(227.21, 27.08) --
	(227.82, 27.08) --
	(228.43, 27.08) --
	(229.04, 27.08) --
	(229.64, 27.08) --
	(230.25, 27.08) --
	(230.86, 27.08) --
	(231.47, 27.08) --
	(232.07, 27.08) --
	(232.68, 27.08) --
	(233.29, 27.08) --
	(233.90, 27.08) --
	(234.51, 27.08) --
	(235.11, 27.08) --
	(235.72, 27.08) --
	(236.33, 27.08) --
	(236.94, 27.08) --
	(237.54, 27.08) --
	(238.15, 27.08) --
	(238.76, 27.08) --
	(239.37, 27.08) --
	(239.98, 27.08) --
	(240.58, 27.08) --
	(241.19, 27.08) --
	(241.80, 27.08) --
	(242.41, 27.08) --
	(243.01, 27.08) --
	(243.62, 27.08) --
	(244.23, 27.08) --
	(244.84, 27.08) --
	(245.45, 27.08) --
	(246.05, 27.08) --
	(246.66, 27.08) --
	(247.27, 27.08) --
	(247.88, 27.08) --
	(248.48, 27.08) --
	(249.09, 27.08) --
	(249.70, 27.08) --
	(250.31, 27.08) --
	(250.91, 27.08) --
	(251.52, 27.08) --
	(252.13, 27.08) --
	(252.74, 27.08) --
	(253.35, 27.08) --
	(253.95, 27.08) --
	(254.56, 27.08) --
	(255.17, 27.08) --
	(255.78, 27.08) --
	(256.38, 27.08) --
	(256.99, 27.08) --
	(257.60, 27.08) --
	(258.21, 27.08) --
	(258.82, 27.08) --
	(259.42, 27.08) --
	(260.03, 27.08) --
	(260.64, 27.08) --
	(261.25, 27.08) --
	(261.85, 27.08) --
	(262.46, 27.08) --
	(263.07, 27.08) --
	(263.68, 27.08) --
	(264.29, 27.08) --
	(264.89, 27.08) --
	(265.50, 27.08) --
	(266.11, 27.08) --
	(266.72, 27.08) --
	(267.32, 27.08) --
	(267.93, 27.08) --
	(268.54, 27.08) --
	(269.15, 27.08) --
	(269.76, 27.08) --
	(270.36, 27.08) --
	(270.97, 27.08) --
	(271.58, 27.08) --
	(272.19, 27.08) --
	(272.79, 27.08) --
	(273.40, 27.08) --
	(274.01, 27.08) --
	(274.62, 27.08) --
	(275.23, 27.08) --
	(275.83, 27.08) --
	(276.44, 27.08) --
	(277.05, 27.08) --
	(277.66, 27.08) --
	(278.26, 27.08) --
	(278.87, 27.08) --
	(279.48, 27.08) --
	(280.09, 27.08) --
	(280.70, 27.08) --
	(281.30, 27.08) --
	(281.91, 27.08) --
	(282.52, 27.08) --
	(283.13, 27.08) --
	(283.73, 27.08) --
	(284.34, 27.08) --
	(284.95, 27.08) --
	(285.56, 27.08) --
	(286.16, 27.08) --
	(286.77, 27.08) --
	(287.38, 27.08) --
	(287.99, 27.08) --
	(288.60, 27.08) --
	(289.20, 27.08) --
	(289.81, 27.08) --
	(290.42, 27.08) --
	(291.03, 27.08) --
	(291.63, 27.08) --
	(292.24, 27.08) --
	(292.85, 27.08) --
	(293.46, 27.08) --
	(294.07, 27.08) --
	(294.67, 27.08) --
	(295.28, 27.08) --
	(295.89, 27.08) --
	(296.50, 27.08) --
	(297.10, 27.08) --
	(297.71, 27.08) --
	(298.32, 27.08) --
	(298.93, 27.08) --
	(299.54, 27.08) --
	(300.14, 27.08) --
	(300.75, 27.08) --
	(301.36, 27.08) --
	(301.97, 27.08) --
	(302.57, 27.08) --
	(303.18, 27.08) --
	(303.79, 27.08) --
	(304.40, 27.08) --
	(305.01, 27.08) --
	(305.61, 27.08) --
	(306.22, 27.08) --
	(306.83, 27.08) --
	(307.44, 27.08) --
	(308.04, 27.08) --
	(308.65, 27.08) --
	(309.26, 27.08) --
	(309.87, 27.08) --
	(310.48, 27.08) --
	(311.08, 27.08) --
	(311.69, 27.08) --
	(312.30, 27.08) --
	(312.91, 27.08) --
	(313.51, 27.08) --
	(314.12, 27.08) --
	(314.73, 27.08) --
	(315.34, 27.08) --
	(315.95, 27.08) --
	(316.55, 27.08) --
	(317.16, 27.08) --
	(317.77, 27.08) --
	(318.38, 27.08) --
	(318.98, 27.08) --
	(319.59, 27.08) --
	(320.20, 27.08) --
	(320.81, 27.08) --
	(321.41, 27.08) --
	(322.02, 27.08) --
	(322.63, 27.08) --
	(323.24, 27.08) --
	(323.85, 27.08) --
	(324.45, 27.08) --
	(325.06, 27.08) --
	(325.67, 27.08) --
	(326.28, 27.08) --
	(326.88, 27.08) --
	(327.49, 27.08) --
	(328.10, 27.08) --
	(328.71, 27.08) --
	(329.32, 27.08) --
	(329.92, 27.08) --
	(330.53, 27.08) --
	(331.14, 27.08) --
	(331.75, 27.08) --
	(332.35, 27.08) --
	(332.96, 27.08) --
	(333.57, 27.08) --
	(334.18, 27.08) --
	(334.79, 27.08) --
	(335.39, 27.08) --
	(336.00, 27.08) --
	(336.61, 27.08) --
	(337.22, 27.08) --
	(337.82, 27.08) --
	(338.43, 27.08) --
	(339.04, 27.08) --
	(339.65, 27.08) --
	(340.26, 27.08) --
	(340.86, 27.08) --
	(341.47, 27.08) --
	(342.08, 27.08) --
	(342.69, 27.08);
\definecolor{drawColor}{RGB}{0,158,115}

\path[draw=drawColor,line width= 0.6pt,dash pattern=on 7pt off 3pt ,line join=round] ( 39.42, 27.08) --
	( 40.02, 27.08) --
	( 40.63, 27.08) --
	( 41.24, 27.08) --
	( 41.85, 27.08) --
	( 42.45, 27.08) --
	( 43.06, 27.08) --
	( 43.67, 27.08) --
	( 44.28, 27.08) --
	( 44.89, 27.08) --
	( 45.49, 27.08) --
	( 46.10, 27.08) --
	( 46.71, 27.08) --
	( 47.32, 27.08) --
	( 47.92, 27.08) --
	( 48.53, 27.08) --
	( 49.14, 27.08) --
	( 49.75, 27.08) --
	( 50.35, 27.08) --
	( 50.96, 27.08) --
	( 51.57, 27.08) --
	( 52.18, 27.08) --
	( 52.79, 27.08) --
	( 53.39, 27.08) --
	( 54.00, 27.08) --
	( 54.61, 27.08) --
	( 55.22, 27.08) --
	( 55.82, 27.08) --
	( 56.43, 27.08) --
	( 57.04, 27.08) --
	( 57.65, 27.08) --
	( 58.26, 27.08) --
	( 58.86, 27.08) --
	( 59.47, 27.08) --
	( 60.08, 27.08) --
	( 60.69, 27.08) --
	( 61.29, 27.08) --
	( 61.90, 27.08) --
	( 62.51, 27.08) --
	( 63.12, 27.08) --
	( 63.73, 27.08) --
	( 64.33, 27.08) --
	( 64.94, 27.08) --
	( 65.55, 27.08) --
	( 66.16, 27.08) --
	( 66.76, 27.08) --
	( 67.37, 27.08) --
	( 67.98, 27.08) --
	( 68.59, 27.08) --
	( 69.20, 27.08) --
	( 69.80, 27.08) --
	( 70.41, 27.08) --
	( 71.02, 27.08) --
	( 71.63, 27.08) --
	( 72.23, 27.08) --
	( 72.84, 27.08) --
	( 73.45, 27.08) --
	( 74.06, 27.08) --
	( 74.67, 27.08) --
	( 75.27, 27.08) --
	( 75.88, 27.08) --
	( 76.49, 27.08) --
	( 77.10, 27.08) --
	( 77.70, 27.08) --
	( 78.31, 27.08) --
	( 78.92, 27.08) --
	( 79.53, 27.08) --
	( 80.13, 27.08) --
	( 80.74, 27.08) --
	( 81.35, 27.08) --
	( 81.96, 27.08) --
	( 82.57, 27.08) --
	( 83.17, 27.08) --
	( 83.78, 27.08) --
	( 84.39, 27.08) --
	( 85.00, 27.08) --
	( 85.60, 27.08) --
	( 86.21, 27.08) --
	( 86.82, 27.08) --
	( 87.43, 27.08) --
	( 88.04, 27.08) --
	( 88.64, 27.08) --
	( 89.25, 27.08) --
	( 89.86, 27.08) --
	( 90.47, 27.08) --
	( 91.07, 27.08) --
	( 91.68, 27.08) --
	( 92.29, 27.08) --
	( 92.90, 27.08) --
	( 93.51, 27.08) --
	( 94.11, 27.08) --
	( 94.72, 27.08) --
	( 95.33, 27.08) --
	( 95.94, 27.08) --
	( 96.54, 27.08) --
	( 97.15, 27.08) --
	( 97.76, 27.08) --
	( 98.37, 27.08) --
	( 98.98, 27.08) --
	( 99.58, 27.08) --
	(100.19, 27.08) --
	(100.80, 27.08) --
	(101.41, 27.08) --
	(102.01, 27.08) --
	(102.62, 27.08) --
	(103.23, 27.08) --
	(103.84, 27.08) --
	(104.45, 27.08) --
	(105.05, 27.08) --
	(105.66, 27.08) --
	(106.27, 27.08) --
	(106.88, 27.08) --
	(107.48, 27.08) --
	(108.09, 27.08) --
	(108.70, 27.08) --
	(109.31, 27.08) --
	(109.92, 27.08) --
	(110.52, 27.08) --
	(111.13, 27.08) --
	(111.74, 27.08) --
	(112.35, 27.08) --
	(112.95, 27.08) --
	(113.56, 27.08) --
	(114.17, 27.08) --
	(114.78, 27.08) --
	(115.38, 27.08) --
	(115.99, 27.08) --
	(116.60, 27.09) --
	(117.21, 27.09) --
	(117.82, 27.09) --
	(118.42, 27.09) --
	(119.03, 27.09) --
	(119.64, 27.09) --
	(120.25, 27.09) --
	(120.85, 27.09) --
	(121.46, 27.09) --
	(122.07, 27.09) --
	(122.68, 27.10) --
	(123.29, 27.10) --
	(123.89, 27.10) --
	(124.50, 27.10) --
	(125.11, 27.11) --
	(125.72, 27.11) --
	(126.32, 27.11) --
	(126.93, 27.11) --
	(127.54, 27.12) --
	(128.15, 27.12) --
	(128.76, 27.13) --
	(129.36, 27.13) --
	(129.97, 27.14) --
	(130.58, 27.14) --
	(131.19, 27.15) --
	(131.79, 27.16) --
	(132.40, 27.17) --
	(133.01, 27.17) --
	(133.62, 27.18) --
	(134.23, 27.19) --
	(134.83, 27.21) --
	(135.44, 27.22) --
	(136.05, 27.23) --
	(136.66, 27.24) --
	(137.26, 27.26) --
	(137.87, 27.27) --
	(138.48, 27.29) --
	(139.09, 27.31) --
	(139.70, 27.33) --
	(140.30, 27.35) --
	(140.91, 27.37) --
	(141.52, 27.40) --
	(142.13, 27.43) --
	(142.73, 27.45) --
	(143.34, 27.48) --
	(143.95, 27.51) --
	(144.56, 27.55) --
	(145.17, 27.58) --
	(145.77, 27.62) --
	(146.38, 27.66) --
	(146.99, 27.70) --
	(147.60, 27.75) --
	(148.20, 27.80) --
	(148.81, 27.85) --
	(149.42, 27.90) --
	(150.03, 27.96) --
	(150.63, 28.02) --
	(151.24, 28.08) --
	(151.85, 28.14) --
	(152.46, 28.21) --
	(153.07, 28.28) --
	(153.67, 28.36) --
	(154.28, 28.44) --
	(154.89, 28.52) --
	(155.50, 28.60) --
	(156.10, 28.69) --
	(156.71, 28.79) --
	(157.32, 28.89) --
	(157.93, 28.99) --
	(158.54, 29.09) --
	(159.14, 29.20) --
	(159.75, 29.32) --
	(160.36, 29.44) --
	(160.97, 29.56) --
	(161.57, 29.69) --
	(162.18, 29.82) --
	(162.79, 29.96) --
	(163.40, 30.10) --
	(164.01, 30.24) --
	(164.61, 30.39) --
	(165.22, 30.55) --
	(165.83, 30.71) --
	(166.44, 30.87) --
	(167.04, 31.04) --
	(167.65, 31.22) --
	(168.26, 31.39) --
	(168.87, 31.58) --
	(169.48, 31.77) --
	(170.08, 31.96) --
	(170.69, 32.15) --
	(171.30, 32.36) --
	(171.91, 32.56) --
	(172.51, 32.77) --
	(173.12, 32.99) --
	(173.73, 33.21) --
	(174.34, 33.43) --
	(174.95, 33.66) --
	(175.55, 33.89) --
	(176.16, 34.13) --
	(176.77, 34.36) --
	(177.38, 34.61) --
	(177.98, 34.85) --
	(178.59, 35.10) --
	(179.20, 35.36) --
	(179.81, 35.61) --
	(180.42, 35.87) --
	(181.02, 36.13) --
	(181.63, 36.40) --
	(182.24, 36.66) --
	(182.85, 36.93) --
	(183.45, 37.20) --
	(184.06, 37.47) --
	(184.67, 37.75) --
	(185.28, 38.02) --
	(185.88, 38.30) --
	(186.49, 38.58) --
	(187.10, 38.85) --
	(187.71, 39.13) --
	(188.32, 39.41) --
	(188.92, 39.69) --
	(189.53, 39.97) --
	(190.14, 40.25) --
	(190.75, 40.52) --
	(191.35, 40.80) --
	(191.96, 41.08) --
	(192.57, 41.35) --
	(193.18, 41.62) --
	(193.79, 41.89) --
	(194.39, 42.16) --
	(195.00, 42.43) --
	(195.61, 42.69) --
	(196.22, 42.95) --
	(196.82, 43.21) --
	(197.43, 43.46) --
	(198.04, 43.72) --
	(198.65, 43.96) --
	(199.26, 44.21) --
	(199.86, 44.44) --
	(200.47, 44.68) --
	(201.08, 44.91) --
	(201.69, 45.13) --
	(202.29, 45.35) --
	(202.90, 45.57) --
	(203.51, 45.78) --
	(204.12, 45.98) --
	(204.73, 46.18) --
	(205.33, 46.37) --
	(205.94, 46.56) --
	(206.55, 46.74) --
	(207.16, 46.91) --
	(207.76, 47.08) --
	(208.37, 47.24) --
	(208.98, 47.39) --
	(209.59, 47.54) --
	(210.20, 47.68) --
	(210.80, 47.81) --
	(211.41, 47.93) --
	(212.02, 48.05) --
	(212.63, 48.16) --
	(213.23, 48.27) --
	(213.84, 48.36) --
	(214.45, 48.45) --
	(215.06, 48.53) --
	(215.67, 48.60) --
	(216.27, 48.67) --
	(216.88, 48.72) --
	(217.49, 48.77) --
	(218.10, 48.82) --
	(218.70, 48.85) --
	(219.31, 48.88) --
	(219.92, 48.90) --
	(220.53, 48.91) --
	(221.13, 48.91) --
	(221.74, 48.91) --
	(222.35, 48.90) --
	(222.96, 48.88) --
	(223.57, 48.85) --
	(224.17, 48.82) --
	(224.78, 48.78) --
	(225.39, 48.73) --
	(226.00, 48.67) --
	(226.60, 48.61) --
	(227.21, 48.54) --
	(227.82, 48.47) --
	(228.43, 48.39) --
	(229.04, 48.30) --
	(229.64, 48.21) --
	(230.25, 48.10) --
	(230.86, 48.00) --
	(231.47, 47.89) --
	(232.07, 47.77) --
	(232.68, 47.64) --
	(233.29, 47.51) --
	(233.90, 47.38) --
	(234.51, 47.24) --
	(235.11, 47.09) --
	(235.72, 46.94) --
	(236.33, 46.79) --
	(236.94, 46.63) --
	(237.54, 46.46) --
	(238.15, 46.29) --
	(238.76, 46.12) --
	(239.37, 45.95) --
	(239.98, 45.77) --
	(240.58, 45.58) --
	(241.19, 45.40) --
	(241.80, 45.21) --
	(242.41, 45.01) --
	(243.01, 44.82) --
	(243.62, 44.62) --
	(244.23, 44.42) --
	(244.84, 44.21) --
	(245.45, 44.01) --
	(246.05, 43.80) --
	(246.66, 43.59) --
	(247.27, 43.38) --
	(247.88, 43.17) --
	(248.48, 42.95) --
	(249.09, 42.73) --
	(249.70, 42.52) --
	(250.31, 42.30) --
	(250.91, 42.08) --
	(251.52, 41.86) --
	(252.13, 41.64) --
	(252.74, 41.42) --
	(253.35, 41.20) --
	(253.95, 40.98) --
	(254.56, 40.76) --
	(255.17, 40.54) --
	(255.78, 40.32) --
	(256.38, 40.10) --
	(256.99, 39.88) --
	(257.60, 39.66) --
	(258.21, 39.44) --
	(258.82, 39.23) --
	(259.42, 39.01) --
	(260.03, 38.80) --
	(260.64, 38.58) --
	(261.25, 38.37) --
	(261.85, 38.16) --
	(262.46, 37.95) --
	(263.07, 37.74) --
	(263.68, 37.53) --
	(264.29, 37.33) --
	(264.89, 37.12) --
	(265.50, 36.92) --
	(266.11, 36.72) --
	(266.72, 36.52) --
	(267.32, 36.33) --
	(267.93, 36.13) --
	(268.54, 35.94) --
	(269.15, 35.75) --
	(269.76, 35.57) --
	(270.36, 35.38) --
	(270.97, 35.20) --
	(271.58, 35.02) --
	(272.19, 34.84) --
	(272.79, 34.66) --
	(273.40, 34.49) --
	(274.01, 34.32) --
	(274.62, 34.15) --
	(275.23, 33.98) --
	(275.83, 33.82) --
	(276.44, 33.66) --
	(277.05, 33.50) --
	(277.66, 33.34) --
	(278.26, 33.19) --
	(278.87, 33.04) --
	(279.48, 32.89) --
	(280.09, 32.75) --
	(280.70, 32.60) --
	(281.30, 32.46) --
	(281.91, 32.32) --
	(282.52, 32.19) --
	(283.13, 32.05) --
	(283.73, 31.92) --
	(284.34, 31.80) --
	(284.95, 31.67) --
	(285.56, 31.55) --
	(286.16, 31.43) --
	(286.77, 31.31) --
	(287.38, 31.19) --
	(287.99, 31.08) --
	(288.60, 30.97) --
	(289.20, 30.86) --
	(289.81, 30.75) --
	(290.42, 30.65) --
	(291.03, 30.55) --
	(291.63, 30.45) --
	(292.24, 30.35) --
	(292.85, 30.26) --
	(293.46, 30.16) --
	(294.07, 30.07) --
	(294.67, 29.98) --
	(295.28, 29.90) --
	(295.89, 29.81) --
	(296.50, 29.73) --
	(297.10, 29.65) --
	(297.71, 29.57) --
	(298.32, 29.50) --
	(298.93, 29.42) --
	(299.54, 29.35) --
	(300.14, 29.28) --
	(300.75, 29.21) --
	(301.36, 29.14) --
	(301.97, 29.08) --
	(302.57, 29.01) --
	(303.18, 28.95) --
	(303.79, 28.89) --
	(304.40, 28.83) --
	(305.01, 28.78) --
	(305.61, 28.72) --
	(306.22, 28.67) --
	(306.83, 28.62) --
	(307.44, 28.57) --
	(308.04, 28.52) --
	(308.65, 28.47) --
	(309.26, 28.42) --
	(309.87, 28.38) --
	(310.48, 28.33) --
	(311.08, 28.29) --
	(311.69, 28.25) --
	(312.30, 28.21) --
	(312.91, 28.17) --
	(313.51, 28.13) --
	(314.12, 28.10) --
	(314.73, 28.06) --
	(315.34, 28.03) --
	(315.95, 27.99) --
	(316.55, 27.96) --
	(317.16, 27.93) --
	(317.77, 27.90) --
	(318.38, 27.87) --
	(318.98, 27.84) --
	(319.59, 27.81) --
	(320.20, 27.79) --
	(320.81, 27.76) --
	(321.41, 27.74) --
	(322.02, 27.71) --
	(322.63, 27.69) --
	(323.24, 27.67) --
	(323.85, 27.64) --
	(324.45, 27.62) --
	(325.06, 27.60) --
	(325.67, 27.58) --
	(326.28, 27.56) --
	(326.88, 27.55) --
	(327.49, 27.53) --
	(328.10, 27.51) --
	(328.71, 27.50) --
	(329.32, 27.48) --
	(329.92, 27.46) --
	(330.53, 27.45) --
	(331.14, 27.43) --
	(331.75, 27.42) --
	(332.35, 27.41) --
	(332.96, 27.40) --
	(333.57, 27.38) --
	(334.18, 27.37) --
	(334.79, 27.36) --
	(335.39, 27.35) --
	(336.00, 27.34) --
	(336.61, 27.33) --
	(337.22, 27.32) --
	(337.82, 27.31) --
	(338.43, 27.30) --
	(339.04, 27.29) --
	(339.65, 27.28) --
	(340.26, 27.27) --
	(340.86, 27.27) --
	(341.47, 27.26) --
	(342.08, 27.25) --
	(342.69, 27.24);
\end{scope}
\begin{scope}
\path[clip] (  0.00,  0.00) rectangle (361.35,144.54);
\definecolor{drawColor}{RGB}{0,0,0}

\path[draw=drawColor,line width= 0.4pt,line join=round] ( 24.25, 21.65) --
	( 24.25,141.04);
\end{scope}
\begin{scope}
\path[clip] (  0.00,  0.00) rectangle (361.35,144.54);
\definecolor{drawColor}{gray}{0.30}

\node[text=drawColor,anchor=base east,inner sep=0pt, outer sep=0pt, scale=  0.56] at ( 21.10, 25.15) {0};

\node[text=drawColor,anchor=base east,inner sep=0pt, outer sep=0pt, scale=  0.56] at ( 21.10, 54.72) {100};

\node[text=drawColor,anchor=base east,inner sep=0pt, outer sep=0pt, scale=  0.56] at ( 21.10, 84.29) {200};

\node[text=drawColor,anchor=base east,inner sep=0pt, outer sep=0pt, scale=  0.56] at ( 21.10,113.85) {300};
\end{scope}
\begin{scope}
\path[clip] (  0.00,  0.00) rectangle (361.35,144.54);
\definecolor{drawColor}{gray}{0.20}

\path[draw=drawColor,line width= 0.4pt,line join=round] ( 22.50, 27.08) --
	( 24.25, 27.08);

\path[draw=drawColor,line width= 0.4pt,line join=round] ( 22.50, 56.65) --
	( 24.25, 56.65);

\path[draw=drawColor,line width= 0.4pt,line join=round] ( 22.50, 86.21) --
	( 24.25, 86.21);

\path[draw=drawColor,line width= 0.4pt,line join=round] ( 22.50,115.78) --
	( 24.25,115.78);
\end{scope}
\begin{scope}
\path[clip] (  0.00,  0.00) rectangle (361.35,144.54);
\definecolor{drawColor}{RGB}{0,0,0}

\path[draw=drawColor,line width= 0.4pt,line join=round] ( 24.25, 21.65) --
	(357.85, 21.65);
\end{scope}
\begin{scope}
\path[clip] (  0.00,  0.00) rectangle (361.35,144.54);
\definecolor{drawColor}{gray}{0.20}

\path[draw=drawColor,line width= 0.4pt,line join=round] ( 39.42, 19.90) --
	( 39.42, 21.65);

\path[draw=drawColor,line width= 0.4pt,line join=round] (100.07, 19.90) --
	(100.07, 21.65);

\path[draw=drawColor,line width= 0.4pt,line join=round] (160.72, 19.90) --
	(160.72, 21.65);

\path[draw=drawColor,line width= 0.4pt,line join=round] (221.38, 19.90) --
	(221.38, 21.65);

\path[draw=drawColor,line width= 0.4pt,line join=round] (282.03, 19.90) --
	(282.03, 21.65);

\path[draw=drawColor,line width= 0.4pt,line join=round] (342.69, 19.90) --
	(342.69, 21.65);
\end{scope}
\begin{scope}
\path[clip] (  0.00,  0.00) rectangle (361.35,144.54);
\definecolor{drawColor}{gray}{0.30}

\node[text=drawColor,anchor=base,inner sep=0pt, outer sep=0pt, scale=  0.56] at ( 39.42, 14.65) {0.00};

\node[text=drawColor,anchor=base,inner sep=0pt, outer sep=0pt, scale=  0.56] at (100.07, 14.65) {0.01};

\node[text=drawColor,anchor=base,inner sep=0pt, outer sep=0pt, scale=  0.56] at (160.72, 14.65) {0.02};

\node[text=drawColor,anchor=base,inner sep=0pt, outer sep=0pt, scale=  0.56] at (221.38, 14.65) {0.03};

\node[text=drawColor,anchor=base,inner sep=0pt, outer sep=0pt, scale=  0.56] at (282.03, 14.65) {0.04};

\node[text=drawColor,anchor=base,inner sep=0pt, outer sep=0pt, scale=  0.56] at (342.69, 14.65) {0.05};
\end{scope}
\begin{scope}
\path[clip] (  0.00,  0.00) rectangle (361.35,144.54);
\definecolor{drawColor}{RGB}{0,0,0}

\node[text=drawColor,anchor=base,inner sep=0pt, outer sep=0pt, scale=  0.80] at (191.05,  5.44) {$p$};
\end{scope}
\begin{scope}
\path[clip] (  0.00,  0.00) rectangle (361.35,144.54);
\definecolor{drawColor}{RGB}{0,0,0}

\node[text=drawColor,rotate= 90.00,anchor=base,inner sep=0pt, outer sep=0pt, scale=  0.80] at (  9.01, 81.35) {Density of $\tilde{p}$ given $p$};
\end{scope}
\end{tikzpicture}

%% file: Figures/ec_plt1.tex
\begin{tikzpicture}[x=1pt,y=1pt]
\definecolor{fillColor}{RGB}{255,255,255}
\path[use as bounding box,fill=fillColor,fill opacity=0.00] (0,0) rectangle (144.54,144.54);
\begin{scope}
\path[clip] (  0.00,  0.00) rectangle (144.54,144.54);
\definecolor{drawColor}{RGB}{255,255,255}
\definecolor{fillColor}{RGB}{255,255,255}

\path[draw=drawColor,line width= 0.4pt,line join=round,line cap=round,fill=fillColor] (  0.00, -0.00) rectangle (144.54,144.54);
\end{scope}
\begin{scope}
\path[clip] ( 20.25, 23.41) rectangle (140.54,129.09);
\definecolor{fillColor}{RGB}{255,255,255}

\path[fill=fillColor] ( 20.25, 23.41) rectangle (140.54,129.09);
\definecolor{drawColor}{RGB}{213,94,0}

\path[draw=drawColor,line width= 0.6pt,dash pattern=on 7pt off 3pt ,line join=round] ( 25.72,124.28) --
	( 26.27, 86.98) --
	( 26.82, 70.00) --
	( 27.37, 62.09) --
	( 27.92, 57.76) --
	( 28.47, 53.45) --
	( 29.02, 52.28) --
	( 29.57, 49.81) --
	( 30.12, 48.30) --
	( 30.67, 47.52) --
	( 31.22, 45.76) --
	( 31.77, 45.25) --
	( 32.31, 44.31) --
	( 32.86, 43.87) --
	( 33.41, 43.24) --
	( 33.96, 42.43) --
	( 34.51, 42.58) --
	( 35.06, 42.13) --
	( 35.61, 41.90) --
	( 36.16, 41.05) --
	( 36.71, 41.04) --
	( 37.26, 41.05) --
	( 37.81, 40.84) --
	( 38.36, 40.68) --
	( 38.91, 40.33) --
	( 39.46, 40.28) --
	( 40.01, 40.08) --
	( 40.56, 39.67) --
	( 41.11, 39.28) --
	( 41.66, 39.57) --
	( 42.21, 39.35) --
	( 42.76, 39.26) --
	( 43.30, 39.17) --
	( 43.85, 38.98) --
	( 44.40, 39.03) --
	( 44.95, 39.09) --
	( 45.50, 38.81) --
	( 46.05, 38.73) --
	( 46.60, 38.55) --
	( 47.15, 38.89) --
	( 47.70, 38.61) --
	( 48.25, 38.68) --
	( 48.80, 38.33) --
	( 49.35, 38.23) --
	( 49.90, 38.16) --
	( 50.45, 38.22) --
	( 51.00, 38.09) --
	( 51.55, 38.18) --
	( 52.10, 37.81) --
	( 52.65, 37.69) --
	( 53.20, 38.24) --
	( 53.75, 38.06) --
	( 54.29, 37.99) --
	( 54.84, 37.90) --
	( 55.39, 37.57) --
	( 55.94, 37.88) --
	( 56.49, 37.73) --
	( 57.04, 37.91) --
	( 57.59, 37.76) --
	( 58.14, 37.76) --
	( 58.69, 37.72) --
	( 59.24, 37.56) --
	( 59.79, 37.64) --
	( 60.34, 37.70) --
	( 60.89, 37.60) --
	( 61.44, 37.59) --
	( 61.99, 37.47) --
	( 62.54, 37.47) --
	( 63.09, 37.61) --
	( 63.64, 37.53) --
	( 64.19, 37.23) --
	( 64.74, 37.28) --
	( 65.29, 37.16) --
	( 65.83, 37.28) --
	( 66.38, 37.28) --
	( 66.93, 37.28) --
	( 67.48, 37.30) --
	( 68.03, 37.21) --
	( 68.58, 37.43) --
	( 69.13, 37.12) --
	( 69.68, 37.15) --
	( 70.23, 37.42) --
	( 70.78, 37.16) --
	( 71.33, 37.13) --
	( 71.88, 37.07) --
	( 72.43, 37.17) --
	( 72.98, 37.20) --
	( 73.53, 36.97) --
	( 74.08, 36.96) --
	( 74.63, 37.16) --
	( 75.18, 37.13) --
	( 75.73, 36.92) --
	( 76.28, 36.92) --
	( 76.82, 36.87) --
	( 77.37, 36.97) --
	( 77.92, 37.07) --
	( 78.47, 36.93) --
	( 79.02, 36.95) --
	( 79.57, 36.86) --
	( 80.12, 36.99) --
	( 80.67, 36.96) --
	( 81.22, 36.93) --
	( 81.77, 36.92) --
	( 82.32, 37.04) --
	( 82.87, 37.10) --
	( 83.42, 36.96) --
	( 83.97, 36.69) --
	( 84.52, 36.98) --
	( 85.07, 36.89) --
	( 85.62, 36.81) --
	( 86.17, 36.80) --
	( 86.72, 36.88) --
	( 87.27, 36.73) --
	( 87.81, 36.66) --
	( 88.36, 36.76) --
	( 88.91, 36.84) --
	( 89.46, 36.68) --
	( 90.01, 36.66) --
	( 90.56, 36.86) --
	( 91.11, 36.63) --
	( 91.66, 36.66) --
	( 92.21, 36.66) --
	( 92.76, 36.77) --
	( 93.31, 36.82) --
	( 93.86, 36.63) --
	( 94.41, 36.42) --
	( 94.96, 36.74) --
	( 95.51, 36.59) --
	( 96.06, 36.55) --
	( 96.61, 36.67) --
	( 97.16, 36.72) --
	( 97.71, 36.80) --
	( 98.26, 36.55) --
	( 98.80, 36.65) --
	( 99.35, 36.64) --
	( 99.90, 36.83) --
	(100.45, 36.52) --
	(101.00, 36.60) --
	(101.55, 36.75) --
	(102.10, 36.69) --
	(102.65, 36.70) --
	(103.20, 36.73) --
	(103.75, 36.55) --
	(104.30, 36.79) --
	(104.85, 36.62) --
	(105.40, 36.59) --
	(105.95, 36.66) --
	(106.50, 36.65) --
	(107.05, 36.64) --
	(107.60, 36.40) --
	(108.15, 36.62) --
	(108.70, 36.56) --
	(109.25, 36.45) --
	(109.80, 36.46) --
	(110.34, 36.60) --
	(110.89, 36.47) --
	(111.44, 36.60) --
	(111.99, 36.39) --
	(112.54, 36.56) --
	(113.09, 36.48) --
	(113.64, 36.53) --
	(114.19, 36.45) --
	(114.74, 36.41) --
	(115.29, 36.53) --
	(115.84, 36.59) --
	(116.39, 36.47) --
	(116.94, 36.54) --
	(117.49, 36.36) --
	(118.04, 36.57) --
	(118.59, 36.38) --
	(119.14, 36.53) --
	(119.69, 36.59) --
	(120.24, 36.43) --
	(120.79, 36.44) --
	(121.33, 36.51) --
	(121.88, 36.41) --
	(122.43, 36.43) --
	(122.98, 36.49) --
	(123.53, 36.45) --
	(124.08, 36.49) --
	(124.63, 36.39) --
	(125.18, 36.40) --
	(125.73, 36.34) --
	(126.28, 36.51) --
	(126.83, 36.33) --
	(127.38, 36.40) --
	(127.93, 36.33) --
	(128.48, 36.44) --
	(129.03, 36.51) --
	(129.58, 36.36) --
	(130.13, 36.31) --
	(130.68, 36.41) --
	(131.23, 36.50) --
	(131.78, 36.30) --
	(132.32, 36.20) --
	(132.87, 36.39) --
	(133.42, 36.39) --
	(133.97, 36.38) --
	(134.52, 36.29) --
	(135.07, 36.25);
\definecolor{drawColor}{RGB}{0,0,0}

\path[draw=drawColor,line width= 0.6pt,line join=round] ( 20.25, 35.87) -- (140.54, 35.87);
\definecolor{drawColor}{RGB}{0,158,115}

\path[draw=drawColor,line width= 0.6pt,line join=round] ( 25.72, 28.21) --
	( 26.27, 28.75) --
	( 26.82, 29.13) --
	( 27.37, 29.69) --
	( 27.92, 29.81) --
	( 28.47, 29.95) --
	( 29.02, 30.38) --
	( 29.57, 30.62) --
	( 30.12, 30.67) --
	( 30.67, 31.14) --
	( 31.22, 31.27) --
	( 31.77, 31.50) --
	( 32.31, 31.63) --
	( 32.86, 31.74) --
	( 33.41, 31.93) --
	( 33.96, 31.99) --
	( 34.51, 32.02) --
	( 35.06, 32.30) --
	( 35.61, 32.41) --
	( 36.16, 32.22) --
	( 36.71, 32.59) --
	( 37.26, 32.63) --
	( 37.81, 32.83) --
	( 38.36, 32.82) --
	( 38.91, 33.00) --
	( 39.46, 32.97) --
	( 40.01, 33.01) --
	( 40.56, 33.09) --
	( 41.11, 33.25) --
	( 41.66, 33.34) --
	( 42.21, 33.32) --
	( 42.76, 33.40) --
	( 43.30, 33.35) --
	( 43.85, 33.49) --
	( 44.40, 33.61) --
	( 44.95, 33.65) --
	( 45.50, 33.66) --
	( 46.05, 33.67) --
	( 46.60, 33.79) --
	( 47.15, 33.62) --
	( 47.70, 33.81) --
	( 48.25, 33.77) --
	( 48.80, 33.83) --
	( 49.35, 33.96) --
	( 49.90, 33.85) --
	( 50.45, 33.95) --
	( 51.00, 34.01) --
	( 51.55, 34.09) --
	( 52.10, 34.05) --
	( 52.65, 34.00) --
	( 53.20, 34.12) --
	( 53.75, 34.10) --
	( 54.29, 34.11) --
	( 54.84, 34.12) --
	( 55.39, 34.26) --
	( 55.94, 34.22) --
	( 56.49, 34.26) --
	( 57.04, 34.34) --
	( 57.59, 34.33) --
	( 58.14, 34.32) --
	( 58.69, 34.38) --
	( 59.24, 34.34) --
	( 59.79, 34.43) --
	( 60.34, 34.43) --
	( 60.89, 34.49) --
	( 61.44, 34.51) --
	( 61.99, 34.41) --
	( 62.54, 34.55) --
	( 63.09, 34.55) --
	( 63.64, 34.53) --
	( 64.19, 34.55) --
	( 64.74, 34.60) --
	( 65.29, 34.55) --
	( 65.83, 34.58) --
	( 66.38, 34.53) --
	( 66.93, 34.60) --
	( 67.48, 34.68) --
	( 68.03, 34.65) --
	( 68.58, 34.68) --
	( 69.13, 34.67) --
	( 69.68, 34.68) --
	( 70.23, 34.69) --
	( 70.78, 34.70) --
	( 71.33, 34.73) --
	( 71.88, 34.78) --
	( 72.43, 34.75) --
	( 72.98, 34.79) --
	( 73.53, 34.77) --
	( 74.08, 34.81) --
	( 74.63, 34.78) --
	( 75.18, 34.82) --
	( 75.73, 34.85) --
	( 76.28, 34.80) --
	( 76.82, 34.83) --
	( 77.37, 34.89) --
	( 77.92, 34.85) --
	( 78.47, 34.87) --
	( 79.02, 34.89) --
	( 79.57, 34.91) --
	( 80.12, 34.90) --
	( 80.67, 34.93) --
	( 81.22, 34.95) --
	( 81.77, 34.95) --
	( 82.32, 34.89) --
	( 82.87, 34.93) --
	( 83.42, 34.95) --
	( 83.97, 35.00) --
	( 84.52, 34.96) --
	( 85.07, 34.98) --
	( 85.62, 35.01) --
	( 86.17, 34.99) --
	( 86.72, 35.01) --
	( 87.27, 35.01) --
	( 87.81, 35.05) --
	( 88.36, 35.04) --
	( 88.91, 34.99) --
	( 89.46, 35.04) --
	( 90.01, 35.07) --
	( 90.56, 35.07) --
	( 91.11, 35.08) --
	( 91.66, 35.06) --
	( 92.21, 35.08) --
	( 92.76, 35.06) --
	( 93.31, 35.07) --
	( 93.86, 35.11) --
	( 94.41, 35.10) --
	( 94.96, 35.10) --
	( 95.51, 35.10) --
	( 96.06, 35.12) --
	( 96.61, 35.13) --
	( 97.16, 35.10) --
	( 97.71, 35.17) --
	( 98.26, 35.13) --
	( 98.80, 35.14) --
	( 99.35, 35.12) --
	( 99.90, 35.13) --
	(100.45, 35.17) --
	(101.00, 35.18) --
	(101.55, 35.16) --
	(102.10, 35.19) --
	(102.65, 35.16) --
	(103.20, 35.14) --
	(103.75, 35.15) --
	(104.30, 35.18) --
	(104.85, 35.19) --
	(105.40, 35.14) --
	(105.95, 35.22) --
	(106.50, 35.20) --
	(107.05, 35.21) --
	(107.60, 35.22) --
	(108.15, 35.22) --
	(108.70, 35.24) --
	(109.25, 35.21) --
	(109.80, 35.22) --
	(110.34, 35.23) --
	(110.89, 35.25) --
	(111.44, 35.24) --
	(111.99, 35.23) --
	(112.54, 35.25) --
	(113.09, 35.27) --
	(113.64, 35.27) --
	(114.19, 35.24) --
	(114.74, 35.27) --
	(115.29, 35.26) --
	(115.84, 35.26) --
	(116.39, 35.28) --
	(116.94, 35.29) --
	(117.49, 35.28) --
	(118.04, 35.27) --
	(118.59, 35.28) --
	(119.14, 35.28) --
	(119.69, 35.29) --
	(120.24, 35.30) --
	(120.79, 35.28) --
	(121.33, 35.32) --
	(121.88, 35.28) --
	(122.43, 35.31) --
	(122.98, 35.32) --
	(123.53, 35.35) --
	(124.08, 35.31) --
	(124.63, 35.32) --
	(125.18, 35.36) --
	(125.73, 35.34) --
	(126.28, 35.32) --
	(126.83, 35.32) --
	(127.38, 35.33) --
	(127.93, 35.33) --
	(128.48, 35.34) --
	(129.03, 35.37) --
	(129.58, 35.38) --
	(130.13, 35.35) --
	(130.68, 35.36) --
	(131.23, 35.35) --
	(131.78, 35.36) --
	(132.32, 35.39) --
	(132.87, 35.33) --
	(133.42, 35.37) --
	(133.97, 35.38) --
	(134.52, 35.36) --
	(135.07, 35.37);
\end{scope}
\begin{scope}
\path[clip] (  0.00,  0.00) rectangle (144.54,144.54);
\definecolor{drawColor}{RGB}{0,0,0}

\path[draw=drawColor,line width= 0.4pt,line join=round] ( 20.25, 23.41) --
	( 20.25,129.09);
\end{scope}
\begin{scope}
\path[clip] (  0.00,  0.00) rectangle (144.54,144.54);
\definecolor{drawColor}{gray}{0.30}

\node[text=drawColor,anchor=base east,inner sep=0pt, outer sep=0pt, scale=  0.64] at ( 16.65, 33.67) {0};

\node[text=drawColor,anchor=base east,inner sep=0pt, outer sep=0pt, scale=  0.64] at ( 16.65, 61.43) {1};

\node[text=drawColor,anchor=base east,inner sep=0pt, outer sep=0pt, scale=  0.64] at ( 16.65, 89.19) {2};

\node[text=drawColor,anchor=base east,inner sep=0pt, outer sep=0pt, scale=  0.64] at ( 16.65,116.94) {3};
\end{scope}
\begin{scope}
\path[clip] (  0.00,  0.00) rectangle (144.54,144.54);
\definecolor{drawColor}{gray}{0.20}

\path[draw=drawColor,line width= 0.4pt,line join=round] ( 18.25, 35.87) --
	( 20.25, 35.87);

\path[draw=drawColor,line width= 0.4pt,line join=round] ( 18.25, 63.63) --
	( 20.25, 63.63);

\path[draw=drawColor,line width= 0.4pt,line join=round] ( 18.25, 91.39) --
	( 20.25, 91.39);

\path[draw=drawColor,line width= 0.4pt,line join=round] ( 18.25,119.15) --
	( 20.25,119.15);
\end{scope}
\begin{scope}
\path[clip] (  0.00,  0.00) rectangle (144.54,144.54);
\definecolor{drawColor}{RGB}{0,0,0}

\path[draw=drawColor,line width= 0.4pt,line join=round] ( 20.25, 23.41) --
	(140.54, 23.41);
\end{scope}
\begin{scope}
\path[clip] (  0.00,  0.00) rectangle (144.54,144.54);
\definecolor{drawColor}{gray}{0.20}

\path[draw=drawColor,line width= 0.4pt,line join=round] ( 25.23, 21.41) --
	( 25.23, 23.41);

\path[draw=drawColor,line width= 0.4pt,line join=round] ( 47.25, 21.41) --
	( 47.25, 23.41);

\path[draw=drawColor,line width= 0.4pt,line join=round] ( 69.27, 21.41) --
	( 69.27, 23.41);

\path[draw=drawColor,line width= 0.4pt,line join=round] ( 91.30, 21.41) --
	( 91.30, 23.41);

\path[draw=drawColor,line width= 0.4pt,line join=round] (113.32, 21.41) --
	(113.32, 23.41);

\path[draw=drawColor,line width= 0.4pt,line join=round] (135.35, 21.41) --
	(135.35, 23.41);
\end{scope}
\begin{scope}
\path[clip] (  0.00,  0.00) rectangle (144.54,144.54);
\definecolor{drawColor}{gray}{0.30}

\node[text=drawColor,anchor=base,inner sep=0pt, outer sep=0pt, scale=  0.64] at ( 25.23, 15.40) {0.00};

\node[text=drawColor,anchor=base,inner sep=0pt, outer sep=0pt, scale=  0.64] at ( 47.25, 15.40) {0.01};

\node[text=drawColor,anchor=base,inner sep=0pt, outer sep=0pt, scale=  0.64] at ( 69.27, 15.40) {0.02};

\node[text=drawColor,anchor=base,inner sep=0pt, outer sep=0pt, scale=  0.64] at ( 91.30, 15.40) {0.03};

\node[text=drawColor,anchor=base,inner sep=0pt, outer sep=0pt, scale=  0.64] at (113.32, 15.40) {0.04};

\node[text=drawColor,anchor=base,inner sep=0pt, outer sep=0pt, scale=  0.64] at (135.35, 15.40) {0.05};
\end{scope}
\begin{scope}
\path[clip] (  0.00,  0.00) rectangle (144.54,144.54);
\definecolor{drawColor}{RGB}{0,0,0}

\node[text=drawColor,anchor=base,inner sep=0pt, outer sep=0pt, scale=  0.80] at ( 80.40,  5.94) {$\tilde{p}$};
\end{scope}
\begin{scope}
\path[clip] (  0.00,  0.00) rectangle (144.54,144.54);
\definecolor{drawColor}{RGB}{0,0,0}

\node[text=drawColor,rotate= 90.00,anchor=base,inner sep=0pt, outer sep=0pt, scale=  0.80] at (  9.51, 76.25) {Excess Certainty};
\end{scope}
\begin{scope}
\path[clip] (  0.00,  0.00) rectangle (144.54,144.54);
\definecolor{drawColor}{RGB}{0,0,0}

\node[text=drawColor,anchor=base,inner sep=0pt, outer sep=0pt, scale=  0.80] at ( 80.40,135.03) {$\gamma^* =$ 0.001};
\end{scope}
\end{tikzpicture}

%% file: Figures/ec_plt2.tex
\begin{tikzpicture}[x=1pt,y=1pt]
\definecolor{fillColor}{RGB}{255,255,255}
\path[use as bounding box,fill=fillColor,fill opacity=0.00] (0,0) rectangle (144.54,144.54);
\begin{scope}
\path[clip] (  0.00,  0.00) rectangle (144.54,144.54);
\definecolor{drawColor}{RGB}{255,255,255}
\definecolor{fillColor}{RGB}{255,255,255}

\path[draw=drawColor,line width= 0.4pt,line join=round,line cap=round,fill=fillColor] (  0.00, -0.00) rectangle (144.54,144.54);
\end{scope}
\begin{scope}
\path[clip] ( 23.45, 23.41) rectangle (140.54,129.09);
\definecolor{fillColor}{RGB}{255,255,255}

\path[fill=fillColor] ( 23.45, 23.41) rectangle (140.54,129.09);
\definecolor{drawColor}{RGB}{213,94,0}

\path[draw=drawColor,line width= 0.6pt,dash pattern=on 7pt off 3pt ,line join=round] ( 28.77,124.28) --
	( 29.31, 81.95) --
	( 29.84, 66.51) --
	( 30.38, 58.51) --
	( 30.91, 54.67) --
	( 31.45, 50.73) --
	( 31.98, 48.44) --
	( 32.52, 46.43) --
	( 33.05, 44.69) --
	( 33.59, 43.48) --
	( 34.12, 41.99) --
	( 34.66, 41.52) --
	( 35.19, 40.90) --
	( 35.73, 40.13) --
	( 36.26, 39.45) --
	( 36.80, 39.11) --
	( 37.33, 38.17) --
	( 37.87, 38.00) --
	( 38.40, 37.58) --
	( 38.94, 37.26) --
	( 39.47, 37.17) --
	( 40.01, 36.47) --
	( 40.54, 36.30) --
	( 41.08, 36.28) --
	( 41.61, 36.13) --
	( 42.15, 35.73) --
	( 42.68, 35.63) --
	( 43.22, 35.52) --
	( 43.75, 35.10) --
	( 44.29, 34.93) --
	( 44.82, 34.96) --
	( 45.36, 34.58) --
	( 45.89, 34.66) --
	( 46.43, 34.26) --
	( 46.96, 34.12) --
	( 47.50, 34.21) --
	( 48.03, 33.96) --
	( 48.57, 34.11) --
	( 49.10, 33.70) --
	( 49.64, 33.92) --
	( 50.17, 33.59) --
	( 50.70, 33.48) --
	( 51.24, 33.54) --
	( 51.77, 33.36) --
	( 52.31, 33.28) --
	( 52.84, 33.28) --
	( 53.38, 33.22) --
	( 53.91, 33.13) --
	( 54.45, 33.08) --
	( 54.98, 32.98) --
	( 55.52, 33.01) --
	( 56.05, 32.93) --
	( 56.59, 32.76) --
	( 57.12, 32.71) --
	( 57.66, 32.66) --
	( 58.19, 32.60) --
	( 58.73, 32.67) --
	( 59.26, 32.63) --
	( 59.80, 32.53) --
	( 60.33, 32.34) --
	( 60.87, 32.29) --
	( 61.40, 32.37) --
	( 61.94, 32.30) --
	( 62.47, 32.19) --
	( 63.01, 32.19) --
	( 63.54, 32.25) --
	( 64.08, 32.03) --
	( 64.61, 32.20) --
	( 65.15, 31.96) --
	( 65.68, 31.83) --
	( 66.22, 32.08) --
	( 66.75, 31.99) --
	( 67.29, 31.92) --
	( 67.82, 31.87) --
	( 68.36, 31.93) --
	( 68.89, 31.84) --
	( 69.43, 31.83) --
	( 69.96, 31.83) --
	( 70.50, 31.83) --
	( 71.03, 31.78) --
	( 71.57, 31.63) --
	( 72.10, 31.68) --
	( 72.64, 31.67) --
	( 73.17, 31.65) --
	( 73.71, 31.71) --
	( 74.24, 31.60) --
	( 74.78, 31.72) --
	( 75.31, 31.49) --
	( 75.84, 31.56) --
	( 76.38, 31.48) --
	( 76.91, 31.45) --
	( 77.45, 31.40) --
	( 77.98, 31.30) --
	( 78.52, 31.36) --
	( 79.05, 31.44) --
	( 79.59, 31.42) --
	( 80.12, 31.38) --
	( 80.66, 31.36) --
	( 81.19, 31.38) --
	( 81.73, 31.31) --
	( 82.26, 31.29) --
	( 82.80, 31.23) --
	( 83.33, 31.29) --
	( 83.87, 31.38) --
	( 84.40, 31.27) --
	( 84.94, 31.24) --
	( 85.47, 31.14) --
	( 86.01, 31.10) --
	( 86.54, 31.23) --
	( 87.08, 31.20) --
	( 87.61, 31.14) --
	( 88.15, 31.16) --
	( 88.68, 31.16) --
	( 89.22, 31.15) --
	( 89.75, 31.02) --
	( 90.29, 31.00) --
	( 90.82, 30.97) --
	( 91.36, 30.93) --
	( 91.89, 31.03) --
	( 92.43, 30.95) --
	( 92.96, 30.95) --
	( 93.50, 30.93) --
	( 94.03, 30.96) --
	( 94.57, 30.95) --
	( 95.10, 30.96) --
	( 95.64, 30.93) --
	( 96.17, 30.80) --
	( 96.71, 30.87) --
	( 97.24, 30.92) --
	( 97.78, 30.90) --
	( 98.31, 30.83) --
	( 98.85, 30.82) --
	( 99.38, 30.83) --
	( 99.91, 30.84) --
	(100.45, 30.80) --
	(100.98, 30.73) --
	(101.52, 30.81) --
	(102.05, 30.77) --
	(102.59, 30.79) --
	(103.12, 30.71) --
	(103.66, 30.75) --
	(104.19, 30.75) --
	(104.73, 30.71) --
	(105.26, 30.76) --
	(105.80, 30.80) --
	(106.33, 30.76) --
	(106.87, 30.61) --
	(107.40, 30.66) --
	(107.94, 30.69) --
	(108.47, 30.66) --
	(109.01, 30.69) --
	(109.54, 30.68) --
	(110.08, 30.62) --
	(110.61, 30.60) --
	(111.15, 30.58) --
	(111.68, 30.61) --
	(112.22, 30.58) --
	(112.75, 30.66) --
	(113.29, 30.50) --
	(113.82, 30.52) --
	(114.36, 30.51) --
	(114.89, 30.55) --
	(115.43, 30.51) --
	(115.96, 30.49) --
	(116.50, 30.53) --
	(117.03, 30.58) --
	(117.57, 30.51) --
	(118.10, 30.45) --
	(118.64, 30.47) --
	(119.17, 30.53) --
	(119.71, 30.50) --
	(120.24, 30.44) --
	(120.78, 30.43) --
	(121.31, 30.46) --
	(121.85, 30.48) --
	(122.38, 30.42) --
	(122.92, 30.41) --
	(123.45, 30.42) --
	(123.99, 30.39) --
	(124.52, 30.41) --
	(125.05, 30.38) --
	(125.59, 30.42) --
	(126.12, 30.38) --
	(126.66, 30.29) --
	(127.19, 30.33) --
	(127.73, 30.31) --
	(128.26, 30.30) --
	(128.80, 30.29) --
	(129.33, 30.32) --
	(129.87, 30.35) --
	(130.40, 30.30) --
	(130.94, 30.29) --
	(131.47, 30.28) --
	(132.01, 30.30) --
	(132.54, 30.30) --
	(133.08, 30.24) --
	(133.61, 30.20) --
	(134.15, 30.22) --
	(134.68, 30.20) --
	(135.22, 30.19);
\definecolor{drawColor}{RGB}{0,0,0}

\path[draw=drawColor,line width= 0.6pt,line join=round] ( 23.45, 29.60) -- (140.54, 29.60);
\definecolor{drawColor}{RGB}{0,158,115}

\path[draw=drawColor,line width= 0.6pt,line join=round] ( 28.77, 28.21) --
	( 29.31, 28.24) --
	( 29.84, 28.23) --
	( 30.38, 28.30) --
	( 30.91, 28.27) --
	( 31.45, 28.33) --
	( 31.98, 28.31) --
	( 32.52, 28.36) --
	( 33.05, 28.34) --
	( 33.59, 28.42) --
	( 34.12, 28.39) --
	( 34.66, 28.40) --
	( 35.19, 28.38) --
	( 35.73, 28.42) --
	( 36.26, 28.40) --
	( 36.80, 28.44) --
	( 37.33, 28.41) --
	( 37.87, 28.40) --
	( 38.40, 28.43) --
	( 38.94, 28.46) --
	( 39.47, 28.43) --
	( 40.01, 28.45) --
	( 40.54, 28.48) --
	( 41.08, 28.45) --
	( 41.61, 28.47) --
	( 42.15, 28.48) --
	( 42.68, 28.50) --
	( 43.22, 28.51) --
	( 43.75, 28.50) --
	( 44.29, 28.51) --
	( 44.82, 28.54) --
	( 45.36, 28.55) --
	( 45.89, 28.49) --
	( 46.43, 28.51) --
	( 46.96, 28.56) --
	( 47.50, 28.56) --
	( 48.03, 28.55) --
	( 48.57, 28.58) --
	( 49.10, 28.58) --
	( 49.64, 28.59) --
	( 50.17, 28.60) --
	( 50.70, 28.62) --
	( 51.24, 28.61) --
	( 51.77, 28.60) --
	( 52.31, 28.63) --
	( 52.84, 28.64) --
	( 53.38, 28.59) --
	( 53.91, 28.63) --
	( 54.45, 28.63) --
	( 54.98, 28.61) --
	( 55.52, 28.67) --
	( 56.05, 28.63) --
	( 56.59, 28.64) --
	( 57.12, 28.67) --
	( 57.66, 28.65) --
	( 58.19, 28.67) --
	( 58.73, 28.65) --
	( 59.26, 28.69) --
	( 59.80, 28.70) --
	( 60.33, 28.65) --
	( 60.87, 28.67) --
	( 61.40, 28.69) --
	( 61.94, 28.71) --
	( 62.47, 28.68) --
	( 63.01, 28.70) --
	( 63.54, 28.71) --
	( 64.08, 28.70) --
	( 64.61, 28.76) --
	( 65.15, 28.72) --
	( 65.68, 28.69) --
	( 66.22, 28.73) --
	( 66.75, 28.76) --
	( 67.29, 28.74) --
	( 67.82, 28.79) --
	( 68.36, 28.75) --
	( 68.89, 28.76) --
	( 69.43, 28.76) --
	( 69.96, 28.78) --
	( 70.50, 28.76) --
	( 71.03, 28.76) --
	( 71.57, 28.74) --
	( 72.10, 28.78) --
	( 72.64, 28.76) --
	( 73.17, 28.76) --
	( 73.71, 28.81) --
	( 74.24, 28.81) --
	( 74.78, 28.77) --
	( 75.31, 28.82) --
	( 75.84, 28.81) --
	( 76.38, 28.77) --
	( 76.91, 28.78) --
	( 77.45, 28.78) --
	( 77.98, 28.80) --
	( 78.52, 28.81) --
	( 79.05, 28.80) --
	( 79.59, 28.83) --
	( 80.12, 28.87) --
	( 80.66, 28.86) --
	( 81.19, 28.85) --
	( 81.73, 28.85) --
	( 82.26, 28.85) --
	( 82.80, 28.84) --
	( 83.33, 28.86) --
	( 83.87, 28.87) --
	( 84.40, 28.86) --
	( 84.94, 28.87) --
	( 85.47, 28.82) --
	( 86.01, 28.88) --
	( 86.54, 28.87) --
	( 87.08, 28.90) --
	( 87.61, 28.88) --
	( 88.15, 28.89) --
	( 88.68, 28.87) --
	( 89.22, 28.91) --
	( 89.75, 28.90) --
	( 90.29, 28.91) --
	( 90.82, 28.89) --
	( 91.36, 28.92) --
	( 91.89, 28.90) --
	( 92.43, 28.93) --
	( 92.96, 28.91) --
	( 93.50, 28.90) --
	( 94.03, 28.90) --
	( 94.57, 28.89) --
	( 95.10, 28.93) --
	( 95.64, 28.93) --
	( 96.17, 28.93) --
	( 96.71, 28.93) --
	( 97.24, 28.94) --
	( 97.78, 28.94) --
	( 98.31, 28.93) --
	( 98.85, 28.93) --
	( 99.38, 28.94) --
	( 99.91, 28.92) --
	(100.45, 28.96) --
	(100.98, 28.94) --
	(101.52, 28.96) --
	(102.05, 28.97) --
	(102.59, 28.97) --
	(103.12, 28.96) --
	(103.66, 28.97) --
	(104.19, 28.95) --
	(104.73, 29.00) --
	(105.26, 29.00) --
	(105.80, 28.97) --
	(106.33, 28.99) --
	(106.87, 28.97) --
	(107.40, 28.97) --
	(107.94, 28.99) --
	(108.47, 28.98) --
	(109.01, 29.00) --
	(109.54, 29.02) --
	(110.08, 28.99) --
	(110.61, 29.01) --
	(111.15, 29.03) --
	(111.68, 29.01) --
	(112.22, 29.01) --
	(112.75, 29.04) --
	(113.29, 29.01) --
	(113.82, 28.99) --
	(114.36, 29.04) --
	(114.89, 29.01) --
	(115.43, 29.05) --
	(115.96, 29.03) --
	(116.50, 29.05) --
	(117.03, 29.04) --
	(117.57, 29.05) --
	(118.10, 29.04) --
	(118.64, 29.06) --
	(119.17, 29.07) --
	(119.71, 29.08) --
	(120.24, 29.05) --
	(120.78, 29.07) --
	(121.31, 29.09) --
	(121.85, 29.07) --
	(122.38, 29.06) --
	(122.92, 29.05) --
	(123.45, 29.09) --
	(123.99, 29.08) --
	(124.52, 29.08) --
	(125.05, 29.11) --
	(125.59, 29.10) --
	(126.12, 29.07) --
	(126.66, 29.08) --
	(127.19, 29.10) --
	(127.73, 29.09) --
	(128.26, 29.08) --
	(128.80, 29.10) --
	(129.33, 29.13) --
	(129.87, 29.11) --
	(130.40, 29.09) --
	(130.94, 29.12) --
	(131.47, 29.13) --
	(132.01, 29.13) --
	(132.54, 29.13) --
	(133.08, 29.11) --
	(133.61, 29.13) --
	(134.15, 29.13) --
	(134.68, 29.13) --
	(135.22, 29.12);
\end{scope}
\begin{scope}
\path[clip] (  0.00,  0.00) rectangle (144.54,144.54);
\definecolor{drawColor}{RGB}{0,0,0}

\path[draw=drawColor,line width= 0.4pt,line join=round] ( 23.45, 23.41) --
	( 23.45,129.09);
\end{scope}
\begin{scope}
\path[clip] (  0.00,  0.00) rectangle (144.54,144.54);
\definecolor{drawColor}{gray}{0.30}

\node[text=drawColor,anchor=base east,inner sep=0pt, outer sep=0pt, scale=  0.64] at ( 19.85, 27.40) {0};

\node[text=drawColor,anchor=base east,inner sep=0pt, outer sep=0pt, scale=  0.64] at ( 19.85, 49.96) {5};

\node[text=drawColor,anchor=base east,inner sep=0pt, outer sep=0pt, scale=  0.64] at ( 19.85, 72.53) {10};

\node[text=drawColor,anchor=base east,inner sep=0pt, outer sep=0pt, scale=  0.64] at ( 19.85, 95.09) {15};

\node[text=drawColor,anchor=base east,inner sep=0pt, outer sep=0pt, scale=  0.64] at ( 19.85,117.66) {20};
\end{scope}
\begin{scope}
\path[clip] (  0.00,  0.00) rectangle (144.54,144.54);
\definecolor{drawColor}{gray}{0.20}

\path[draw=drawColor,line width= 0.4pt,line join=round] ( 21.45, 29.60) --
	( 23.45, 29.60);

\path[draw=drawColor,line width= 0.4pt,line join=round] ( 21.45, 52.16) --
	( 23.45, 52.16);

\path[draw=drawColor,line width= 0.4pt,line join=round] ( 21.45, 74.73) --
	( 23.45, 74.73);

\path[draw=drawColor,line width= 0.4pt,line join=round] ( 21.45, 97.29) --
	( 23.45, 97.29);

\path[draw=drawColor,line width= 0.4pt,line join=round] ( 21.45,119.86) --
	( 23.45,119.86);
\end{scope}
\begin{scope}
\path[clip] (  0.00,  0.00) rectangle (144.54,144.54);
\definecolor{drawColor}{RGB}{0,0,0}

\path[draw=drawColor,line width= 0.4pt,line join=round] ( 23.45, 23.41) --
	(140.54, 23.41);
\end{scope}
\begin{scope}
\path[clip] (  0.00,  0.00) rectangle (144.54,144.54);
\definecolor{drawColor}{gray}{0.20}

\path[draw=drawColor,line width= 0.4pt,line join=round] ( 28.29, 21.41) --
	( 28.29, 23.41);

\path[draw=drawColor,line width= 0.4pt,line join=round] ( 49.73, 21.41) --
	( 49.73, 23.41);

\path[draw=drawColor,line width= 0.4pt,line join=round] ( 71.17, 21.41) --
	( 71.17, 23.41);

\path[draw=drawColor,line width= 0.4pt,line join=round] ( 92.61, 21.41) --
	( 92.61, 23.41);

\path[draw=drawColor,line width= 0.4pt,line join=round] (114.05, 21.41) --
	(114.05, 23.41);

\path[draw=drawColor,line width= 0.4pt,line join=round] (135.49, 21.41) --
	(135.49, 23.41);
\end{scope}
\begin{scope}
\path[clip] (  0.00,  0.00) rectangle (144.54,144.54);
\definecolor{drawColor}{gray}{0.30}

\node[text=drawColor,anchor=base,inner sep=0pt, outer sep=0pt, scale=  0.64] at ( 28.29, 15.40) {0.00};

\node[text=drawColor,anchor=base,inner sep=0pt, outer sep=0pt, scale=  0.64] at ( 49.73, 15.40) {0.01};

\node[text=drawColor,anchor=base,inner sep=0pt, outer sep=0pt, scale=  0.64] at ( 71.17, 15.40) {0.02};

\node[text=drawColor,anchor=base,inner sep=0pt, outer sep=0pt, scale=  0.64] at ( 92.61, 15.40) {0.03};

\node[text=drawColor,anchor=base,inner sep=0pt, outer sep=0pt, scale=  0.64] at (114.05, 15.40) {0.04};

\node[text=drawColor,anchor=base,inner sep=0pt, outer sep=0pt, scale=  0.64] at (135.49, 15.40) {0.05};
\end{scope}
\begin{scope}
\path[clip] (  0.00,  0.00) rectangle (144.54,144.54);
\definecolor{drawColor}{RGB}{0,0,0}

\node[text=drawColor,anchor=base,inner sep=0pt, outer sep=0pt, scale=  0.80] at ( 82.00,  5.94) {$\tilde{p}$};
\end{scope}
\begin{scope}
\path[clip] (  0.00,  0.00) rectangle (144.54,144.54);
\definecolor{drawColor}{RGB}{0,0,0}

\node[text=drawColor,rotate= 90.00,anchor=base,inner sep=0pt, outer sep=0pt, scale=  0.80] at (  9.51, 76.25) {Excess Certainty};
\end{scope}
\begin{scope}
\path[clip] (  0.00,  0.00) rectangle (144.54,144.54);
\definecolor{drawColor}{RGB}{0,0,0}

\node[text=drawColor,anchor=base,inner sep=0pt, outer sep=0pt, scale=  0.80] at ( 82.00,135.03) {$\gamma^* =$ 0.01};
\end{scope}
\end{tikzpicture}

%% file: Figures/ec_plt3.tex
\begin{tikzpicture}[x=1pt,y=1pt]
\definecolor{fillColor}{RGB}{255,255,255}
\path[use as bounding box,fill=fillColor,fill opacity=0.00] (0,0) rectangle (144.54,144.54);
\begin{scope}
\path[clip] (  0.00,  0.00) rectangle (144.54,144.54);
\definecolor{drawColor}{RGB}{255,255,255}
\definecolor{fillColor}{RGB}{255,255,255}

\path[draw=drawColor,line width= 0.4pt,line join=round,line cap=round,fill=fillColor] (  0.00, -0.00) rectangle (144.54,144.54);
\end{scope}
\begin{scope}
\path[clip] ( 23.45, 23.41) rectangle (140.54,129.09);
\definecolor{fillColor}{RGB}{255,255,255}

\path[fill=fillColor] ( 23.45, 23.41) rectangle (140.54,129.09);
\definecolor{drawColor}{RGB}{213,94,0}

\path[draw=drawColor,line width= 0.6pt,dash pattern=on 7pt off 3pt ,line join=round] ( 28.77,124.28) --
	( 29.31, 79.80) --
	( 29.84, 64.69) --
	( 30.38, 57.14) --
	( 30.91, 52.52) --
	( 31.45, 48.76) --
	( 31.98, 46.39) --
	( 32.52, 44.57) --
	( 33.05, 42.84) --
	( 33.59, 41.90) --
	( 34.12, 40.63) --
	( 34.66, 39.85) --
	( 35.19, 39.13) --
	( 35.73, 38.42) --
	( 36.26, 38.19) --
	( 36.80, 37.56) --
	( 37.33, 37.00) --
	( 37.87, 36.69) --
	( 38.40, 36.20) --
	( 38.94, 35.90) --
	( 39.47, 35.58) --
	( 40.01, 35.25) --
	( 40.54, 35.04) --
	( 41.08, 34.75) --
	( 41.61, 34.58) --
	( 42.15, 34.46) --
	( 42.68, 34.12) --
	( 43.22, 33.96) --
	( 43.75, 33.83) --
	( 44.29, 33.72) --
	( 44.82, 33.53) --
	( 45.36, 33.30) --
	( 45.89, 33.35) --
	( 46.43, 33.17) --
	( 46.96, 32.96) --
	( 47.50, 32.91) --
	( 48.03, 32.68) --
	( 48.57, 32.64) --
	( 49.10, 32.52) --
	( 49.64, 32.50) --
	( 50.17, 32.28) --
	( 50.70, 32.24) --
	( 51.24, 32.12) --
	( 51.77, 32.12) --
	( 52.31, 32.01) --
	( 52.84, 32.02) --
	( 53.38, 31.91) --
	( 53.91, 31.79) --
	( 54.45, 31.73) --
	( 54.98, 31.63) --
	( 55.52, 31.64) --
	( 56.05, 31.56) --
	( 56.59, 31.42) --
	( 57.12, 31.45) --
	( 57.66, 31.45) --
	( 58.19, 31.30) --
	( 58.73, 31.27) --
	( 59.26, 31.23) --
	( 59.80, 31.16) --
	( 60.33, 31.13) --
	( 60.87, 31.06) --
	( 61.40, 30.99) --
	( 61.94, 30.98) --
	( 62.47, 30.95) --
	( 63.01, 30.91) --
	( 63.54, 30.87) --
	( 64.08, 30.83) --
	( 64.61, 30.80) --
	( 65.15, 30.74) --
	( 65.68, 30.70) --
	( 66.22, 30.67) --
	( 66.75, 30.63) --
	( 67.29, 30.63) --
	( 67.82, 30.56) --
	( 68.36, 30.53) --
	( 68.89, 30.56) --
	( 69.43, 30.42) --
	( 69.96, 30.39) --
	( 70.50, 30.40) --
	( 71.03, 30.42) --
	( 71.57, 30.40) --
	( 72.10, 30.36) --
	( 72.64, 30.33) --
	( 73.17, 30.26) --
	( 73.71, 30.26) --
	( 74.24, 30.21) --
	( 74.78, 30.24) --
	( 75.31, 30.20) --
	( 75.84, 30.16) --
	( 76.38, 30.12) --
	( 76.91, 30.08) --
	( 77.45, 30.08) --
	( 77.98, 30.05) --
	( 78.52, 30.03) --
	( 79.05, 29.96) --
	( 79.59, 29.99) --
	( 80.12, 29.93) --
	( 80.66, 29.95) --
	( 81.19, 29.92) --
	( 81.73, 29.95) --
	( 82.26, 29.93) --
	( 82.80, 29.86) --
	( 83.33, 29.84) --
	( 83.87, 29.84) --
	( 84.40, 29.81) --
	( 84.94, 29.83) --
	( 85.47, 29.75) --
	( 86.01, 29.75) --
	( 86.54, 29.72) --
	( 87.08, 29.73) --
	( 87.61, 29.73) --
	( 88.15, 29.68) --
	( 88.68, 29.70) --
	( 89.22, 29.64) --
	( 89.75, 29.68) --
	( 90.29, 29.61) --
	( 90.82, 29.65) --
	( 91.36, 29.60) --
	( 91.89, 29.57) --
	( 92.43, 29.54) --
	( 92.96, 29.62) --
	( 93.50, 29.54) --
	( 94.03, 29.56) --
	( 94.57, 29.55) --
	( 95.10, 29.51) --
	( 95.64, 29.50) --
	( 96.17, 29.47) --
	( 96.71, 29.48) --
	( 97.24, 29.45) --
	( 97.78, 29.45) --
	( 98.31, 29.40) --
	( 98.85, 29.39) --
	( 99.38, 29.37) --
	( 99.91, 29.41) --
	(100.45, 29.37) --
	(100.98, 29.36) --
	(101.52, 29.38) --
	(102.05, 29.36) --
	(102.59, 29.34) --
	(103.12, 29.32) --
	(103.66, 29.31) --
	(104.19, 29.31) --
	(104.73, 29.29) --
	(105.26, 29.30) --
	(105.80, 29.32) --
	(106.33, 29.29) --
	(106.87, 29.29) --
	(107.40, 29.24) --
	(107.94, 29.22) --
	(108.47, 29.21) --
	(109.01, 29.24) --
	(109.54, 29.20) --
	(110.08, 29.18) --
	(110.61, 29.16) --
	(111.15, 29.19) --
	(111.68, 29.17) --
	(112.22, 29.18) --
	(112.75, 29.12) --
	(113.29, 29.14) --
	(113.82, 29.14) --
	(114.36, 29.11) --
	(114.89, 29.13) --
	(115.43, 29.12) --
	(115.96, 29.13) --
	(116.50, 29.08) --
	(117.03, 29.08) --
	(117.57, 29.10) --
	(118.10, 29.06) --
	(118.64, 29.07) --
	(119.17, 29.08) --
	(119.71, 29.04) --
	(120.24, 29.01) --
	(120.78, 29.05) --
	(121.31, 29.05) --
	(121.85, 29.00) --
	(122.38, 29.02) --
	(122.92, 29.00) --
	(123.45, 29.02) --
	(123.99, 29.01) --
	(124.52, 29.00) --
	(125.05, 28.98) --
	(125.59, 28.94) --
	(126.12, 28.94) --
	(126.66, 28.96) --
	(127.19, 28.94) --
	(127.73, 28.95) --
	(128.26, 28.92) --
	(128.80, 28.94) --
	(129.33, 28.94) --
	(129.87, 28.89) --
	(130.40, 28.90) --
	(130.94, 28.91) --
	(131.47, 28.92) --
	(132.01, 28.89) --
	(132.54, 28.90) --
	(133.08, 28.90) --
	(133.61, 28.87) --
	(134.15, 28.89) --
	(134.68, 28.83) --
	(135.22, 28.87);
\definecolor{drawColor}{RGB}{0,0,0}

\path[draw=drawColor,line width= 0.6pt,line join=round] ( 23.45, 28.59) -- (140.54, 28.59);
\definecolor{drawColor}{RGB}{0,158,115}

\path[draw=drawColor,line width= 0.6pt,line join=round] ( 28.77, 28.21) --
	( 29.31, 28.21) --
	( 29.84, 28.22) --
	( 30.38, 28.23) --
	( 30.91, 28.23) --
	( 31.45, 28.23) --
	( 31.98, 28.23) --
	( 32.52, 28.24) --
	( 33.05, 28.24) --
	( 33.59, 28.25) --
	( 34.12, 28.24) --
	( 34.66, 28.25) --
	( 35.19, 28.25) --
	( 35.73, 28.25) --
	( 36.26, 28.27) --
	( 36.80, 28.26) --
	( 37.33, 28.27) --
	( 37.87, 28.26) --
	( 38.40, 28.27) --
	( 38.94, 28.26) --
	( 39.47, 28.28) --
	( 40.01, 28.28) --
	( 40.54, 28.28) --
	( 41.08, 28.28) --
	( 41.61, 28.28) --
	( 42.15, 28.29) --
	( 42.68, 28.29) --
	( 43.22, 28.29) --
	( 43.75, 28.29) --
	( 44.29, 28.30) --
	( 44.82, 28.30) --
	( 45.36, 28.29) --
	( 45.89, 28.30) --
	( 46.43, 28.30) --
	( 46.96, 28.30) --
	( 47.50, 28.31) --
	( 48.03, 28.31) --
	( 48.57, 28.30) --
	( 49.10, 28.30) --
	( 49.64, 28.30) --
	( 50.17, 28.31) --
	( 50.70, 28.31) --
	( 51.24, 28.31) --
	( 51.77, 28.32) --
	( 52.31, 28.31) --
	( 52.84, 28.32) --
	( 53.38, 28.32) --
	( 53.91, 28.33) --
	( 54.45, 28.33) --
	( 54.98, 28.32) --
	( 55.52, 28.34) --
	( 56.05, 28.33) --
	( 56.59, 28.32) --
	( 57.12, 28.34) --
	( 57.66, 28.34) --
	( 58.19, 28.33) --
	( 58.73, 28.33) --
	( 59.26, 28.34) --
	( 59.80, 28.33) --
	( 60.33, 28.34) --
	( 60.87, 28.34) --
	( 61.40, 28.34) --
	( 61.94, 28.35) --
	( 62.47, 28.34) --
	( 63.01, 28.35) --
	( 63.54, 28.35) --
	( 64.08, 28.34) --
	( 64.61, 28.34) --
	( 65.15, 28.35) --
	( 65.68, 28.35) --
	( 66.22, 28.35) --
	( 66.75, 28.36) --
	( 67.29, 28.35) --
	( 67.82, 28.35) --
	( 68.36, 28.36) --
	( 68.89, 28.36) --
	( 69.43, 28.36) --
	( 69.96, 28.35) --
	( 70.50, 28.36) --
	( 71.03, 28.36) --
	( 71.57, 28.37) --
	( 72.10, 28.37) --
	( 72.64, 28.36) --
	( 73.17, 28.35) --
	( 73.71, 28.36) --
	( 74.24, 28.37) --
	( 74.78, 28.37) --
	( 75.31, 28.38) --
	( 75.84, 28.37) --
	( 76.38, 28.37) --
	( 76.91, 28.36) --
	( 77.45, 28.37) --
	( 77.98, 28.37) --
	( 78.52, 28.37) --
	( 79.05, 28.37) --
	( 79.59, 28.37) --
	( 80.12, 28.36) --
	( 80.66, 28.38) --
	( 81.19, 28.38) --
	( 81.73, 28.38) --
	( 82.26, 28.38) --
	( 82.80, 28.37) --
	( 83.33, 28.37) --
	( 83.87, 28.38) --
	( 84.40, 28.39) --
	( 84.94, 28.38) --
	( 85.47, 28.37) --
	( 86.01, 28.38) --
	( 86.54, 28.38) --
	( 87.08, 28.38) --
	( 87.61, 28.39) --
	( 88.15, 28.38) --
	( 88.68, 28.38) --
	( 89.22, 28.38) --
	( 89.75, 28.40) --
	( 90.29, 28.38) --
	( 90.82, 28.39) --
	( 91.36, 28.39) --
	( 91.89, 28.38) --
	( 92.43, 28.38) --
	( 92.96, 28.40) --
	( 93.50, 28.40) --
	( 94.03, 28.40) --
	( 94.57, 28.39) --
	( 95.10, 28.39) --
	( 95.64, 28.40) --
	( 96.17, 28.39) --
	( 96.71, 28.40) --
	( 97.24, 28.39) --
	( 97.78, 28.40) --
	( 98.31, 28.39) --
	( 98.85, 28.39) --
	( 99.38, 28.39) --
	( 99.91, 28.40) --
	(100.45, 28.39) --
	(100.98, 28.40) --
	(101.52, 28.40) --
	(102.05, 28.40) --
	(102.59, 28.40) --
	(103.12, 28.40) --
	(103.66, 28.40) --
	(104.19, 28.39) --
	(104.73, 28.40) --
	(105.26, 28.40) --
	(105.80, 28.41) --
	(106.33, 28.41) --
	(106.87, 28.41) --
	(107.40, 28.40) --
	(107.94, 28.41) --
	(108.47, 28.40) --
	(109.01, 28.41) --
	(109.54, 28.40) --
	(110.08, 28.40) --
	(110.61, 28.40) --
	(111.15, 28.41) --
	(111.68, 28.41) --
	(112.22, 28.42) --
	(112.75, 28.40) --
	(113.29, 28.41) --
	(113.82, 28.41) --
	(114.36, 28.41) --
	(114.89, 28.42) --
	(115.43, 28.42) --
	(115.96, 28.41) --
	(116.50, 28.41) --
	(117.03, 28.42) --
	(117.57, 28.42) --
	(118.10, 28.42) --
	(118.64, 28.42) --
	(119.17, 28.42) --
	(119.71, 28.41) --
	(120.24, 28.41) --
	(120.78, 28.42) --
	(121.31, 28.43) --
	(121.85, 28.42) --
	(122.38, 28.42) --
	(122.92, 28.42) --
	(123.45, 28.42) --
	(123.99, 28.42) --
	(124.52, 28.43) --
	(125.05, 28.42) --
	(125.59, 28.42) --
	(126.12, 28.42) --
	(126.66, 28.43) --
	(127.19, 28.42) --
	(127.73, 28.43) --
	(128.26, 28.42) --
	(128.80, 28.43) --
	(129.33, 28.43) --
	(129.87, 28.41) --
	(130.40, 28.43) --
	(130.94, 28.43) --
	(131.47, 28.43) --
	(132.01, 28.44) --
	(132.54, 28.43) --
	(133.08, 28.44) --
	(133.61, 28.43) --
	(134.15, 28.43) --
	(134.68, 28.42) --
	(135.22, 28.44);
\end{scope}
\begin{scope}
\path[clip] (  0.00,  0.00) rectangle (144.54,144.54);
\definecolor{drawColor}{RGB}{0,0,0}

\path[draw=drawColor,line width= 0.4pt,line join=round] ( 23.45, 23.41) --
	( 23.45,129.09);
\end{scope}
\begin{scope}
\path[clip] (  0.00,  0.00) rectangle (144.54,144.54);
\definecolor{drawColor}{gray}{0.30}

\node[text=drawColor,anchor=base east,inner sep=0pt, outer sep=0pt, scale=  0.64] at ( 19.85, 26.39) {0};

\node[text=drawColor,anchor=base east,inner sep=0pt, outer sep=0pt, scale=  0.64] at ( 19.85, 50.67) {20};

\node[text=drawColor,anchor=base east,inner sep=0pt, outer sep=0pt, scale=  0.64] at ( 19.85, 74.95) {40};

\node[text=drawColor,anchor=base east,inner sep=0pt, outer sep=0pt, scale=  0.64] at ( 19.85, 99.23) {60};

\node[text=drawColor,anchor=base east,inner sep=0pt, outer sep=0pt, scale=  0.64] at ( 19.85,123.51) {80};
\end{scope}
\begin{scope}
\path[clip] (  0.00,  0.00) rectangle (144.54,144.54);
\definecolor{drawColor}{gray}{0.20}

\path[draw=drawColor,line width= 0.4pt,line join=round] ( 21.45, 28.59) --
	( 23.45, 28.59);

\path[draw=drawColor,line width= 0.4pt,line join=round] ( 21.45, 52.87) --
	( 23.45, 52.87);

\path[draw=drawColor,line width= 0.4pt,line join=round] ( 21.45, 77.16) --
	( 23.45, 77.16);

\path[draw=drawColor,line width= 0.4pt,line join=round] ( 21.45,101.44) --
	( 23.45,101.44);

\path[draw=drawColor,line width= 0.4pt,line join=round] ( 21.45,125.72) --
	( 23.45,125.72);
\end{scope}
\begin{scope}
\path[clip] (  0.00,  0.00) rectangle (144.54,144.54);
\definecolor{drawColor}{RGB}{0,0,0}

\path[draw=drawColor,line width= 0.4pt,line join=round] ( 23.45, 23.41) --
	(140.54, 23.41);
\end{scope}
\begin{scope}
\path[clip] (  0.00,  0.00) rectangle (144.54,144.54);
\definecolor{drawColor}{gray}{0.20}

\path[draw=drawColor,line width= 0.4pt,line join=round] ( 28.29, 21.41) --
	( 28.29, 23.41);

\path[draw=drawColor,line width= 0.4pt,line join=round] ( 49.73, 21.41) --
	( 49.73, 23.41);

\path[draw=drawColor,line width= 0.4pt,line join=round] ( 71.17, 21.41) --
	( 71.17, 23.41);

\path[draw=drawColor,line width= 0.4pt,line join=round] ( 92.61, 21.41) --
	( 92.61, 23.41);

\path[draw=drawColor,line width= 0.4pt,line join=round] (114.05, 21.41) --
	(114.05, 23.41);

\path[draw=drawColor,line width= 0.4pt,line join=round] (135.49, 21.41) --
	(135.49, 23.41);
\end{scope}
\begin{scope}
\path[clip] (  0.00,  0.00) rectangle (144.54,144.54);
\definecolor{drawColor}{gray}{0.30}

\node[text=drawColor,anchor=base,inner sep=0pt, outer sep=0pt, scale=  0.64] at ( 28.29, 15.40) {0.00};

\node[text=drawColor,anchor=base,inner sep=0pt, outer sep=0pt, scale=  0.64] at ( 49.73, 15.40) {0.01};

\node[text=drawColor,anchor=base,inner sep=0pt, outer sep=0pt, scale=  0.64] at ( 71.17, 15.40) {0.02};

\node[text=drawColor,anchor=base,inner sep=0pt, outer sep=0pt, scale=  0.64] at ( 92.61, 15.40) {0.03};

\node[text=drawColor,anchor=base,inner sep=0pt, outer sep=0pt, scale=  0.64] at (114.05, 15.40) {0.04};

\node[text=drawColor,anchor=base,inner sep=0pt, outer sep=0pt, scale=  0.64] at (135.49, 15.40) {0.05};
\end{scope}
\begin{scope}
\path[clip] (  0.00,  0.00) rectangle (144.54,144.54);
\definecolor{drawColor}{RGB}{0,0,0}

\node[text=drawColor,anchor=base,inner sep=0pt, outer sep=0pt, scale=  0.80] at ( 82.00,  5.94) {$\tilde{p}$};
\end{scope}
\begin{scope}
\path[clip] (  0.00,  0.00) rectangle (144.54,144.54);
\definecolor{drawColor}{RGB}{0,0,0}

\node[text=drawColor,rotate= 90.00,anchor=base,inner sep=0pt, outer sep=0pt, scale=  0.80] at (  9.51, 76.25) {Excess Certainty};
\end{scope}
\begin{scope}
\path[clip] (  0.00,  0.00) rectangle (144.54,144.54);
\definecolor{drawColor}{RGB}{0,0,0}

\node[text=drawColor,anchor=base,inner sep=0pt, outer sep=0pt, scale=  0.80] at ( 82.00,135.03) {$\gamma^* =$ 0.05};
\end{scope}
\end{tikzpicture}

%% file: Figures/p_expdiff.tex
\begin{tikzpicture}[x=1pt,y=1pt]
\definecolor{fillColor}{RGB}{255,255,255}
\path[use as bounding box,fill=fillColor,fill opacity=0.00] (0,0) rectangle (144.54,144.54);
\begin{scope}
\path[clip] (  0.00,  0.00) rectangle (144.54,144.54);
\definecolor{drawColor}{RGB}{255,255,255}
\definecolor{fillColor}{RGB}{255,255,255}

\path[draw=drawColor,line width= 0.3pt,line join=round,line cap=round,fill=fillColor] (  0.00,  0.00) rectangle (144.54,144.54);
\end{scope}
\begin{scope}
\path[clip] ( 17.19, 16.78) rectangle (142.04,142.04);
\definecolor{fillColor}{RGB}{255,255,255}

\path[fill=fillColor] ( 17.19, 16.78) rectangle (142.04,142.04);
\definecolor{drawColor}{RGB}{0,0,0}

\path[draw=drawColor,line width= 0.6pt,line join=round] ( 22.86, 22.47) --
	( 23.43, 23.04) --
	( 24.00, 23.61) --
	( 24.57, 24.19) --
	( 25.14, 24.76) --
	( 25.71, 25.33) --
	( 26.28, 25.90) --
	( 26.85, 26.47) --
	( 27.42, 27.05) --
	( 27.99, 27.62) --
	( 28.56, 28.19) --
	( 29.14, 28.76) --
	( 29.71, 29.34) --
	( 30.28, 29.91) --
	( 30.85, 30.48) --
	( 31.42, 31.05) --
	( 31.99, 31.62) --
	( 32.56, 32.20) --
	( 33.13, 32.77) --
	( 33.70, 33.34) --
	( 34.27, 33.91) --
	( 34.84, 34.49) --
	( 35.41, 35.06) --
	( 35.98, 35.63) --
	( 36.55, 36.20) --
	( 37.12, 36.78) --
	( 37.69, 37.35) --
	( 38.26, 37.92) --
	( 38.83, 38.49) --
	( 39.40, 39.06) --
	( 39.97, 39.64) --
	( 40.54, 40.21) --
	( 41.11, 40.78) --
	( 41.68, 41.35) --
	( 42.25, 41.93) --
	( 42.82, 42.50) --
	( 43.39, 43.07) --
	( 43.96, 43.64) --
	( 44.54, 44.21) --
	( 45.11, 44.79) --
	( 45.68, 45.36) --
	( 46.25, 45.93) --
	( 46.82, 46.50) --
	( 47.39, 47.08) --
	( 47.96, 47.65) --
	( 48.53, 48.22) --
	( 49.10, 48.79) --
	( 49.67, 49.36) --
	( 50.24, 49.94) --
	( 50.81, 50.51) --
	( 51.38, 51.08) --
	( 51.95, 51.65) --
	( 52.52, 52.23) --
	( 53.09, 52.80) --
	( 53.66, 53.37) --
	( 54.23, 53.94) --
	( 54.80, 54.51) --
	( 55.37, 55.09) --
	( 55.94, 55.66) --
	( 56.51, 56.23) --
	( 57.08, 56.80) --
	( 57.65, 57.38) --
	( 58.22, 57.95) --
	( 58.79, 58.52) --
	( 59.36, 59.09) --
	( 59.94, 59.67) --
	( 60.51, 60.24) --
	( 61.08, 60.81) --
	( 61.65, 61.38) --
	( 62.22, 61.95) --
	( 62.79, 62.53) --
	( 63.36, 63.10) --
	( 63.93, 63.67) --
	( 64.50, 64.24) --
	( 65.07, 64.82) --
	( 65.64, 65.39) --
	( 66.21, 65.96) --
	( 66.78, 66.53) --
	( 67.35, 67.10) --
	( 67.92, 67.68) --
	( 68.49, 68.25) --
	( 69.06, 68.82) --
	( 69.63, 69.39) --
	( 70.20, 69.97) --
	( 70.77, 70.54) --
	( 71.34, 71.11) --
	( 71.91, 71.68) --
	( 72.48, 72.25) --
	( 73.05, 72.83) --
	( 73.62, 73.40) --
	( 74.19, 73.97) --
	( 74.76, 74.54) --
	( 75.34, 75.12) --
	( 75.91, 75.69) --
	( 76.48, 76.26) --
	( 77.05, 76.83) --
	( 77.62, 77.40) --
	( 78.19, 77.98) --
	( 78.76, 78.55) --
	( 79.33, 79.12) --
	( 79.90, 79.69) --
	( 80.47, 80.27) --
	( 81.04, 80.84) --
	( 81.61, 81.41) --
	( 82.18, 81.98) --
	( 82.75, 82.55) --
	( 83.32, 83.13) --
	( 83.89, 83.70) --
	( 84.46, 84.27) --
	( 85.03, 84.84) --
	( 85.60, 85.42) --
	( 86.17, 85.99) --
	( 86.74, 86.56) --
	( 87.31, 87.13) --
	( 87.88, 87.71) --
	( 88.45, 88.28) --
	( 89.02, 88.85) --
	( 89.59, 89.42) --
	( 90.16, 89.99) --
	( 90.74, 90.57) --
	( 91.31, 91.14) --
	( 91.88, 91.71) --
	( 92.45, 92.28) --
	( 93.02, 92.86) --
	( 93.59, 93.43) --
	( 94.16, 94.00) --
	( 94.73, 94.57) --
	( 95.30, 95.14) --
	( 95.87, 95.72) --
	( 96.44, 96.29) --
	( 97.01, 96.86) --
	( 97.58, 97.43) --
	( 98.15, 98.01) --
	( 98.72, 98.58) --
	( 99.29, 99.15) --
	( 99.86, 99.72) --
	(100.43,100.29) --
	(101.00,100.87) --
	(101.57,101.44) --
	(102.14,102.01) --
	(102.71,102.58) --
	(103.28,103.16) --
	(103.85,103.73) --
	(104.42,104.30) --
	(104.99,104.87) --
	(105.56,105.44) --
	(106.14,106.02) --
	(106.71,106.59) --
	(107.28,107.16) --
	(107.85,107.73) --
	(108.42,108.31) --
	(108.99,108.88) --
	(109.56,109.45) --
	(110.13,110.02) --
	(110.70,110.60) --
	(111.27,111.17) --
	(111.84,111.74) --
	(112.41,112.31) --
	(112.98,112.88) --
	(113.55,113.46) --
	(114.12,114.03) --
	(114.69,114.60) --
	(115.26,115.17) --
	(115.83,115.75) --
	(116.40,116.32) --
	(116.97,116.89) --
	(117.54,117.46) --
	(118.11,118.03) --
	(118.68,118.61) --
	(119.25,119.18) --
	(119.82,119.75) --
	(120.39,120.32) --
	(120.96,120.90) --
	(121.54,121.47) --
	(122.11,122.04) --
	(122.68,122.61) --
	(123.25,123.18) --
	(123.82,123.76) --
	(124.39,124.33) --
	(124.96,124.90) --
	(125.53,125.47) --
	(126.10,126.05) --
	(126.67,126.62) --
	(127.24,127.19) --
	(127.81,127.76) --
	(128.38,128.33) --
	(128.95,128.91) --
	(129.52,129.48) --
	(130.09,130.05) --
	(130.66,130.62) --
	(131.23,131.20) --
	(131.80,131.77) --
	(132.37,132.34) --
	(132.94,132.91) --
	(133.51,133.48) --
	(134.08,134.06) --
	(134.65,134.63) --
	(135.22,135.20) --
	(135.79,135.77) --
	(136.36,136.35);
\definecolor{drawColor}{RGB}{230,159,0}

\path[draw=drawColor,line width= 0.6pt,dash pattern=on 7pt off 3pt ,line join=round] ( 22.86, 22.47) --
	( 22.87, 23.04) --
	( 22.90, 23.61) --
	( 22.94, 24.19) --
	( 23.00, 24.76) --
	( 23.07, 25.33) --
	( 23.16, 25.90) --
	( 23.27, 26.47) --
	( 23.40, 27.05) --
	( 23.54, 27.62) --
	( 23.69, 28.19) --
	( 23.86, 28.76) --
	( 24.05, 29.34) --
	( 24.25, 29.91) --
	( 24.47, 30.48) --
	( 24.70, 31.05) --
	( 24.94, 31.62) --
	( 25.20, 32.20) --
	( 25.48, 32.77) --
	( 25.77, 33.34) --
	( 26.07, 33.91) --
	( 26.39, 34.49) --
	( 26.72, 35.06) --
	( 27.06, 35.63) --
	( 27.42, 36.20) --
	( 27.79, 36.78) --
	( 28.17, 37.35) --
	( 28.56, 37.92) --
	( 28.97, 38.49) --
	( 29.39, 39.06) --
	( 29.82, 39.64) --
	( 30.27, 40.21) --
	( 30.72, 40.78) --
	( 31.19, 41.35) --
	( 31.67, 41.93) --
	( 32.16, 42.50) --
	( 32.66, 43.07) --
	( 33.17, 43.64) --
	( 33.70, 44.21) --
	( 34.23, 44.79) --
	( 34.78, 45.36) --
	( 35.33, 45.93) --
	( 35.89, 46.50) --
	( 36.47, 47.08) --
	( 37.05, 47.65) --
	( 37.65, 48.22) --
	( 38.25, 48.79) --
	( 38.86, 49.36) --
	( 39.49, 49.94) --
	( 40.12, 50.51) --
	( 40.76, 51.08) --
	( 41.40, 51.65) --
	( 42.06, 52.23) --
	( 42.73, 52.80) --
	( 43.40, 53.37) --
	( 44.08, 53.94) --
	( 44.77, 54.51) --
	( 45.46, 55.09) --
	( 46.17, 55.66) --
	( 46.88, 56.23) --
	( 47.59, 56.80) --
	( 48.32, 57.38) --
	( 49.05, 57.95) --
	( 49.79, 58.52) --
	( 50.53, 59.09) --
	( 51.28, 59.67) --
	( 52.03, 60.24) --
	( 52.80, 60.81) --
	( 53.56, 61.38) --
	( 54.34, 61.95) --
	( 55.11, 62.53) --
	( 55.90, 63.10) --
	( 56.68, 63.67) --
	( 57.48, 64.24) --
	( 58.27, 64.82) --
	( 59.08, 65.39) --
	( 59.88, 65.96) --
	( 60.69, 66.53) --
	( 61.50, 67.10) --
	( 62.32, 67.68) --
	( 63.14, 68.25) --
	( 63.97, 68.82) --
	( 64.80, 69.39) --
	( 65.63, 69.97) --
	( 66.46, 70.54) --
	( 67.30, 71.11) --
	( 68.13, 71.68) --
	( 68.97, 72.25) --
	( 69.82, 72.83) --
	( 70.66, 73.40) --
	( 71.51, 73.97) --
	( 72.36, 74.54) --
	( 73.21, 75.12) --
	( 74.06, 75.69) --
	( 74.91, 76.26) --
	( 75.77, 76.83) --
	( 76.62, 77.40) --
	( 77.47, 77.98) --
	( 78.33, 78.55) --
	( 79.19, 79.12) --
	( 80.04, 79.69) --
	( 80.90, 80.27) --
	( 81.75, 80.84) --
	( 82.61, 81.41) --
	( 83.46, 81.98) --
	( 84.31, 82.55) --
	( 85.17, 83.13) --
	( 86.02, 83.70) --
	( 86.87, 84.27) --
	( 87.72, 84.84) --
	( 88.56, 85.42) --
	( 89.41, 85.99) --
	( 90.25, 86.56) --
	( 91.09, 87.13) --
	( 91.93, 87.71) --
	( 92.77, 88.28) --
	( 93.60, 88.85) --
	( 94.43, 89.42) --
	( 95.26, 89.99) --
	( 96.08, 90.57) --
	( 96.90, 91.14) --
	( 97.72, 91.71) --
	( 98.53, 92.28) --
	( 99.34, 92.86) --
	(100.15, 93.43) --
	(100.95, 94.00) --
	(101.75, 94.57) --
	(102.54, 95.14) --
	(103.33, 95.72) --
	(104.11, 96.29) --
	(104.89, 96.86) --
	(105.66, 97.43) --
	(106.43, 98.01) --
	(107.19, 98.58) --
	(107.95, 99.15) --
	(108.70, 99.72) --
	(109.44,100.29) --
	(110.18,100.87) --
	(110.91,101.44) --
	(111.63,102.01) --
	(112.35,102.58) --
	(113.06,103.16) --
	(113.76,103.73) --
	(114.46,104.30) --
	(115.15,104.87) --
	(115.83,105.44) --
	(116.50,106.02) --
	(117.16,106.59) --
	(117.82,107.16) --
	(118.47,107.73) --
	(119.11,108.31) --
	(119.74,108.88) --
	(120.36,109.45) --
	(120.97,110.02) --
	(121.58,110.60) --
	(122.17,111.17) --
	(122.76,111.74) --
	(123.33,112.31) --
	(123.90,112.88) --
	(124.45,113.46) --
	(125.00,114.03) --
	(125.53,114.60) --
	(126.05,115.17) --
	(126.57,115.75) --
	(127.07,116.32) --
	(127.56,116.89) --
	(128.04,117.46) --
	(128.50,118.03) --
	(128.96,118.61) --
	(129.40,119.18) --
	(129.84,119.75) --
	(130.26,120.32) --
	(130.66,120.90) --
	(131.06,121.47) --
	(131.44,122.04) --
	(131.81,122.61) --
	(132.17,123.18) --
	(132.51,123.76) --
	(132.84,124.33) --
	(133.16,124.90) --
	(133.46,125.47) --
	(133.75,126.05) --
	(134.02,126.62) --
	(134.28,127.19) --
	(134.53,127.76) --
	(134.76,128.33) --
	(134.97,128.91) --
	(135.18,129.48) --
	(135.36,130.05) --
	(135.53,130.62) --
	(135.69,131.20) --
	(135.83,131.77) --
	(135.95,132.34) --
	(136.06,132.91) --
	(136.15,133.48) --
	(136.23,134.06) --
	(136.29,134.63) --
	(136.33,135.20) --
	(136.36,135.77) --
	(136.36,136.35);
\definecolor{drawColor}{RGB}{0,158,115}

\path[draw=drawColor,line width= 0.6pt,dash pattern=on 2pt off 2pt on 6pt off 2pt ,line join=round] ( 22.86, 22.47) --
	( 24.27, 23.04) --
	( 25.66, 23.61) --
	( 27.02, 24.19) --
	( 28.36, 24.76) --
	( 29.67, 25.33) --
	( 30.96, 25.90) --
	( 32.23, 26.47) --
	( 33.47, 27.05) --
	( 34.68, 27.62) --
	( 35.87, 28.19) --
	( 37.04, 28.76) --
	( 38.19, 29.34) --
	( 39.31, 29.91) --
	( 40.41, 30.48) --
	( 41.49, 31.05) --
	( 42.55, 31.62) --
	( 43.59, 32.20) --
	( 44.60, 32.77) --
	( 45.59, 33.34) --
	( 46.57, 33.91) --
	( 47.52, 34.49) --
	( 48.45, 35.06) --
	( 49.36, 35.63) --
	( 50.25, 36.20) --
	( 51.12, 36.78) --
	( 51.98, 37.35) --
	( 52.81, 37.92) --
	( 53.62, 38.49) --
	( 54.42, 39.06) --
	( 55.20, 39.64) --
	( 55.96, 40.21) --
	( 56.70, 40.78) --
	( 57.42, 41.35) --
	( 58.13, 41.93) --
	( 58.82, 42.50) --
	( 59.49, 43.07) --
	( 60.15, 43.64) --
	( 60.79, 44.21) --
	( 61.42, 44.79) --
	( 62.03, 45.36) --
	( 62.62, 45.93) --
	( 63.20, 46.50) --
	( 63.76, 47.08) --
	( 64.31, 47.65) --
	( 64.85, 48.22) --
	( 65.37, 48.79) --
	( 65.87, 49.36) --
	( 66.37, 49.94) --
	( 66.85, 50.51) --
	( 67.31, 51.08) --
	( 67.77, 51.65) --
	( 68.21, 52.23) --
	( 68.64, 52.80) --
	( 69.05, 53.37) --
	( 69.46, 53.94) --
	( 69.85, 54.51) --
	( 70.24, 55.09) --
	( 70.61, 55.66) --
	( 70.97, 56.23) --
	( 71.32, 56.80) --
	( 71.66, 57.38) --
	( 71.99, 57.95) --
	( 72.31, 58.52) --
	( 72.62, 59.09) --
	( 72.92, 59.67) --
	( 73.21, 60.24) --
	( 73.50, 60.81) --
	( 73.77, 61.38) --
	( 74.04, 61.95) --
	( 74.30, 62.53) --
	( 74.55, 63.10) --
	( 74.79, 63.67) --
	( 75.03, 64.24) --
	( 75.26, 64.82) --
	( 75.48, 65.39) --
	( 75.70, 65.96) --
	( 75.91, 66.53) --
	( 76.12, 67.10) --
	( 76.32, 67.68) --
	( 76.51, 68.25) --
	( 76.70, 68.82) --
	( 76.89, 69.39) --
	( 77.07, 69.97) --
	( 77.24, 70.54) --
	( 77.41, 71.11) --
	( 77.58, 71.68) --
	( 77.75, 72.25) --
	( 77.91, 72.83) --
	( 78.07, 73.40) --
	( 78.22, 73.97) --
	( 78.37, 74.54) --
	( 78.53, 75.12) --
	( 78.67, 75.69) --
	( 78.82, 76.26) --
	( 78.97, 76.83) --
	( 79.11, 77.40) --
	( 79.26, 77.98) --
	( 79.40, 78.55) --
	( 79.54, 79.12) --
	( 79.68, 79.69) --
	( 79.83, 80.27) --
	( 79.97, 80.84) --
	( 80.11, 81.41) --
	( 80.26, 81.98) --
	( 80.40, 82.55) --
	( 80.55, 83.13) --
	( 80.70, 83.70) --
	( 80.85, 84.27) --
	( 81.00, 84.84) --
	( 81.16, 85.42) --
	( 81.32, 85.99) --
	( 81.48, 86.56) --
	( 81.64, 87.13) --
	( 81.81, 87.71) --
	( 81.98, 88.28) --
	( 82.16, 88.85) --
	( 82.34, 89.42) --
	( 82.52, 89.99) --
	( 82.71, 90.57) --
	( 82.91, 91.14) --
	( 83.11, 91.71) --
	( 83.31, 92.28) --
	( 83.52, 92.86) --
	( 83.74, 93.43) --
	( 83.97, 94.00) --
	( 84.20, 94.57) --
	( 84.43, 95.14) --
	( 84.68, 95.72) --
	( 84.93, 96.29) --
	( 85.19, 96.86) --
	( 85.46, 97.43) --
	( 85.73, 98.01) --
	( 86.01, 98.58) --
	( 86.31, 99.15) --
	( 86.61, 99.72) --
	( 86.92,100.29) --
	( 87.24,100.87) --
	( 87.57,101.44) --
	( 87.91,102.01) --
	( 88.26,102.58) --
	( 88.62,103.16) --
	( 88.99,103.73) --
	( 89.37,104.30) --
	( 89.77,104.87) --
	( 90.17,105.44) --
	( 90.59,106.02) --
	( 91.02,106.59) --
	( 91.46,107.16) --
	( 91.91,107.73) --
	( 92.38,108.31) --
	( 92.86,108.88) --
	( 93.35,109.45) --
	( 93.86,110.02) --
	( 94.38,110.60) --
	( 94.91,111.17) --
	( 95.46,111.74) --
	( 96.03,112.31) --
	( 96.61,112.88) --
	( 97.20,113.46) --
	( 97.81,114.03) --
	( 98.43,114.60) --
	( 99.07,115.17) --
	( 99.73,115.75) --
	(100.40,116.32) --
	(101.09,116.89) --
	(101.80,117.46) --
	(102.53,118.03) --
	(103.27,118.61) --
	(104.03,119.18) --
	(104.81,119.75) --
	(105.60,120.32) --
	(106.42,120.90) --
	(107.25,121.47) --
	(108.10,122.04) --
	(108.97,122.61) --
	(109.87,123.18) --
	(110.78,123.76) --
	(111.71,124.33) --
	(112.66,124.90) --
	(113.63,125.47) --
	(114.63,126.05) --
	(115.64,126.62) --
	(116.67,127.19) --
	(117.73,127.76) --
	(118.81,128.33) --
	(119.91,128.91) --
	(121.04,129.48) --
	(122.18,130.05) --
	(123.35,130.62) --
	(124.54,131.20) --
	(125.76,131.77) --
	(127.00,132.34) --
	(128.26,132.91) --
	(129.55,133.48) --
	(130.86,134.06) --
	(132.20,134.63) --
	(133.56,135.20) --
	(134.95,135.77) --
	(136.36,136.35);
\end{scope}
\begin{scope}
\path[clip] (  0.00,  0.00) rectangle (144.54,144.54);
\definecolor{drawColor}{RGB}{0,0,0}

\path[draw=drawColor,line width= 0.3pt,line join=round] ( 17.19, 16.78) --
	( 17.19,142.04);
\end{scope}
\begin{scope}
\path[clip] (  0.00,  0.00) rectangle (144.54,144.54);
\definecolor{drawColor}{gray}{0.30}

\node[text=drawColor,anchor=base east,inner sep=0pt, outer sep=0pt, scale=  0.40] at ( 14.94, 21.09) {0.0};

\node[text=drawColor,anchor=base east,inner sep=0pt, outer sep=0pt, scale=  0.40] at ( 14.94, 43.87) {0.2};

\node[text=drawColor,anchor=base east,inner sep=0pt, outer sep=0pt, scale=  0.40] at ( 14.94, 66.64) {0.4};

\node[text=drawColor,anchor=base east,inner sep=0pt, outer sep=0pt, scale=  0.40] at ( 14.94, 89.42) {0.6};

\node[text=drawColor,anchor=base east,inner sep=0pt, outer sep=0pt, scale=  0.40] at ( 14.94,112.19) {0.8};

\node[text=drawColor,anchor=base east,inner sep=0pt, outer sep=0pt, scale=  0.40] at ( 14.94,134.97) {1.0};
\end{scope}
\begin{scope}
\path[clip] (  0.00,  0.00) rectangle (144.54,144.54);
\definecolor{drawColor}{gray}{0.20}

\path[draw=drawColor,line width= 0.3pt,line join=round] ( 15.94, 22.47) --
	( 17.19, 22.47);

\path[draw=drawColor,line width= 0.3pt,line join=round] ( 15.94, 45.24) --
	( 17.19, 45.24);

\path[draw=drawColor,line width= 0.3pt,line join=round] ( 15.94, 68.02) --
	( 17.19, 68.02);

\path[draw=drawColor,line width= 0.3pt,line join=round] ( 15.94, 90.80) --
	( 17.19, 90.80);

\path[draw=drawColor,line width= 0.3pt,line join=round] ( 15.94,113.57) --
	( 17.19,113.57);

\path[draw=drawColor,line width= 0.3pt,line join=round] ( 15.94,136.35) --
	( 17.19,136.35);
\end{scope}
\begin{scope}
\path[clip] (  0.00,  0.00) rectangle (144.54,144.54);
\definecolor{drawColor}{RGB}{0,0,0}

\path[draw=drawColor,line width= 0.3pt,line join=round] ( 17.19, 16.78) --
	(142.04, 16.78);
\end{scope}
\begin{scope}
\path[clip] (  0.00,  0.00) rectangle (144.54,144.54);
\definecolor{drawColor}{gray}{0.20}

\path[draw=drawColor,line width= 0.3pt,line join=round] ( 22.86, 15.53) --
	( 22.86, 16.78);

\path[draw=drawColor,line width= 0.3pt,line join=round] ( 51.24, 15.53) --
	( 51.24, 16.78);

\path[draw=drawColor,line width= 0.3pt,line join=round] ( 79.61, 15.53) --
	( 79.61, 16.78);

\path[draw=drawColor,line width= 0.3pt,line join=round] (107.99, 15.53) --
	(107.99, 16.78);

\path[draw=drawColor,line width= 0.3pt,line join=round] (136.36, 15.53) --
	(136.36, 16.78);
\end{scope}
\begin{scope}
\path[clip] (  0.00,  0.00) rectangle (144.54,144.54);
\definecolor{drawColor}{gray}{0.30}

\node[text=drawColor,anchor=base,inner sep=0pt, outer sep=0pt, scale=  0.40] at ( 22.86, 11.77) {0.00};

\node[text=drawColor,anchor=base,inner sep=0pt, outer sep=0pt, scale=  0.40] at ( 51.24, 11.77) {0.25};

\node[text=drawColor,anchor=base,inner sep=0pt, outer sep=0pt, scale=  0.40] at ( 79.61, 11.77) {0.50};

\node[text=drawColor,anchor=base,inner sep=0pt, outer sep=0pt, scale=  0.40] at (107.99, 11.77) {0.75};

\node[text=drawColor,anchor=base,inner sep=0pt, outer sep=0pt, scale=  0.40] at (136.36, 11.77) {1.00};
\end{scope}
\begin{scope}
\path[clip] (  0.00,  0.00) rectangle (144.54,144.54);
\definecolor{drawColor}{RGB}{0,0,0}

\node[text=drawColor,anchor=base,inner sep=0pt, outer sep=0pt, scale=  0.60] at ( 79.61,  4.44) {$p$};
\end{scope}
\begin{scope}
\path[clip] (  0.00,  0.00) rectangle (144.54,144.54);
\definecolor{drawColor}{RGB}{0,0,0}

\node[text=drawColor,rotate= 90.00,anchor=base,inner sep=0pt, outer sep=0pt, scale=  0.60] at (  6.63, 79.41) {$E(\tilde{p}|p)$};
\end{scope}
\end{tikzpicture}

%% file: Figures/ESPN_Revision_Plot.tex
\begin{tikzpicture}[x=1pt,y=1pt]
\definecolor{fillColor}{RGB}{255,255,255}
\path[use as bounding box,fill=fillColor,fill opacity=0.00] (0,0) rectangle (397.48,202.36);
\begin{scope}
\path[clip] (  0.00,  0.00) rectangle (397.48,202.36);
\definecolor{drawColor}{RGB}{255,255,255}
\definecolor{fillColor}{RGB}{255,255,255}

\path[draw=drawColor,line width= 0.5pt,line join=round,line cap=round,fill=fillColor] (  0.00,  0.00) rectangle (397.48,202.36);
\end{scope}
\begin{scope}
\path[clip] ( 23.45, 26.91) rectangle (392.48,197.36);
\definecolor{fillColor}{RGB}{255,255,255}

\path[fill=fillColor] ( 23.45, 26.91) rectangle (392.48,197.36);
\definecolor{drawColor}{RGB}{0,158,115}

\path[draw=drawColor,line width= 1.7pt,line join=round] ( 40.23, 56.22) --
	(107.32, 48.69) --
	(174.42, 46.30) --
	(241.52, 47.14) --
	(308.61, 49.15) --
	(375.71, 52.39);
\definecolor{drawColor}{RGB}{213,94,0}

\path[draw=drawColor,line width= 1.7pt,line join=round] ( 40.23,115.05) --
	(107.32,107.26) --
	(174.42,107.16) --
	(241.52, 84.91) --
	(308.61, 76.51) --
	(375.71,123.73);
\definecolor{drawColor}{RGB}{0,158,115}

\path[draw=drawColor,draw opacity=0.90,line width= 0.5pt,dash pattern=on 7pt off 3pt ,line join=round] ( 40.23, 45.12) --
	(107.32, 39.06) --
	(174.42, 38.47) --
	(241.52, 39.90) --
	(308.61, 34.66) --
	(375.71, 41.86);
\definecolor{drawColor}{RGB}{213,94,0}

\path[draw=drawColor,draw opacity=0.90,line width= 0.5pt,dash pattern=on 7pt off 3pt ,line join=round] ( 40.23, 76.13) --
	(107.32, 57.96) --
	(174.42, 59.94) --
	(241.52, 59.07) --
	(308.61, 34.66) --
	(375.71, 74.72);
\definecolor{drawColor}{RGB}{0,158,115}

\path[draw=drawColor,draw opacity=0.90,line width= 0.5pt,dash pattern=on 7pt off 3pt ,line join=round] ( 40.23, 67.49) --
	(107.32, 57.09) --
	(174.42, 51.96) --
	(241.52, 54.90) --
	(308.61, 59.34) --
	(375.71, 65.49);
\definecolor{drawColor}{RGB}{213,94,0}

\path[draw=drawColor,draw opacity=0.90,line width= 0.5pt,dash pattern=on 7pt off 3pt ,line join=round] ( 40.23,154.06) --
	(107.32,146.17) --
	(174.42,152.16) --
	(241.52,119.65) --
	(308.61,108.20) --
	(375.71,189.61);
\definecolor{drawColor}{RGB}{0,0,0}

\path[draw=drawColor,draw opacity=0.82,line width= 0.9pt,dash pattern=on 1pt off 3pt ,line join=round] ( 23.45, 55.66) -- (392.48, 55.66);
\end{scope}
\begin{scope}
\path[clip] (  0.00,  0.00) rectangle (397.48,202.36);
\definecolor{drawColor}{RGB}{0,0,0}

\path[draw=drawColor,line width= 0.5pt,line join=round] ( 23.45, 26.91) --
	( 23.45,197.36);
\end{scope}
\begin{scope}
\path[clip] (  0.00,  0.00) rectangle (397.48,202.36);
\definecolor{drawColor}{gray}{0.30}

\node[text=drawColor,anchor=base east,inner sep=0pt, outer sep=0pt, scale=  0.80] at ( 18.95, 52.90) {0};

\node[text=drawColor,anchor=base east,inner sep=0pt, outer sep=0pt, scale=  0.80] at ( 18.95, 94.91) {2};

\node[text=drawColor,anchor=base east,inner sep=0pt, outer sep=0pt, scale=  0.80] at ( 18.95,136.92) {4};

\node[text=drawColor,anchor=base east,inner sep=0pt, outer sep=0pt, scale=  0.80] at ( 18.95,178.92) {6};
\end{scope}
\begin{scope}
\path[clip] (  0.00,  0.00) rectangle (397.48,202.36);
\definecolor{drawColor}{gray}{0.20}

\path[draw=drawColor,line width= 0.5pt,line join=round] ( 20.95, 55.66) --
	( 23.45, 55.66);

\path[draw=drawColor,line width= 0.5pt,line join=round] ( 20.95, 97.67) --
	( 23.45, 97.67);

\path[draw=drawColor,line width= 0.5pt,line join=round] ( 20.95,139.67) --
	( 23.45,139.67);

\path[draw=drawColor,line width= 0.5pt,line join=round] ( 20.95,181.68) --
	( 23.45,181.68);
\end{scope}
\begin{scope}
\path[clip] (  0.00,  0.00) rectangle (397.48,202.36);
\definecolor{drawColor}{RGB}{0,0,0}

\path[draw=drawColor,line width= 0.5pt,line join=round] ( 23.45, 26.91) --
	(392.48, 26.91);
\end{scope}
\begin{scope}
\path[clip] (  0.00,  0.00) rectangle (397.48,202.36);
\definecolor{drawColor}{gray}{0.20}

\path[draw=drawColor,line width= 0.5pt,line join=round] ( 40.23, 24.41) --
	( 40.23, 26.91);

\path[draw=drawColor,line width= 0.5pt,line join=round] (107.32, 24.41) --
	(107.32, 26.91);

\path[draw=drawColor,line width= 0.5pt,line join=round] (174.42, 24.41) --
	(174.42, 26.91);

\path[draw=drawColor,line width= 0.5pt,line join=round] (241.52, 24.41) --
	(241.52, 26.91);

\path[draw=drawColor,line width= 0.5pt,line join=round] (308.61, 24.41) --
	(308.61, 26.91);

\path[draw=drawColor,line width= 0.5pt,line join=round] (375.71, 24.41) --
	(375.71, 26.91);
\end{scope}
\begin{scope}
\path[clip] (  0.00,  0.00) rectangle (397.48,202.36);
\definecolor{drawColor}{gray}{0.30}

\node[text=drawColor,anchor=base,inner sep=0pt, outer sep=0pt, scale=  0.80] at ( 40.23, 16.90) {01:00};

\node[text=drawColor,anchor=base,inner sep=0pt, outer sep=0pt, scale=  0.80] at (107.32, 16.90) {00:50};

\node[text=drawColor,anchor=base,inner sep=0pt, outer sep=0pt, scale=  0.80] at (174.42, 16.90) {00:40};

\node[text=drawColor,anchor=base,inner sep=0pt, outer sep=0pt, scale=  0.80] at (241.52, 16.90) {00:30};

\node[text=drawColor,anchor=base,inner sep=0pt, outer sep=0pt, scale=  0.80] at (308.61, 16.90) {00:20};

\node[text=drawColor,anchor=base,inner sep=0pt, outer sep=0pt, scale=  0.80] at (375.71, 16.90) {00:10};
\end{scope}
\begin{scope}
\path[clip] (  0.00,  0.00) rectangle (397.48,202.36);
\definecolor{drawColor}{RGB}{0,0,0}

\node[text=drawColor,anchor=base,inner sep=0pt, outer sep=0pt, scale=  0.80] at (207.97,  6.94) {Seconds left in game};
\end{scope}
\begin{scope}
\path[clip] (  0.00,  0.00) rectangle (397.48,202.36);
\definecolor{drawColor}{RGB}{0,0,0}

\node[text=drawColor,rotate= 90.00,anchor=base,inner sep=0pt, outer sep=0pt, scale=  0.80] at ( 10.51,112.13) {$\widehat{EC_\delta}$};
\end{scope}
\begin{scope}
\path[clip] (  0.00,  0.00) rectangle (397.48,202.36);
\definecolor{fillColor}{RGB}{255,255,255}

\path[fill=fillColor] ( 46.77,152.83) rectangle (147.75,204.38);
\end{scope}
\begin{scope}
\path[clip] (  0.00,  0.00) rectangle (397.48,202.36);
\definecolor{drawColor}{RGB}{0,0,0}

\node[text=drawColor,anchor=base west,inner sep=0pt, outer sep=0pt, scale=  0.90] at ( 51.77,192.21) {Method Type};
\end{scope}
\begin{scope}
\path[clip] (  0.00,  0.00) rectangle (397.48,202.36);
\definecolor{drawColor}{RGB}{0,158,115}

\path[draw=drawColor,line width= 1.7pt,line join=round] ( 53.21,179.51) -- ( 64.77,179.51);
\end{scope}
\begin{scope}
\path[clip] (  0.00,  0.00) rectangle (397.48,202.36);
\definecolor{drawColor}{RGB}{0,158,115}

\path[draw=drawColor,draw opacity=0.90,line width= 0.5pt,dash pattern=on 7pt off 3pt ,line join=round] ( 53.21,179.51) -- ( 64.77,179.51);
\end{scope}
\begin{scope}
\path[clip] (  0.00,  0.00) rectangle (397.48,202.36);
\definecolor{drawColor}{RGB}{0,158,115}

\path[draw=drawColor,draw opacity=0.90,line width= 0.5pt,dash pattern=on 7pt off 3pt ,line join=round] ( 53.21,179.51) -- ( 64.77,179.51);
\end{scope}
\begin{scope}
\path[clip] (  0.00,  0.00) rectangle (397.48,202.36);
\definecolor{drawColor}{RGB}{213,94,0}

\path[draw=drawColor,line width= 1.7pt,line join=round] ( 53.21,165.06) -- ( 64.77,165.06);
\end{scope}
\begin{scope}
\path[clip] (  0.00,  0.00) rectangle (397.48,202.36);
\definecolor{drawColor}{RGB}{213,94,0}

\path[draw=drawColor,draw opacity=0.90,line width= 0.5pt,dash pattern=on 7pt off 3pt ,line join=round] ( 53.21,165.06) -- ( 64.77,165.06);
\end{scope}
\begin{scope}
\path[clip] (  0.00,  0.00) rectangle (397.48,202.36);
\definecolor{drawColor}{RGB}{213,94,0}

\path[draw=drawColor,draw opacity=0.90,line width= 0.5pt,dash pattern=on 7pt off 3pt ,line join=round] ( 53.21,165.06) -- ( 64.77,165.06);
\end{scope}
\begin{scope}
\path[clip] (  0.00,  0.00) rectangle (397.48,202.36);
\definecolor{drawColor}{RGB}{0,0,0}

\node[text=drawColor,anchor=base west,inner sep=0pt, outer sep=0pt, scale=  0.90] at ( 70.72,176.41) {ECAP};
\end{scope}
\begin{scope}
\path[clip] (  0.00,  0.00) rectangle (397.48,202.36);
\definecolor{drawColor}{RGB}{0,0,0}

\node[text=drawColor,anchor=base west,inner sep=0pt, outer sep=0pt, scale=  0.90] at ( 70.72,161.96) {ESPN Unadjusted};
\end{scope}
\end{tikzpicture}

%% file: Figures/cap_tilde.tex
\begin{tikzpicture}[x=1pt,y=1pt]
\definecolor{fillColor}{RGB}{255,255,255}
\path[use as bounding box,fill=fillColor,fill opacity=0.00] (0,0) rectangle (216.81,216.81);
\begin{scope}
\path[clip] (  0.00,  0.00) rectangle (216.81,216.81);
\definecolor{drawColor}{RGB}{255,255,255}
\definecolor{fillColor}{RGB}{255,255,255}

\path[draw=drawColor,line width= 0.4pt,line join=round,line cap=round,fill=fillColor] (  0.00,  0.00) rectangle (216.81,216.81);
\end{scope}
\begin{scope}
\path[clip] ( 25.81, 21.65) rectangle (213.31,213.31);
\definecolor{fillColor}{RGB}{255,255,255}

\path[fill=fillColor] ( 25.81, 21.65) rectangle (213.31,213.31);
\definecolor{drawColor}{RGB}{0,158,115}

\path[draw=drawColor,line width= 0.6pt,line join=round] ( 34.33, 30.37) --
	( 34.33, 30.37) --
	( 34.33, 30.37) --
	( 34.33, 30.37) --
	( 34.33, 30.37) --
	( 34.33, 30.37) --
	( 34.33, 30.37) --
	( 34.33, 30.37) --
	( 34.33, 30.37) --
	( 34.33, 30.37) --
	( 34.33, 30.37) --
	( 34.33, 30.37) --
	( 34.33, 30.37) --
	( 34.33, 30.37) --
	( 34.33, 30.37) --
	( 34.33, 30.37) --
	( 34.33, 30.37) --
	( 34.33, 30.37) --
	( 34.33, 30.37) --
	( 34.33, 30.37) --
	( 34.33, 30.37) --
	( 34.33, 30.37) --
	( 34.33, 30.37) --
	( 34.33, 30.37) --
	( 34.33, 30.37) --
	( 34.33, 30.37) --
	( 34.33, 30.37) --
	( 34.33, 30.37) --
	( 34.33, 30.37) --
	( 34.33, 30.37) --
	( 34.33, 30.37) --
	( 34.33, 30.37) --
	( 34.33, 30.37) --
	( 34.33, 30.37) --
	( 34.33, 30.37) --
	( 34.33, 30.37) --
	( 34.33, 30.37) --
	( 34.33, 30.37) --
	( 34.33, 30.37) --
	( 34.33, 30.37) --
	( 34.33, 30.37) --
	( 34.33, 30.37) --
	( 34.33, 30.37) --
	( 34.33, 30.37) --
	( 34.33, 30.37) --
	( 34.33, 30.37) --
	( 34.34, 30.37) --
	( 34.34, 30.37) --
	( 34.34, 30.37) --
	( 34.34, 30.37) --
	( 34.34, 30.37) --
	( 34.34, 30.37) --
	( 34.34, 30.37) --
	( 34.34, 30.37) --
	( 34.34, 30.37) --
	( 34.34, 30.37) --
	( 34.34, 30.37) --
	( 34.34, 30.37) --
	( 34.34, 30.37) --
	( 34.34, 30.37) --
	( 34.34, 30.37) --
	( 34.34, 30.37) --
	( 34.34, 30.37) --
	( 34.34, 30.37) --
	( 34.34, 30.37) --
	( 34.34, 30.37) --
	( 34.35, 30.37) --
	( 34.35, 30.37) --
	( 34.35, 30.37) --
	( 34.35, 30.37) --
	( 34.35, 30.37) --
	( 34.35, 30.37) --
	( 34.35, 30.37) --
	( 34.35, 30.37) --
	( 34.35, 30.37) --
	( 34.35, 30.37) --
	( 34.35, 30.37) --
	( 34.35, 30.37) --
	( 34.35, 30.37) --
	( 34.35, 30.37) --
	( 34.36, 30.37) --
	( 34.36, 30.37) --
	( 34.36, 30.37) --
	( 34.36, 30.37) --
	( 34.36, 30.37) --
	( 34.36, 30.37) --
	( 34.37, 30.37) --
	( 34.37, 30.37) --
	( 34.37, 30.37) --
	( 34.37, 30.37) --
	( 34.37, 30.37) --
	( 34.37, 30.37) --
	( 34.37, 30.37) --
	( 34.37, 30.37) --
	( 34.37, 30.37) --
	( 34.37, 30.37) --
	( 34.38, 30.37) --
	( 34.38, 30.37) --
	( 34.38, 30.37) --
	( 34.38, 30.37) --
	( 34.38, 30.37) --
	( 34.38, 30.37) --
	( 34.39, 30.37) --
	( 34.39, 30.37) --
	( 34.39, 30.37) --
	( 34.39, 30.37) --
	( 34.40, 30.37) --
	( 34.40, 30.37) --
	( 34.40, 30.37) --
	( 34.40, 30.37) --
	( 34.40, 30.37) --
	( 34.40, 30.37) --
	( 34.40, 30.37) --
	( 34.40, 30.37) --
	( 34.40, 30.37) --
	( 34.40, 30.37) --
	( 34.41, 30.37) --
	( 34.41, 30.37) --
	( 34.41, 30.37) --
	( 34.41, 30.37) --
	( 34.41, 30.37) --
	( 34.41, 30.37) --
	( 34.41, 30.37) --
	( 34.41, 30.37) --
	( 34.43, 30.37) --
	( 34.43, 30.37) --
	( 34.43, 30.37) --
	( 34.43, 30.37) --
	( 34.43, 30.37) --
	( 34.43, 30.37) --
	( 34.43, 30.37) --
	( 34.43, 30.37) --
	( 34.44, 30.37) --
	( 34.44, 30.37) --
	( 34.44, 30.37) --
	( 34.44, 30.37) --
	( 34.44, 30.37) --
	( 34.44, 30.37) --
	( 34.45, 30.37) --
	( 34.45, 30.37) --
	( 34.45, 30.37) --
	( 34.45, 30.37) --
	( 34.47, 30.37) --
	( 34.47, 30.37) --
	( 34.48, 30.37) --
	( 34.48, 30.37) --
	( 34.49, 30.37) --
	( 34.49, 30.37) --
	( 34.51, 30.37) --
	( 34.51, 30.37) --
	( 34.51, 30.37) --
	( 34.51, 30.37) --
	( 34.51, 30.37) --
	( 34.51, 30.37) --
	( 34.52, 30.37) --
	( 34.52, 30.37) --
	( 34.52, 30.37) --
	( 34.52, 30.37) --
	( 34.54, 30.37) --
	( 34.54, 30.37) --
	( 34.55, 30.37) --
	( 34.55, 30.37) --
	( 34.56, 30.37) --
	( 34.56, 30.37) --
	( 34.58, 30.37) --
	( 34.58, 30.37) --
	( 34.58, 30.37) --
	( 34.58, 30.37) --
	( 34.59, 30.37) --
	( 34.59, 30.37) --
	( 34.62, 30.37) --
	( 34.62, 30.37) --
	( 34.62, 30.37) --
	( 34.62, 30.37) --
	( 34.62, 30.37) --
	( 34.62, 30.37) --
	( 34.63, 30.37) --
	( 34.63, 30.37) --
	( 34.63, 30.37) --
	( 34.63, 30.37) --
	( 34.64, 30.37) --
	( 34.64, 30.37) --
	( 34.67, 30.37) --
	( 34.67, 30.37) --
	( 34.74, 30.37) --
	( 34.74, 30.37) --
	( 34.74, 30.37) --
	( 34.74, 30.37) --
	( 34.76, 30.37) --
	( 34.76, 30.37) --
	( 34.81, 30.38) --
	( 34.81, 30.38) --
	( 34.88, 30.38) --
	( 34.88, 30.38) --
	( 34.94, 30.38) --
	( 34.94, 30.38) --
	( 35.01, 30.38) --
	( 35.01, 30.38) --
	( 35.13, 30.39) --
	( 35.13, 30.39) --
	( 35.13, 30.39) --
	( 35.13, 30.39) --
	( 35.14, 30.39) --
	( 35.14, 30.39) --
	( 35.15, 30.39) --
	( 35.15, 30.39) --
	( 35.16, 30.39) --
	( 35.16, 30.39) --
	( 35.27, 30.40) --
	( 35.27, 30.40) --
	( 35.30, 30.40) --
	( 35.30, 30.40) --
	( 35.35, 30.40) --
	( 35.35, 30.40) --
	( 35.35, 30.40) --
	( 35.35, 30.40) --
	( 35.41, 30.40) --
	( 35.41, 30.40) --
	( 35.43, 30.40) --
	( 35.43, 30.40) --
	( 35.54, 30.41) --
	( 35.54, 30.41) --
	( 35.58, 30.41) --
	( 35.58, 30.41) --
	( 35.60, 30.41) --
	( 35.60, 30.41) --
	( 35.65, 30.42) --
	( 35.65, 30.42) --
	( 35.68, 30.42) --
	( 35.68, 30.42) --
	( 35.77, 30.42) --
	( 35.77, 30.42) --
	( 35.88, 30.43) --
	( 35.88, 30.43) --
	( 35.88, 30.43) --
	( 35.88, 30.43) --
	( 35.88, 30.43) --
	( 35.88, 30.43) --
	( 35.88, 30.43) --
	( 35.88, 30.43) --
	( 35.94, 30.44) --
	( 35.94, 30.44) --
	( 36.03, 30.44) --
	( 36.03, 30.44) --
	( 36.03, 30.44) --
	( 36.03, 30.44) --
	( 36.13, 30.45) --
	( 36.13, 30.45) --
	( 36.46, 30.48) --
	( 36.46, 30.48) --
	( 36.51, 30.48) --
	( 36.51, 30.48) --
	( 36.56, 30.49) --
	( 36.56, 30.49) --
	( 36.57, 30.49) --
	( 36.57, 30.49) --
	( 36.69, 30.50) --
	( 36.69, 30.50) --
	( 37.01, 30.53) --
	( 37.01, 30.53) --
	( 37.21, 30.56) --
	( 37.21, 30.56) --
	( 37.30, 30.57) --
	( 37.30, 30.57) --
	( 37.60, 30.60) --
	( 37.60, 30.60) --
	( 37.85, 30.64) --
	( 37.85, 30.64) --
	( 37.97, 30.66) --
	( 37.97, 30.66) --
	( 38.11, 30.67) --
	( 38.11, 30.67) --
	( 38.17, 30.68) --
	( 38.17, 30.68) --
	( 38.20, 30.69) --
	( 38.20, 30.69) --
	( 38.27, 30.70) --
	( 38.27, 30.70) --
	( 38.31, 30.71) --
	( 38.31, 30.71) --
	( 38.69, 30.77) --
	( 38.69, 30.77) --
	( 39.48, 30.91) --
	( 39.48, 30.91) --
	( 39.94, 31.00) --
	( 39.94, 31.00) --
	( 40.49, 31.12) --
	( 40.49, 31.12) --
	( 40.79, 31.19) --
	( 40.79, 31.19) --
	( 40.84, 31.20) --
	( 40.84, 31.20) --
	( 41.02, 31.25) --
	( 41.02, 31.25) --
	( 41.43, 31.35) --
	( 41.43, 31.35) --
	( 42.11, 31.53) --
	( 42.11, 31.53) --
	( 42.16, 31.55) --
	( 42.16, 31.55) --
	( 44.14, 32.16) --
	( 44.14, 32.16) --
	( 44.72, 32.37) --
	( 44.72, 32.37) --
	( 44.84, 32.42) --
	( 44.84, 32.42) --
	( 45.03, 32.48) --
	( 45.03, 32.48) --
	( 45.48, 32.65) --
	( 45.48, 32.65) --
	( 45.49, 32.66) --
	( 45.49, 32.66) --
	( 45.52, 32.67) --
	( 45.52, 32.67) --
	( 45.79, 32.78) --
	( 45.79, 32.78) --
	( 46.13, 32.91) --
	( 46.13, 32.91) --
	( 46.89, 33.23) --
	( 46.89, 33.23) --
	( 47.17, 33.35) --
	( 47.17, 33.35) --
	( 48.85, 34.13) --
	( 48.85, 34.13) --
	( 49.11, 34.26) --
	( 49.11, 34.26) --
	( 50.04, 34.73) --
	( 50.04, 34.73) --
	( 50.32, 34.88) --
	( 50.32, 34.88) --
	( 51.95, 35.78) --
	( 51.95, 35.78) --
	( 52.49, 36.10) --
	( 52.49, 36.10) --
	( 52.53, 36.12) --
	( 52.53, 36.12) --
	( 54.69, 37.47) --
	( 54.69, 37.47) --
	( 55.38, 37.93) --
	( 55.38, 37.93) --
	( 56.06, 38.39) --
	( 56.06, 38.39) --
	( 57.19, 39.18) --
	( 57.19, 39.18) --
	( 58.07, 39.82) --
	( 58.07, 39.82) --
	( 58.17, 39.90) --
	( 58.17, 39.90) --
	( 58.43, 40.09) --
	( 58.43, 40.09) --
	( 58.63, 40.24) --
	( 58.63, 40.24) --
	( 58.72, 40.31) --
	( 58.72, 40.31) --
	( 58.76, 40.34) --
	( 58.76, 40.34) --
	( 60.33, 41.57) --
	( 60.33, 41.57) --
	( 60.66, 41.83) --
	( 60.66, 41.83) --
	( 60.68, 41.84) --
	( 60.68, 41.84) --
	( 60.97, 42.08) --
	( 60.97, 42.08) --
	( 62.01, 42.94) --
	( 62.01, 42.94) --
	( 62.20, 43.10) --
	( 62.20, 43.10) --
	( 62.77, 43.58) --
	( 62.77, 43.58) --
	( 62.81, 43.62) --
	( 62.81, 43.62) --
	( 64.95, 45.51) --
	( 64.95, 45.51) --
	( 66.18, 46.64) --
	( 66.18, 46.64) --
	( 66.78, 47.21) --
	( 66.78, 47.21) --
	( 67.27, 47.67) --
	( 67.27, 47.67) --
	( 67.82, 48.21) --
	( 67.82, 48.21) --
	( 68.06, 48.44) --
	( 68.06, 48.44) --
	( 68.90, 49.26) --
	( 68.90, 49.26) --
	( 70.16, 50.54) --
	( 70.16, 50.54) --
	( 70.48, 50.87) --
	( 70.48, 50.87) --
	( 70.74, 51.14) --
	( 70.74, 51.14) --
	( 72.05, 52.51) --
	( 72.05, 52.51) --
	( 72.15, 52.61) --
	( 72.15, 52.61) --
	( 72.65, 53.15) --
	( 72.65, 53.15) --
	( 72.70, 53.20) --
	( 72.70, 53.20) --
	( 73.23, 53.77) --
	( 73.23, 53.77) --
	( 74.65, 55.33) --
	( 74.65, 55.33) --
	( 75.00, 55.71) --
	( 75.00, 55.71) --
	( 75.80, 56.61) --
	( 75.80, 56.61) --
	( 76.00, 56.84) --
	( 76.00, 56.84) --
	( 77.34, 58.37) --
	( 77.34, 58.37) --
	( 78.50, 59.73) --
	( 78.50, 59.73) --
	( 79.76, 61.23) --
	( 79.76, 61.23) --
	( 79.95, 61.45) --
	( 79.95, 61.45) --
	( 81.60, 63.46) --
	( 81.60, 63.46) --
	( 81.76, 63.66) --
	( 81.76, 63.66) --
	( 82.44, 64.49) --
	( 82.44, 64.49) --
	( 82.70, 64.82) --
	( 82.70, 64.82) --
	( 87.34, 70.75) --
	( 87.34, 70.75) --
	( 89.29, 73.34) --
	( 89.29, 73.34) --
	( 89.72, 73.92) --
	( 89.72, 73.92) --
	( 90.09, 74.42) --
	( 90.09, 74.42) --
	( 93.49, 79.07) --
	( 93.49, 79.07) --
	( 93.56, 79.17) --
	( 93.56, 79.17) --
	( 97.27, 84.38) --
	( 97.27, 84.38) --
	( 97.69, 84.99) --
	( 97.69, 84.99) --
	(100.28, 88.71) --
	(100.28, 88.71) --
	(105.06, 95.74) --
	(105.06, 95.74) --
	(105.83, 96.88) --
	(105.83, 96.88) --
	(106.07, 97.23) --
	(106.07, 97.23) --
	(107.31, 99.09) --
	(107.31, 99.09) --
	(107.31, 99.09) --
	(107.31, 99.09) --
	(108.74,101.23) --
	(108.74,101.23) --
	(110.04,103.19) --
	(110.04,103.19) --
	(112.30,106.62) --
	(112.30,106.62) --
	(113.02,107.71) --
	(113.02,107.71) --
	(115.48,111.46) --
	(115.48,111.46) --
	(117.18,114.04) --
	(117.18,114.04) --
	(118.10,115.45) --
	(118.10,115.45) --
	(122.15,121.25) --
	(122.15,121.25) --
	(123.36,123.09) --
	(123.36,123.09) --
	(123.36,123.09) --
	(123.36,123.09) --
	(127.84,129.89) --
	(127.84,129.89) --
	(129.50,132.41) --
	(129.50,132.41) --
	(129.53,132.45) --
	(129.53,132.45) --
	(130.76,134.31) --
	(130.76,134.31) --
	(131.31,135.12) --
	(131.31,135.12) --
	(131.53,135.46) --
	(131.53,135.46) --
	(132.03,136.20) --
	(132.03,136.20) --
	(133.59,138.53) --
	(133.59,138.53) --
	(135.74,141.71) --
	(135.74,141.71) --
	(136.01,142.12) --
	(136.01,142.12) --
	(138.59,145.89) --
	(138.59,145.89) --
	(139.06,146.56) --
	(139.06,146.56) --
	(139.59,147.34) --
	(139.59,147.34) --
	(140.48,148.63) --
	(140.48,148.63) --
	(142.06,150.88) --
	(142.06,150.88) --
	(142.44,151.42) --
	(142.44,151.42) --
	(142.85,152.00) --
	(142.85,152.00) --
	(148.36,159.65) --
	(148.36,159.65) --
	(148.51,159.85) --
	(148.51,159.85) --
	(152.69,165.40) --
	(152.69,165.40) --
	(154.07,167.18) --
	(154.07,167.18) --
	(154.30,167.48) --
	(154.30,167.48) --
	(156.80,170.62) --
	(156.80,170.62) --
	(163.91,179.02) --
	(163.91,179.02) --
	(165.95,181.26) --
	(165.95,181.26) --
	(166.40,181.75) --
	(166.40,181.75) --
	(167.72,183.14) --
	(167.72,183.14) --
	(168.98,184.45) --
	(168.98,184.45) --
	(169.38,184.86) --
	(169.38,184.86) --
	(174.68,189.92) --
	(174.68,189.92) --
	(177.46,192.32) --
	(177.46,192.32) --
	(178.92,193.50) --
	(178.92,193.50) --
	(179.09,193.64) --
	(179.09,193.64) --
	(179.21,193.73) --
	(179.21,193.73) --
	(179.84,194.23) --
	(179.84,194.23) --
	(180.22,194.52) --
	(180.22,194.52) --
	(181.61,195.56) --
	(181.61,195.56) --
	(183.85,197.12) --
	(183.85,197.12) --
	(184.11,197.29) --
	(184.11,197.29) --
	(185.01,197.87) --
	(185.01,197.87) --
	(185.62,198.26) --
	(185.62,198.26) --
	(186.39,198.73) --
	(186.39,198.73) --
	(186.42,198.74) --
	(186.42,198.74) --
	(187.07,199.13) --
	(187.07,199.13) --
	(191.86,201.58) --
	(191.86,201.58) --
	(192.98,202.05) --
	(192.98,202.05) --
	(193.08,202.09) --
	(193.08,202.09) --
	(193.53,202.27) --
	(193.53,202.27) --
	(194.61,202.67) --
	(194.61,202.67) --
	(195.60,203.01) --
	(195.60,203.01) --
	(195.67,203.03) --
	(195.67,203.03) --
	(196.01,203.14) --
	(196.01,203.14) --
	(196.21,203.20) --
	(196.21,203.20) --
	(196.58,203.31) --
	(196.58,203.31) --
	(197.17,203.48) --
	(197.17,203.48) --
	(197.51,203.57) --
	(197.51,203.57) --
	(197.65,203.61) --
	(197.65,203.61) --
	(197.97,203.69) --
	(197.97,203.69) --
	(198.46,203.81) --
	(198.46,203.81) --
	(199.11,203.95) --
	(199.11,203.95) --
	(199.23,203.97) --
	(199.23,203.97) --
	(199.44,204.02) --
	(199.44,204.02) --
	(200.50,204.21) --
	(200.50,204.21) --
	(200.90,204.27) --
	(200.90,204.27) --
	(201.16,204.31) --
	(201.16,204.31) --
	(201.37,204.34) --
	(201.37,204.34) --
	(201.39,204.34) --
	(201.39,204.34) --
	(201.69,204.38) --
	(201.69,204.38) --
	(202.27,204.45) --
	(202.27,204.45) --
	(202.28,204.45) --
	(202.28,204.45) --
	(202.34,204.46) --
	(202.34,204.46) --
	(202.62,204.48) --
	(202.62,204.48) --
	(202.73,204.49) --
	(202.73,204.49) --
	(202.76,204.50) --
	(202.76,204.50) --
	(203.16,204.53) --
	(203.16,204.53) --
	(203.16,204.53) --
	(203.16,204.53) --
	(203.17,204.53) --
	(203.17,204.53) --
	(203.24,204.53) --
	(203.24,204.53) --
	(203.24,204.53) --
	(203.24,204.53) --
	(203.33,204.54) --
	(203.33,204.54) --
	(203.54,204.55) --
	(203.54,204.55) --
	(203.80,204.57) --
	(203.80,204.57) --
	(203.94,204.57) --
	(203.94,204.57) --
	(204.06,204.58) --
	(204.06,204.58) --
	(204.06,204.58) --
	(204.06,204.58) --
	(204.12,204.58) --
	(204.12,204.58) --
	(204.16,204.58) --
	(204.16,204.58) --
	(204.17,204.58) --
	(204.17,204.58) --
	(204.18,204.58) --
	(204.18,204.58) --
	(204.24,204.59) --
	(204.24,204.59) --
	(204.33,204.59) --
	(204.33,204.59) --
	(204.34,204.59) --
	(204.34,204.59) --
	(204.40,204.59) --
	(204.40,204.59) --
	(204.41,204.59) --
	(204.41,204.59) --
	(204.46,204.59) --
	(204.46,204.59) --
	(204.50,204.59) --
	(204.50,204.59) --
	(204.53,204.59) --
	(204.53,204.59) --
	(204.54,204.59) --
	(204.54,204.59) --
	(204.58,204.60) --
	(204.58,204.60) --
	(204.59,204.60) --
	(204.59,204.60) --
	(204.60,204.60) --
	(204.60,204.60) --
	(204.62,204.60) --
	(204.62,204.60) --
	(204.65,204.60) --
	(204.65,204.60) --
	(204.66,204.60) --
	(204.66,204.60) --
	(204.67,204.60) --
	(204.67,204.60) --
	(204.67,204.60) --
	(204.67,204.60) --
	(204.69,204.60) --
	(204.69,204.60) --
	(204.69,204.60) --
	(204.69,204.60) --
	(204.69,204.60) --
	(204.69,204.60) --
	(204.70,204.60) --
	(204.70,204.60) --
	(204.70,204.60) --
	(204.70,204.60) --
	(204.70,204.60) --
	(204.70,204.60) --
	(204.71,204.60) --
	(204.71,204.60) --
	(204.72,204.60) --
	(204.72,204.60) --
	(204.72,204.60) --
	(204.72,204.60) --
	(204.72,204.60) --
	(204.72,204.60) --
	(204.72,204.60) --
	(204.72,204.60) --
	(204.73,204.60) --
	(204.73,204.60) --
	(204.73,204.60) --
	(204.73,204.60) --
	(204.73,204.60) --
	(204.73,204.60) --
	(204.74,204.60) --
	(204.74,204.60) --
	(204.75,204.60) --
	(204.75,204.60) --
	(204.75,204.60) --
	(204.75,204.60) --
	(204.75,204.60) --
	(204.75,204.60) --
	(204.76,204.60) --
	(204.76,204.60) --
	(204.76,204.60) --
	(204.76,204.60) --
	(204.76,204.60) --
	(204.76,204.60) --
	(204.76,204.60) --
	(204.76,204.60) --
	(204.76,204.60) --
	(204.76,204.60) --
	(204.76,204.60) --
	(204.76,204.60) --
	(204.76,204.60) --
	(204.76,204.60) --
	(204.77,204.60) --
	(204.77,204.60) --
	(204.77,204.60) --
	(204.77,204.60) --
	(204.77,204.60) --
	(204.77,204.60) --
	(204.77,204.60) --
	(204.77,204.60) --
	(204.77,204.60) --
	(204.77,204.60) --
	(204.77,204.60) --
	(204.77,204.60) --
	(204.77,204.60) --
	(204.77,204.60) --
	(204.77,204.60) --
	(204.77,204.60) --
	(204.77,204.60) --
	(204.77,204.60) --
	(204.77,204.60) --
	(204.77,204.60) --
	(204.77,204.60) --
	(204.77,204.60) --
	(204.77,204.60) --
	(204.77,204.60) --
	(204.77,204.60) --
	(204.77,204.60) --
	(204.77,204.60) --
	(204.77,204.60) --
	(204.77,204.60) --
	(204.77,204.60) --
	(204.77,204.60) --
	(204.77,204.60) --
	(204.77,204.60) --
	(204.77,204.60) --
	(204.78,204.60) --
	(204.78,204.60) --
	(204.78,204.60) --
	(204.78,204.60) --
	(204.78,204.60) --
	(204.78,204.60) --
	(204.78,204.60) --
	(204.78,204.60) --
	(204.78,204.60) --
	(204.78,204.60) --
	(204.78,204.60) --
	(204.78,204.60) --
	(204.78,204.60) --
	(204.78,204.60) --
	(204.78,204.60) --
	(204.78,204.60) --
	(204.78,204.60) --
	(204.78,204.60) --
	(204.78,204.60) --
	(204.78,204.60) --
	(204.78,204.60) --
	(204.78,204.60) --
	(204.78,204.60) --
	(204.78,204.60) --
	(204.78,204.60) --
	(204.78,204.60) --
	(204.78,204.60) --
	(204.78,204.60) --
	(204.78,204.60) --
	(204.78,204.60) --
	(204.78,204.60) --
	(204.78,204.60) --
	(204.78,204.60) --
	(204.78,204.60) --
	(204.78,204.60) --
	(204.78,204.60) --
	(204.78,204.60) --
	(204.78,204.60) --
	(204.78,204.60) --
	(204.78,204.60) --
	(204.78,204.60) --
	(204.78,204.60) --
	(204.78,204.60) --
	(204.78,204.60) --
	(204.78,204.60) --
	(204.78,204.60) --
	(204.78,204.60) --
	(204.78,204.60) --
	(204.78,204.60) --
	(204.78,204.60) --
	(204.78,204.60) --
	(204.78,204.60) --
	(204.78,204.60) --
	(204.78,204.60) --
	(204.78,204.60) --
	(204.78,204.60) --
	(204.78,204.60) --
	(204.78,204.60) --
	(204.78,204.60) --
	(204.78,204.60) --
	(204.78,204.60) --
	(204.78,204.60) --
	(204.78,204.60) --
	(204.78,204.60) --
	(204.78,204.60) --
	(204.78,204.60) --
	(204.79,204.60) --
	(204.79,204.60) --
	(204.79,204.60) --
	(204.79,204.60) --
	(204.79,204.60) --
	(204.79,204.60) --
	(204.79,204.60) --
	(204.79,204.60) --
	(204.79,204.60) --
	(204.79,204.60) --
	(204.79,204.60) --
	(204.79,204.60) --
	(204.79,204.60) --
	(204.79,204.60) --
	(204.79,204.60) --
	(204.79,204.60) --
	(204.79,204.60) --
	(204.79,204.60) --
	(204.79,204.60) --
	(204.79,204.60) --
	(204.79,204.60) --
	(204.79,204.60) --
	(204.79,204.60) --
	(204.79,204.60) --
	(204.79,204.60) --
	(204.79,204.60) --
	(204.79,204.60) --
	(204.79,204.60) --
	(204.79,204.60) --
	(204.79,204.60) --
	(204.79,204.60) --
	(204.79,204.60) --
	(204.79,204.60) --
	(204.79,204.60) --
	(204.79,204.60) --
	(204.79,204.60) --
	(204.79,204.60) --
	(204.79,204.60) --
	(204.79,204.60) --
	(204.79,204.60) --
	(204.79,204.60) --
	(204.79,204.60) --
	(204.79,204.60) --
	(204.79,204.60) --
	(204.79,204.60) --
	(204.79,204.60) --
	(204.79,204.60) --
	(204.79,204.60) --
	(204.79,204.60) --
	(204.79,204.60) --
	(204.79,204.60) --
	(204.79,204.60) --
	(204.79,204.60) --
	(204.79,204.60) --
	(204.79,204.60) --
	(204.79,204.60) --
	(204.79,204.60) --
	(204.79,204.60) --
	(204.79,204.60) --
	(204.79,204.60) --
	(204.79,204.60) --
	(204.79,204.60) --
	(204.79,204.60) --
	(204.79,204.60) --
	(204.79,204.60) --
	(204.79,204.60) --
	(204.79,204.60) --
	(204.79,204.60) --
	(204.79,204.60) --
	(204.79,204.60) --
	(204.79,204.60) --
	(204.79,204.60) --
	(204.79,204.60) --
	(204.79,204.60) --
	(204.79,204.60) --
	(204.79,204.60) --
	(204.79,204.60) --
	(204.79,204.60) --
	(204.79,204.60) --
	(204.79,204.60) --
	(204.79,204.60) --
	(204.79,204.60) --
	(204.79,204.60) --
	(204.79,204.60) --
	(204.79,204.60) --
	(204.79,204.60) --
	(204.79,204.60) --
	(204.79,204.60) --
	(204.79,204.60) --
	(204.79,204.60) --
	(204.79,204.60) --
	(204.79,204.60) --
	(204.79,204.60) --
	(204.79,204.60) --
	(204.79,204.60) --
	(204.79,204.60) --
	(204.79,204.60) --
	(204.79,204.60) --
	(204.79,204.60) --
	(204.79,204.60) --
	(204.79,204.60) --
	(204.79,204.60) --
	(204.79,204.60) --
	(204.79,204.60) --
	(204.79,204.60) --
	(204.79,204.60) --
	(204.79,204.60) --
	(204.79,204.60) --
	(204.79,204.60) --
	(204.79,204.60) --
	(204.79,204.60) --
	(204.79,204.60) --
	(204.79,204.60) --
	(204.79,204.60) --
	(204.79,204.60) --
	(204.79,204.60) --
	(204.79,204.60) --
	(204.79,204.60) --
	(204.79,204.60) --
	(204.79,204.60) --
	(204.79,204.60) --
	(204.79,204.60) --
	(204.79,204.60) --
	(204.79,204.60) --
	(204.79,204.60) --
	(204.79,204.60) --
	(204.79,204.60) --
	(204.79,204.60) --
	(204.79,204.60) --
	(204.79,204.60) --
	(204.79,204.60) --
	(204.79,204.60) --
	(204.79,204.60) --
	(204.79,204.60) --
	(204.79,204.60) --
	(204.79,204.60) --
	(204.79,204.60) --
	(204.79,204.60) --
	(204.79,204.60) --
	(204.79,204.60) --
	(204.79,204.60) --
	(204.79,204.60) --
	(204.79,204.60) --
	(204.79,204.60) --
	(204.79,204.60) --
	(204.79,204.60) --
	(204.79,204.60) --
	(204.79,204.60) --
	(204.79,204.60) --
	(204.79,204.60) --
	(204.79,204.60) --
	(204.79,204.60) --
	(204.79,204.60) --
	(204.79,204.60) --
	(204.79,204.60) --
	(204.79,204.60) --
	(204.79,204.60) --
	(204.79,204.60) --
	(204.79,204.60) --
	(204.79,204.60) --
	(204.79,204.60) --
	(204.79,204.60) --
	(204.79,204.60) --
	(204.79,204.60) --
	(204.79,204.60) --
	(204.79,204.60) --
	(204.79,204.60) --
	(204.79,204.60) --
	(204.79,204.60) --
	(204.79,204.60) --
	(204.79,204.60) --
	(204.79,204.60) --
	(204.79,204.60) --
	(204.79,204.60) --
	(204.79,204.60) --
	(204.79,204.60);
\definecolor{drawColor}{RGB}{230,159,0}

\path[draw=drawColor,line width= 0.6pt,dash pattern=on 4pt off 4pt ,line join=round] ( 25.81, 21.65) -- (213.31,213.31);
\end{scope}
\begin{scope}
\path[clip] (  0.00,  0.00) rectangle (216.81,216.81);
\definecolor{drawColor}{RGB}{0,0,0}

\path[draw=drawColor,line width= 0.4pt,line join=round] ( 25.81, 21.65) --
	( 25.81,213.31);
\end{scope}
\begin{scope}
\path[clip] (  0.00,  0.00) rectangle (216.81,216.81);
\definecolor{drawColor}{gray}{0.30}

\node[text=drawColor,anchor=base east,inner sep=0pt, outer sep=0pt, scale=  0.56] at ( 22.66, 28.44) {0.00};

\node[text=drawColor,anchor=base east,inner sep=0pt, outer sep=0pt, scale=  0.56] at ( 22.66, 72.00) {0.25};

\node[text=drawColor,anchor=base east,inner sep=0pt, outer sep=0pt, scale=  0.56] at ( 22.66,115.55) {0.50};

\node[text=drawColor,anchor=base east,inner sep=0pt, outer sep=0pt, scale=  0.56] at ( 22.66,159.11) {0.75};

\node[text=drawColor,anchor=base east,inner sep=0pt, outer sep=0pt, scale=  0.56] at ( 22.66,202.67) {1.00};
\end{scope}
\begin{scope}
\path[clip] (  0.00,  0.00) rectangle (216.81,216.81);
\definecolor{drawColor}{gray}{0.20}

\path[draw=drawColor,line width= 0.4pt,line join=round] ( 24.06, 30.37) --
	( 25.81, 30.37);

\path[draw=drawColor,line width= 0.4pt,line join=round] ( 24.06, 73.92) --
	( 25.81, 73.92);

\path[draw=drawColor,line width= 0.4pt,line join=round] ( 24.06,117.48) --
	( 25.81,117.48);

\path[draw=drawColor,line width= 0.4pt,line join=round] ( 24.06,161.04) --
	( 25.81,161.04);

\path[draw=drawColor,line width= 0.4pt,line join=round] ( 24.06,204.60) --
	( 25.81,204.60);
\end{scope}
\begin{scope}
\path[clip] (  0.00,  0.00) rectangle (216.81,216.81);
\definecolor{drawColor}{RGB}{0,0,0}

\path[draw=drawColor,line width= 0.4pt,line join=round] ( 25.81, 21.65) --
	(213.31, 21.65);
\end{scope}
\begin{scope}
\path[clip] (  0.00,  0.00) rectangle (216.81,216.81);
\definecolor{drawColor}{gray}{0.20}

\path[draw=drawColor,line width= 0.4pt,line join=round] ( 34.33, 19.90) --
	( 34.33, 21.65);

\path[draw=drawColor,line width= 0.4pt,line join=round] ( 76.94, 19.90) --
	( 76.94, 21.65);

\path[draw=drawColor,line width= 0.4pt,line join=round] (119.56, 19.90) --
	(119.56, 21.65);

\path[draw=drawColor,line width= 0.4pt,line join=round] (162.17, 19.90) --
	(162.17, 21.65);

\path[draw=drawColor,line width= 0.4pt,line join=round] (204.79, 19.90) --
	(204.79, 21.65);
\end{scope}
\begin{scope}
\path[clip] (  0.00,  0.00) rectangle (216.81,216.81);
\definecolor{drawColor}{gray}{0.30}

\node[text=drawColor,anchor=base,inner sep=0pt, outer sep=0pt, scale=  0.56] at ( 34.33, 14.65) {0.00};

\node[text=drawColor,anchor=base,inner sep=0pt, outer sep=0pt, scale=  0.56] at ( 76.94, 14.65) {0.25};

\node[text=drawColor,anchor=base,inner sep=0pt, outer sep=0pt, scale=  0.56] at (119.56, 14.65) {0.50};

\node[text=drawColor,anchor=base,inner sep=0pt, outer sep=0pt, scale=  0.56] at (162.17, 14.65) {0.75};

\node[text=drawColor,anchor=base,inner sep=0pt, outer sep=0pt, scale=  0.56] at (204.79, 14.65) {1.00};
\end{scope}
\begin{scope}
\path[clip] (  0.00,  0.00) rectangle (216.81,216.81);
\definecolor{drawColor}{RGB}{0,0,0}

\node[text=drawColor,anchor=base,inner sep=0pt, outer sep=0pt, scale=  0.80] at (119.56,  5.44) {Unadjusted Probability (538 Data)};
\end{scope}
\begin{scope}
\path[clip] (  0.00,  0.00) rectangle (216.81,216.81);
\definecolor{drawColor}{RGB}{0,0,0}

\node[text=drawColor,rotate= 90.00,anchor=base,inner sep=0pt, outer sep=0pt, scale=  0.80] at (  9.01,117.48) {ECAP};
\end{scope}
\end{tikzpicture}